\documentclass[twocolumn,english,prx,floatfix]{revtex4-2}
\usepackage{color}
\usepackage[utf8]{inputenc}
\usepackage{graphicx}
\usepackage[T1]{fontenc}
\usepackage{float}
\usepackage{natbib}
\usepackage{textcomp}
\usepackage{amsmath}
\usepackage{amssymb}
\usepackage{graphicx}
\usepackage{esint}
\usepackage{natbib}
\usepackage{xcolor}
\usepackage{multirow}
\usepackage{ulem}

\DeclareGraphicsExtensions{.png,.pdf}

\graphicspath{{./Figures/}}

\begin{document}
\title{Wigner molecules in phosphorene quantum dots}

\author{Tanmay Thakur and Bart\l{}omiej Szafran}

\affiliation{AGH University of Science and Technology, Faculty of Physics and
Applied Computer Science,\\
 al. Mickiewicza 30, 30-059 Kraków, Poland}

\begin{abstract}
We study Wigner crystallization of electron systems in phosphorene quantum dots
with confinement of an electrostatic origin with both circular and elongated geometry. The large effective masses in phosphorene promote the separation of the electron charges
already for quantum dots of relatively small size.
The anisotropy of the effective mass allows for the formation
of Wigner molecules 
 in the laboratory frame with a confined charge density
that has lower symmetry than the  confinement potential. We find that in circular quantum dots 
separate single-electron islands are formed for two and four confined electrons but not
for three trapped carriers. 
The spectral signatures of the Wigner crystallization to be resolved by transport
spectroscopy are discussed. Systems with Wigner molecule states are characterized by a nearly 
degenerate ground state at $B=0$ and are easily spin-polarized by the external magnetic field.
In electron systems for which the single-electron islands
are not formed,  a more even distribution of excited states at $B=0$ is observed, and the confined
system undergoes ground state symmetry transitions at magnetic fields of the order of 1 Tesla. 
The system of five electrons in a circular quantum dot is indicated as a special case with two charge configurations that appear
in the ground-state as the magnetic field is changed: one with the single electron islands formed in the laboratory frame and the other
where only the pair-correlation function in the inner coordinates of the system has a molecular form as for three electrons.
 The formation of Wigner molecules of quasi-1D form is easier for orientation of elongated quantum dots along the zigzag direction with heavier electron mass.
 The smaller electron effective mass along the armchair direction allows for freezing out
 the transverse degree of freedom in the electron motion. 
Calculations are performed with a version of the configuration interaction approach that 
uses a single-electron basis that is pre-optimized to account for the relatively large area
occupied by strongly interacting electrons allowing for convergence speed-up.
\end{abstract}
\maketitle
\section{Introduction}


Electron gas with Coulomb interactions dominating over
the kinetic energies forms a Wigner crystal \cite{wigner,w791,win2,win3}. 
Its finite counterparts, e.g. Wigner molecules \cite{bryla,jaula1d,eggla,shapir,diaz1d,pecker,corrigan,modaresi,yanla,reimla,fila,uzi1,szafran1d,kala,jose,cus1d,sarma1d}
are formed in quantum dots
at low electron numbers in spatially large systems \cite{bryla}
or in a strong magnetic
field that promotes the single-electron localization \cite{reimann,jain}. 

The confined charge density in quantum dots defined in materials with isotropic effective mass reproduces the symmetry of confinement potential. For this reason
in circular quantum dots, separation of the electrons in the Wigner phase occurs only in the inner coordinates
of the system spanned by relative electron-electron distances \cite{reimann}. For lowered 
symmetry, the Wigner molecules can appear in the laboratory frame \cite{wimabrok}, 
with the special case of one-dimensional systems that is studied with much of attention \cite{jaula1d,diaz1d,cus1d,szafran1d,sarma1d}. 

Phosphorene \cite{fosf14,bp,review0,rev} is a particularly interesting material for Wigner-molecule physics due to the large
electron effective masses and their strong anisotropy \cite{anisobp,18,tibikast1,tibikast2,kp,zhou,szafran1}
Large masses reduce the kinetic energy as compared to the electron-electron interaction energy.
 Lowering the Hamiltonian symmetry by the effective mass anisotropy is promising
for observation of the Wigner molecules in the laboratory frame. 

Phosphorene quantum dots \cite{Lee17,Sun15,Wang18,BPQD18,Saroka,s2,s3,pet1} 
in the form of small flakes have been extensively studied, in particular from the point of view of 
optical properties. In this work we consider a clean electrostatic confinement that keeps the confined
electrons off the edge of the flake.  In finite sheets of graphene, the edges inhibit the Wigner crystallization \cite{modaresi}.
Advanced phosphorene gating techniques have been developed \cite{rev,l20,l21,l22,gatetune}  for e.g. fabrication
of the field-effect transistors \cite{bp,he19,bl19} and experimental studies of the quantum Hall effects \cite{natyang,hhe,ehe,qhe} are carried out. Therefore, the formation of clean electrostatic quantum dots \cite{eqd} in phosphorene is 
within experimental reach.

Ordering of the electron charge in Wigner molecules of single-electron islands in quasi 1D systems \cite{jaula1d,diaz1d,cus1d,szafran1d,sarma1d}
reduces the electron-electron interaction energy at the cost of increasing the kinetic energy due to the electron localization. 
In GaAs systems with low electron band effective mass of $0.067m_0$  conditions for Wigner molecule formation occur only in very long systems
of hundreds of nanometers \cite{szafran1d} already for four electrons.
On the other hand, the light electron mass in GaAs favors the reduction of the 2D confinement to an
effectively 1D form with all the electrons occupying the same state of quantization 
for the transverse motion. Hence, the large effective masses in phosphorene are promising for producing the Wigner molecules in systems of relatively small sizes, but may inhibit formation of 1D confinement.

In this paper, we consider the formation of Wigner molecules in the laboratory frame for a few electrons confined
in circular and elongated quantum dots for varied confinement orientation and look for spectral signatures
of Wigner crystallization to be experimentally resolved.
We use the configuration interaction approach  \cite{ci,ci2,ci3,ci4,ci5} that requires an optimized single-electron basis  \cite{ci6,hfci,hfci0} for convergence
due to the strong electron-electron interaction effects \cite{fyang} in phosphorene. 

This paper is organized as follows. In the Theory section we describe the applied computational approach.
In the Results section we first describe the results for circular quantum dots and next the Wigner molecules in quasi one-dimensional confinement 
oriented along the zigzag and armchair crystal directions. Section IV contains the discussion of the experimentally accessible signatures of
the Wigner molecule formation in the laboratory frame. Summary and conclusions are given in Section V. In the Appendix we include details on the single-electron
wave functions used for optimization of the basis, the choice of the computational box, and the spectra without the Zeeman interaction.

\begin{figure*}
 \begin{tabular}{ll}
 \includegraphics[width=0.8\columnwidth]{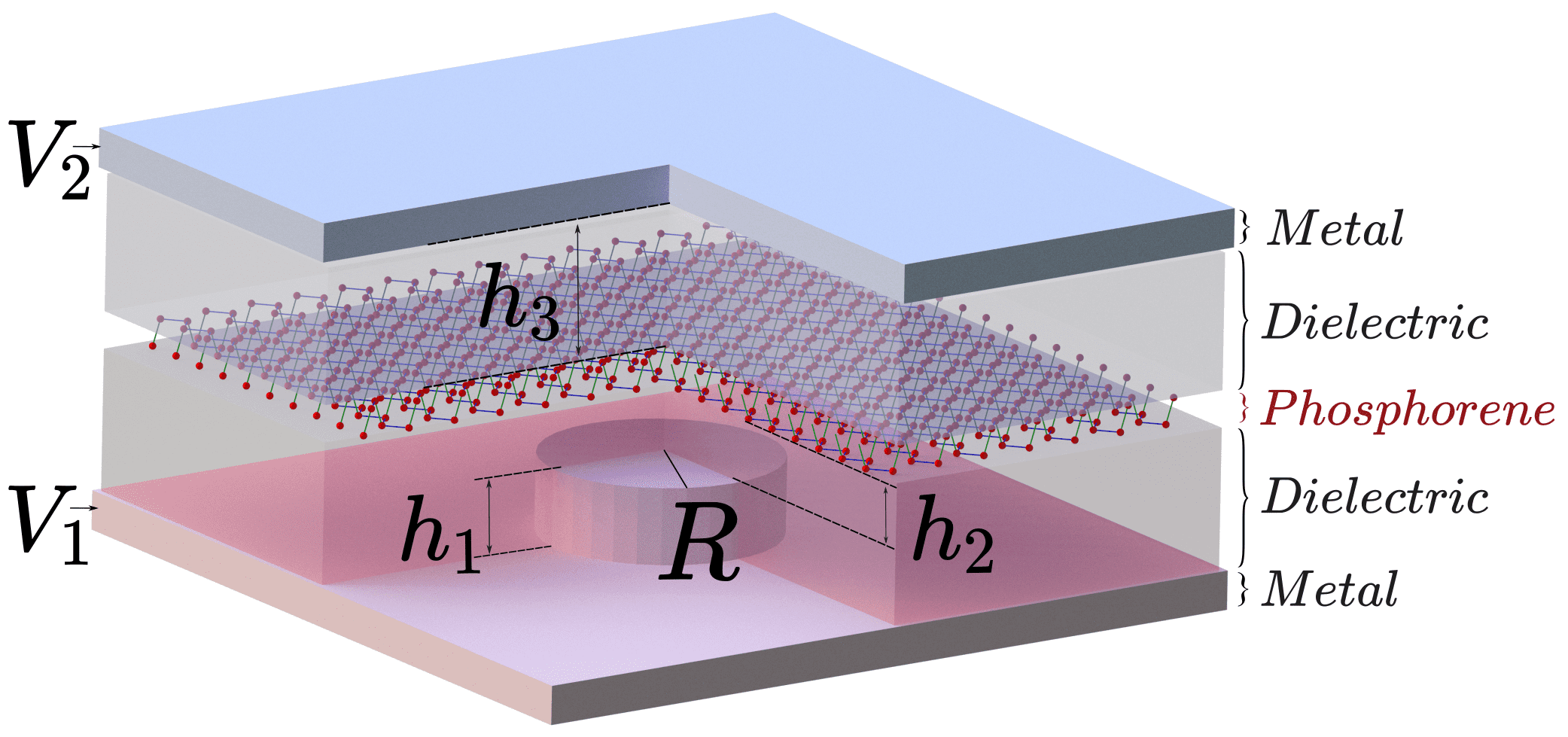}  \put(-20,15){(a)} 
& \includegraphics[width=0.8\columnwidth]{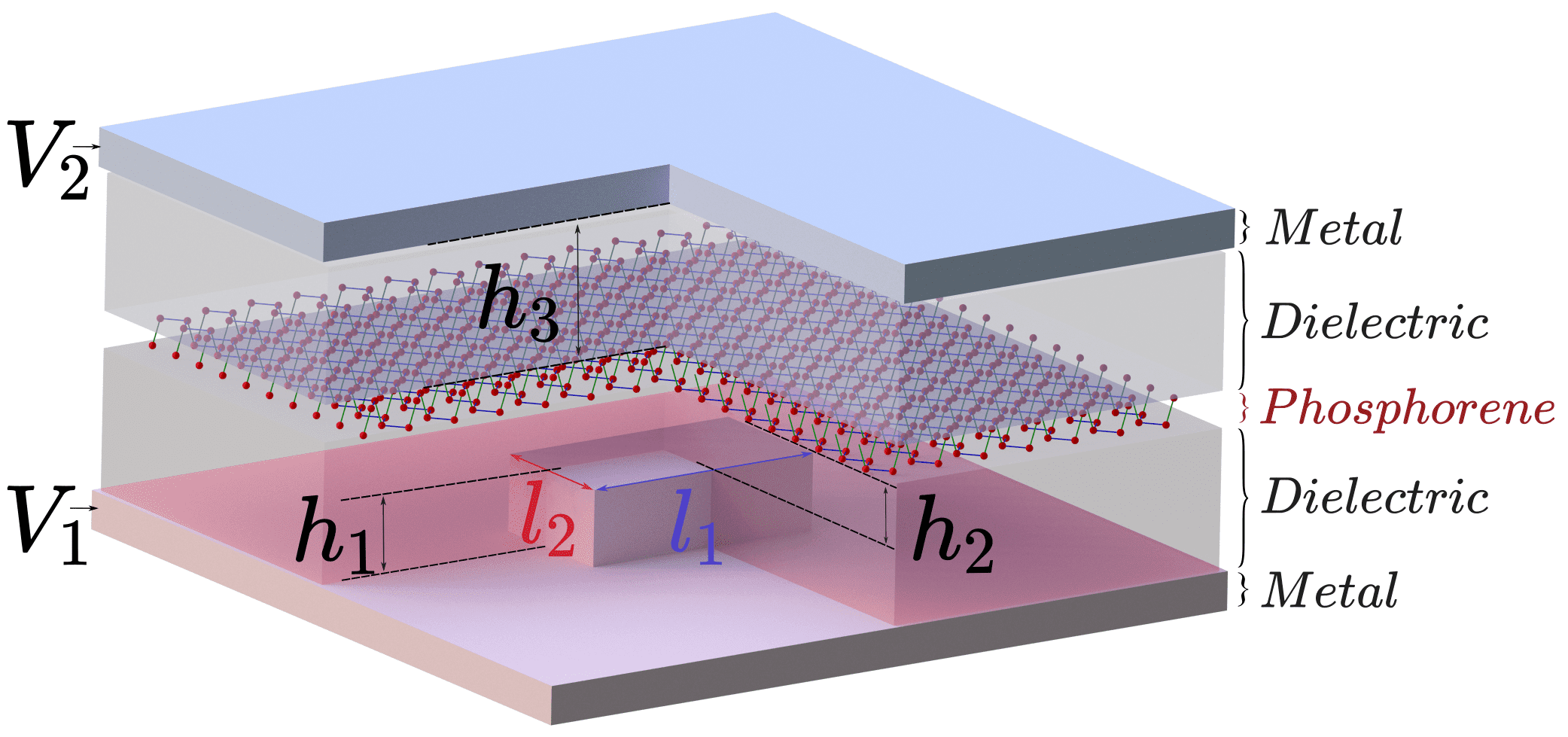}  \put(-20,15){(b)}\\
  \includegraphics[width=0.6\columnwidth]{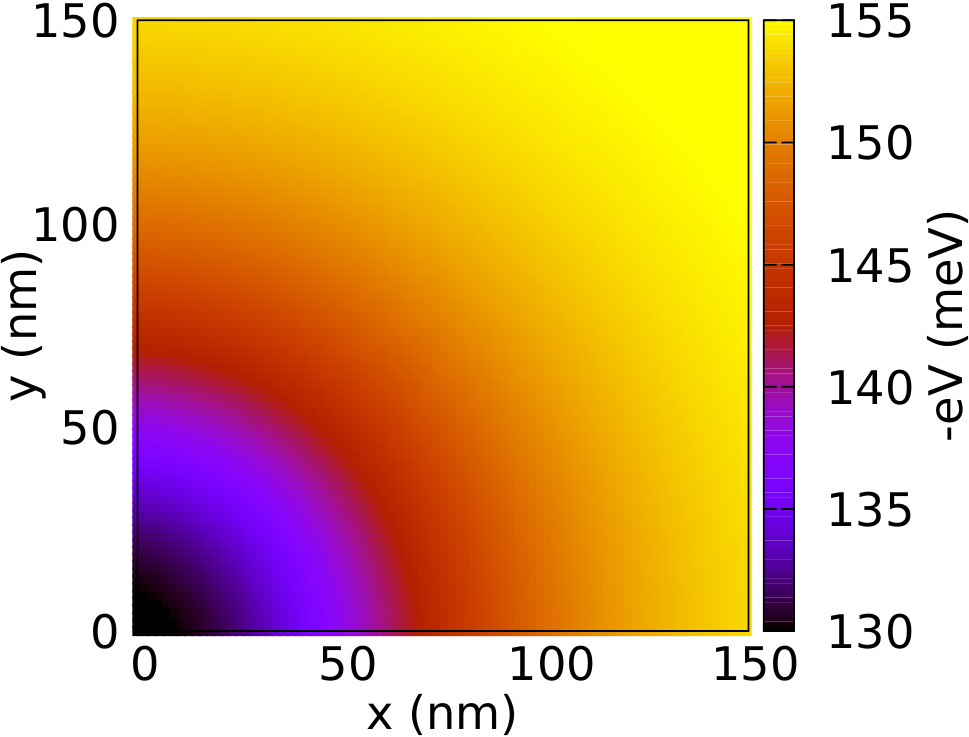} \put(-2,15){(c)} & \includegraphics[width=0.6\columnwidth]{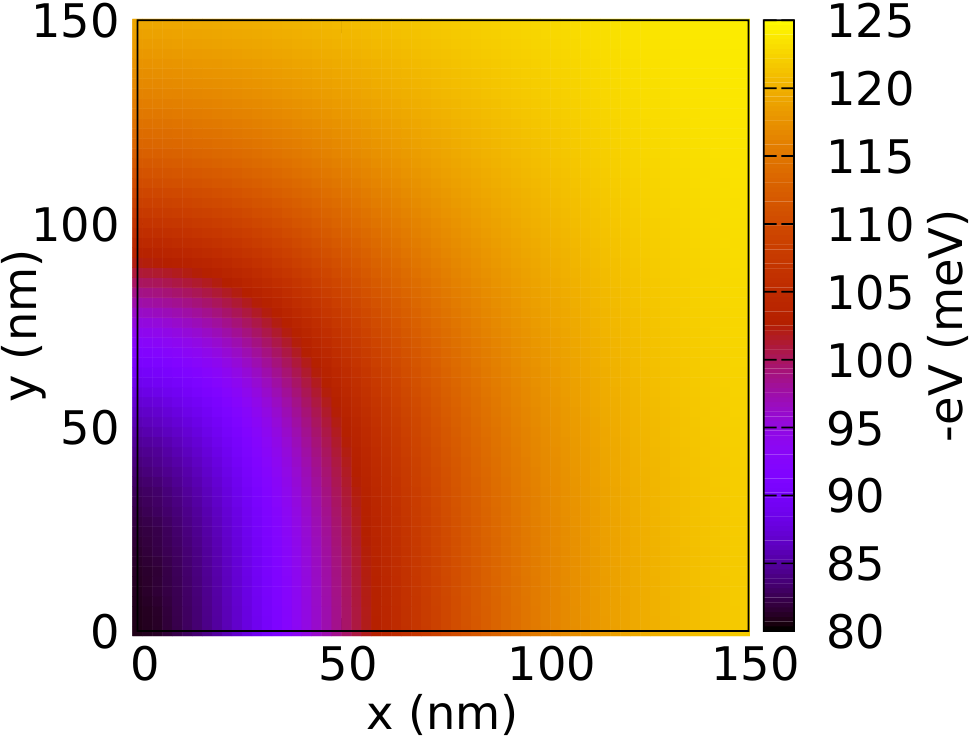} \put(-2,15){(d)}\\
 \end{tabular}
\caption{
Schematics of model systems for evaluation of the confinement potential. The phosphorene monolayer is embedded in a dielectric that fills the space between electrodes of the plane capacitor
configuration. The lower metal electrode contains a circular (a) or rectangular (b) protrusion.
The upper electrode is kept at a higher potential energy $-eV_2=0.25$ eV for the electrons than the lower one $-eV_1=0$. The protrusion introduces an inhomogeneity of the electric field within the capacitor that forms the confinement potential for the electrons of the conduction band. 
We use $h_1+h_2+h_3=150$ nm and $h_1=50$ nm. 
For the circular protrusion (a) we take $h_2=50$ nm and for the rectangular one (b) $h_2=30$ nm.
The radius of the protrusion in (a) is $R=50$ nm, and the sides of the rectangle in (b) have lengths
$l_1=80$ nm and $l_2=30$ nm. 
The confinement potential on the phosphorene plane is plotted in (c) and (d) for the circular
and rectangular protrusions, respectively. The origin in (c,d) is the symmetry center of the protrusion.
}
 \label{sys}
\end{figure*}
 
\section{Theory}
In this section we describe the finite difference method applied to the continuum Hamiltonian of a single
electron in phosporene (subsection II.A), the model potential (II.B) and the configuration interaction approach (II.C) with optimization of the single-electron basis allowing for faster convergence of the configuration interaction approach. Subsection II.D describes the formula for extraction of the charge density and pair correlation functions for discussion of the Wigner crystallization of the confined system.
\subsection{Single-electron Hamiltonian}
We use the single-band approximation for Hamiltonian describing the electrons of the conduction band
of monolayer phosphorene \cite{szafran1,tanmay} 
\begin{eqnarray}
H_0&=&
 {\left(-i\hbar \frac{\partial }{\partial x}+e{A_x}\right)^2}/\, {2m_x}
+{\left(-i\hbar \frac{\partial }{\partial y}+e{A_y}\right)^2}/\, {2m_y} \nonumber \\ &+&W(x,y)+{g\mu_B B\sigma_z}/{2} \, , \label{1eh}
\end{eqnarray}
where $W(x,y)$ is the confinement potential.
In Eq. (\ref{1eh}) we use the effective masses
$m_x=0.17037m_0$ for the armchair crystal direction ($x$) and $m_y=0.85327m_0$ for the zigzag direction ($y$). The values for the masses  were determined in Ref. \cite{szafran1} by fitting the confined energy spectra of the continuum single-band Hamiltonian to the results of the tight-binding method. 
A detailed comparison of the spectra as obtained by the continuum model to the tight-binding ones is given in Ref. \cite {szafran1} for the harmonic oscillator potential and in Ref. \cite{tanmay} for 
the annular confinement. 
In Eq. (\ref{1eh}) we take the Land\'e factor $g=g_0=2$ after the k$\cdot p$ theory of Ref. \cite{Fabbian}.
The values of experimentally extracted $g$-factors vary; in particular an increase with respect to $g_0$ was reported \cite{fyang}
at low filling factors, which is attributed  \cite{Fabbian,fyang}
to strong electron-electron interaction effects in black phosphorus. The electron-electron interaction in this work is treated
in an exact manner. The spin Zeeman term leads to the spin polarization of
the confined system. The exact value of the magnetic field producing the spin polarization
is affected by the adopted $g$-factor value, but no qualitative effect for the Wigner crystallization of the charge density is expected as long as the spin-orbit coupling is absent.
The spectral features of Wigner crystallization for $g=0$ are discussed in the Appendix.

We work with a square mesh with a spacing $\Delta x$ in both the $x$ and $y$ directions.
The Hamiltonian acting on the wave function $\Psi_{\mu,\eta}=\Psi(x_\mu, x_\eta)=\Psi(\mu \Delta x, \eta \Delta x)$
in the finite-difference approach reads
\begin{eqnarray}
H_0 \Psi_{\mu,\eta}&\equiv& \frac{\hbar^2}{2m_x \Delta x^2}\left(2\Psi_{\mu,\eta}-C_y \Psi_{\mu,\eta-1}-C_y^*\psi_{\mu,\eta+1}\right)\nonumber \\
&+&\frac{\hbar^2}{2m_y \Delta x^2}\left(2\Psi_{\mu,\eta}-C_x \Psi_{\mu-1,\eta}-C_x^*\psi_{\mu+1,\eta}\right)\nonumber \\ &+& W_{\mu,\eta}\Psi_{\mu,\eta}+\frac{g\mu_B B}{2}\sigma_z \, , \label{numei}
\end{eqnarray}
where $C_x=\exp(-i\frac{e}{\hbar}\Delta x A_x)$ and $C_y=\exp(-i\frac{e}{\hbar}\Delta x A_y)$
introduce the Peierls phases \cite{governale} for the description of the orbital effects of the perpendicular magnetic field $(0,0,B)$. For calculation of the phase shifts, we use the symmetric gauge ${\bf A}=(A_x,A_y,A_z)=(-By/2,Bx/2,0)$.
Hamiltonian (2) is diagonalized in a finite computational box with the infinite quantum well set
at the end of the box (see the Appendix).

\subsection{Model potential}
For evaluation of a realistic confinement potential $W$ we use a simple model with a phosphorene plane
embedded in a Al$_2$O$_3$ dielectric that fills the area between two parallel electrodes [Fig. \ref{sys}(a,b)]. 
A higher (lower) potential energy for electrons is introduced at the top (bottom) electrode. 
The bottom electrode is grounded and contains a protrusion that approaches the phosphorene layer. 
As a result, the electrostatic potential within phosphorene forms a cavity that traps the electrons of the conduction band.
The model is a variation \cite{bednarek,szafran2} of a gated GaAs quantum dot of Ref. \cite{Ashoori}. 
Below we use two models:  one with a circular protrusion (Fig. \ref{sys}(a)) and the other with a rectangular one (Fig. \ref{sys}(b)). The latter is used in the following to study the case close to the 1D confinement.
The confinement potential to be used in the Hamiltonian is given by $W(x,y)=-eV(x,y,z_p)$, with the electrostatic potential $V$ that  we evaluate by solving the Laplace equation $-\nabla^2 V=0$ and $z_p$ is the coordinate of the phosphorene layer.
For evaluation of the potential we use the finite element method similar to the one applied in Ref. \cite{szafran2} for a charge-neutral phosphorene plane. 
The confinement potential at the monolayer is plotted in Fig. \ref{sys}(c) for the circular protrusion of Fig. \ref{sys}(a)
and in Fig. \ref{sys}(d) for the rectangular protrusion of Fig. \ref{sys}(b).

 \begin{figure}
 \begin{tabular}{l}
 \includegraphics[width=0.79\columnwidth]{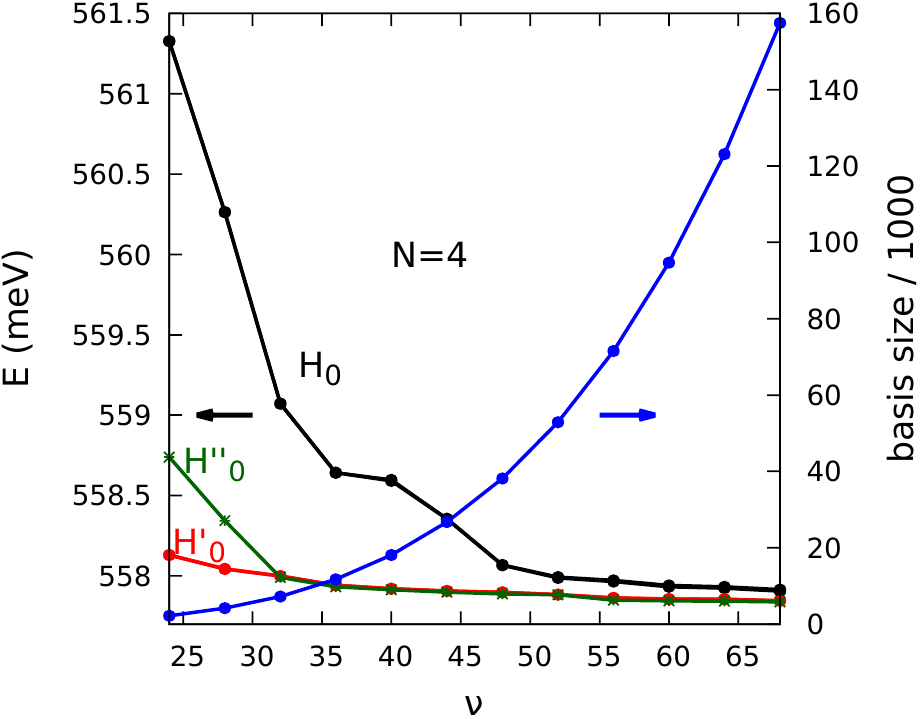} \put(-50,35){(a)} \\ \;\;\; \includegraphics[width=0.63\columnwidth]{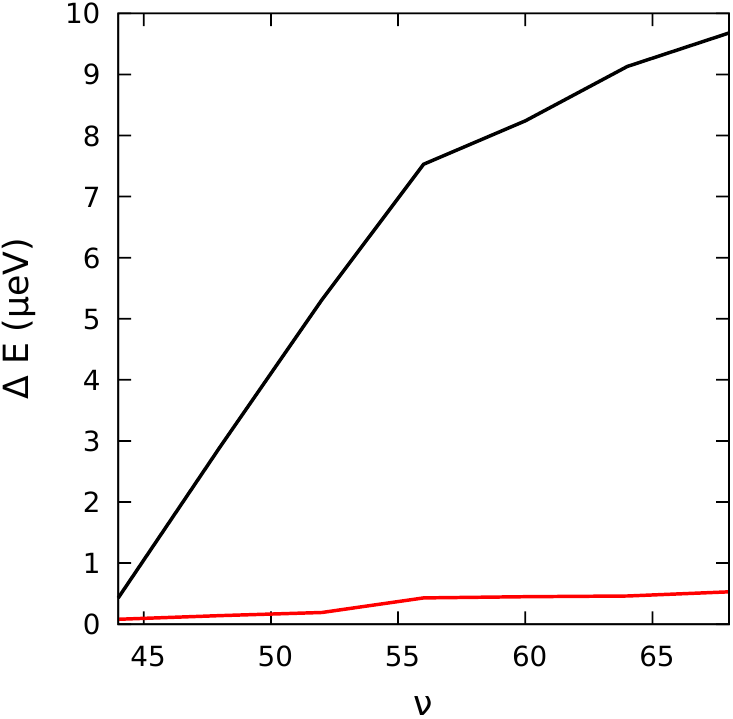} \put(-30,35){(b)} 
 \end{tabular}
  \begin{tabular}{lll}
 \includegraphics[width=0.34\columnwidth]{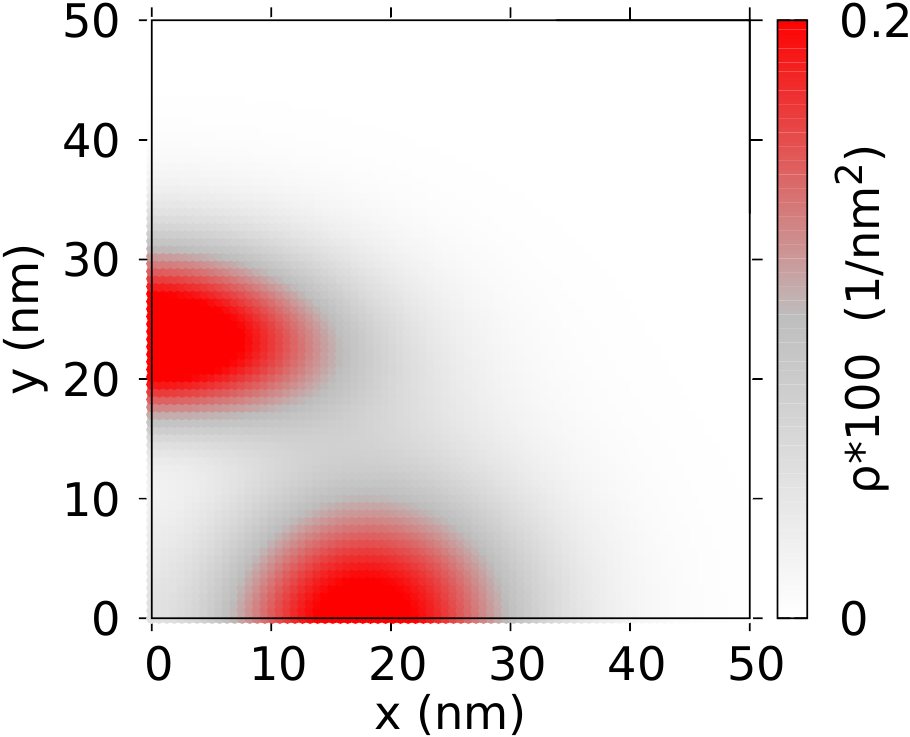} \put(-55,55){$H_0/24$   (c)}  &  \includegraphics[width=0.34\columnwidth]{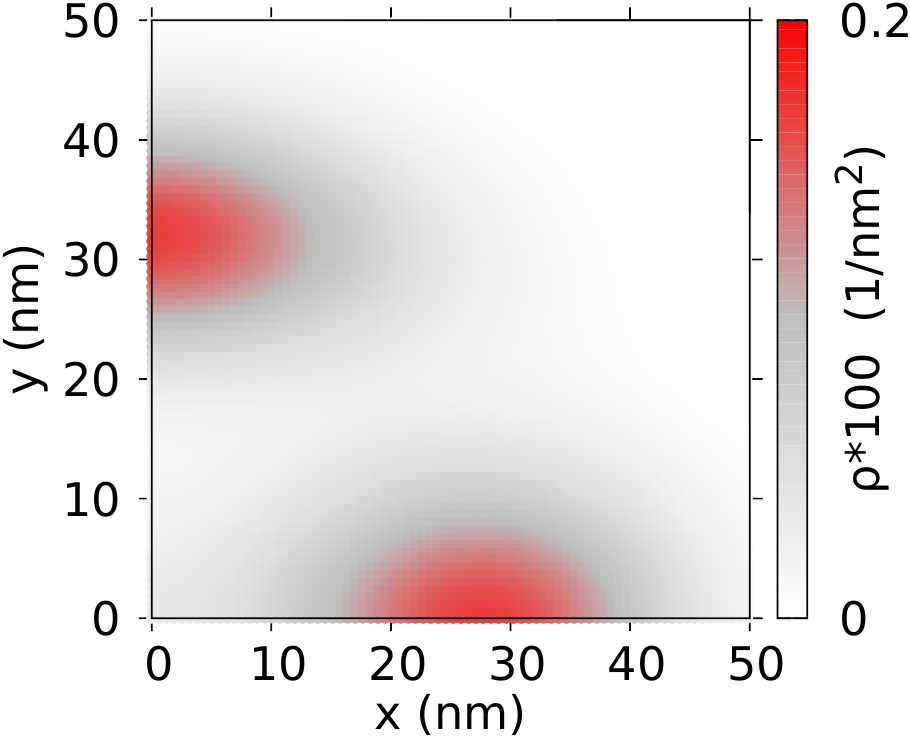}\put(-57,55){$H_0'/24$  (d)}&  \includegraphics[width=0.34\columnwidth]{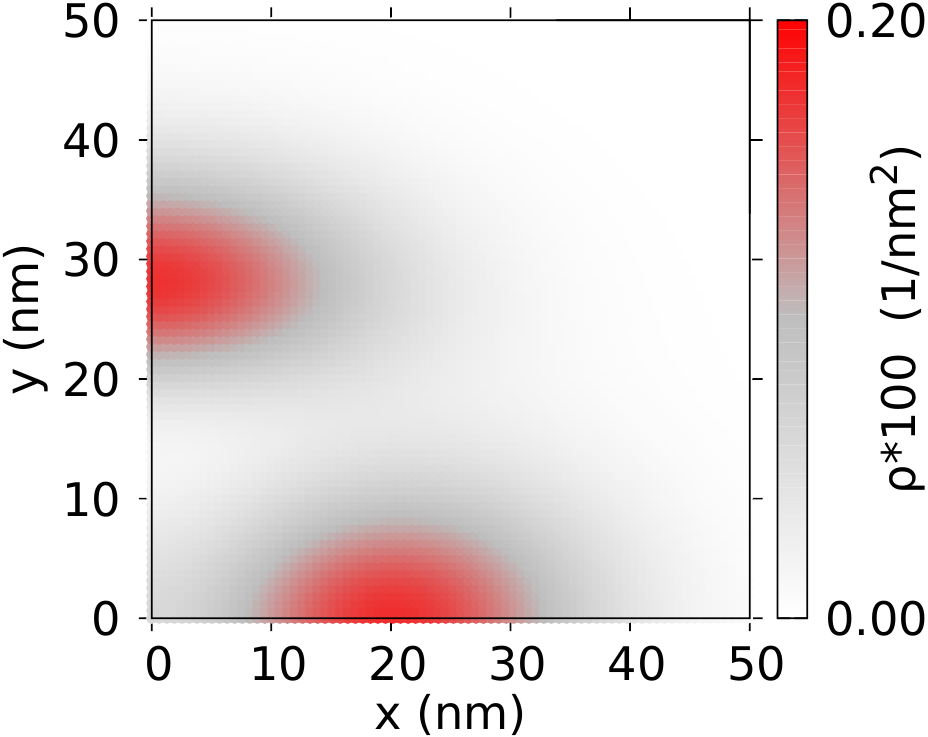}\put(-60,55){$H_0''/24$ (e)}\\
  \includegraphics[width=0.34\columnwidth]{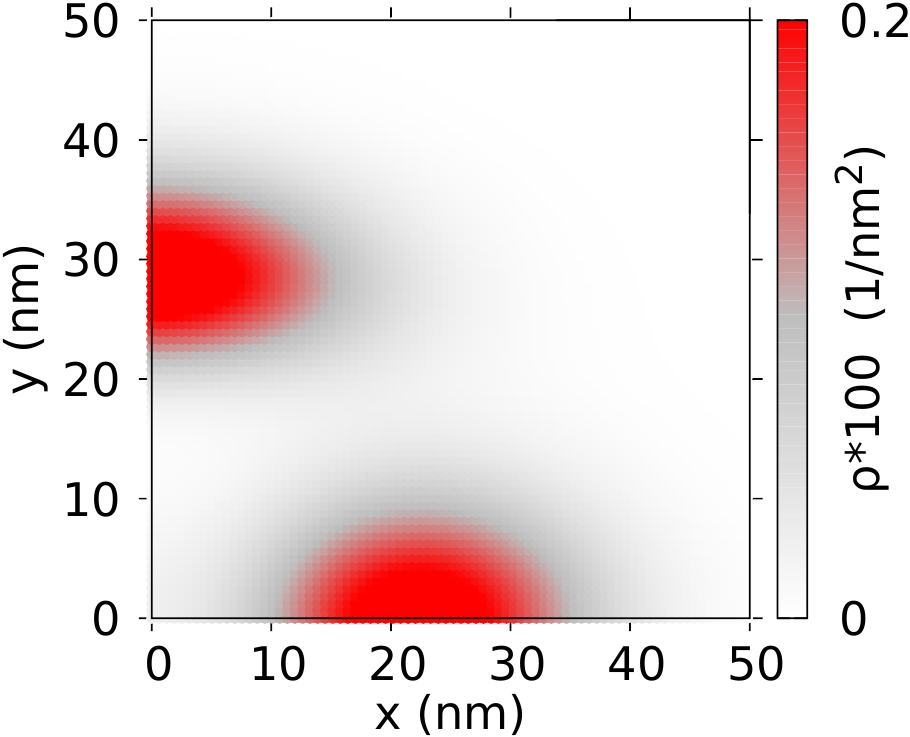}\put(-55,55){$H_0/40$   (f)} &  \includegraphics[width=0.34\columnwidth]{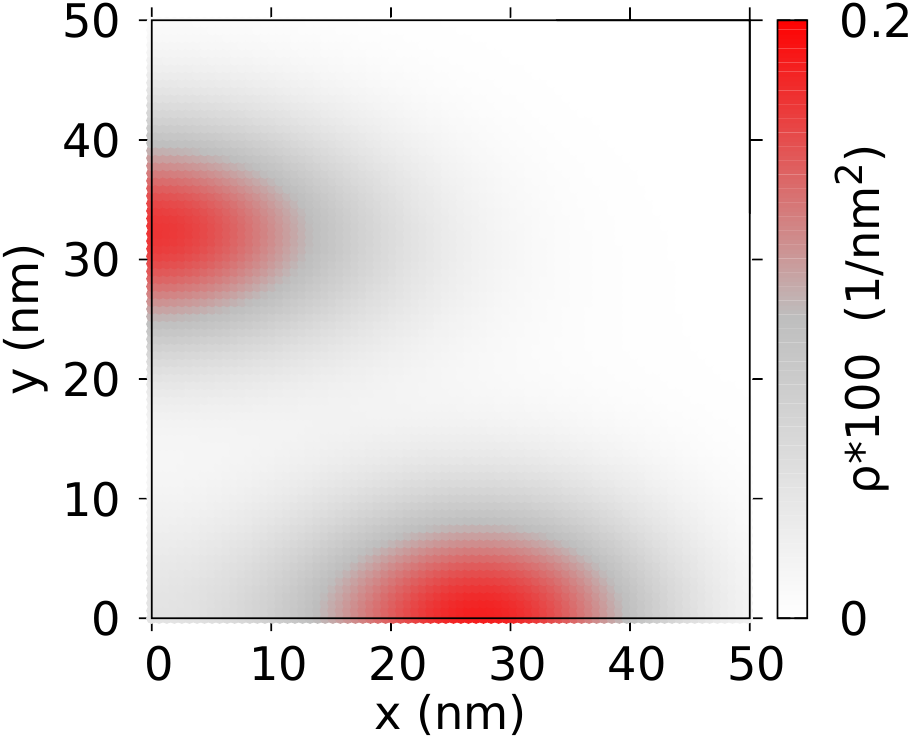} \put(-57,55){$H_0'/40$  (g)}&  \includegraphics[width=0.34\columnwidth]{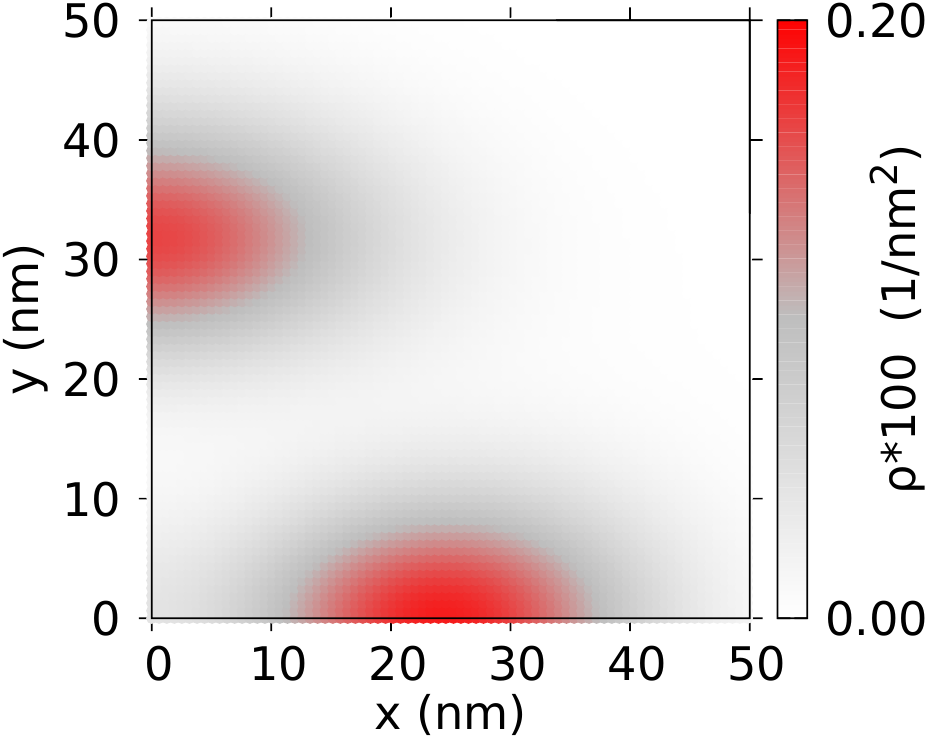}\put(-60,55){$H_0''/40$ (h)}\\
   \includegraphics[width=0.34\columnwidth]{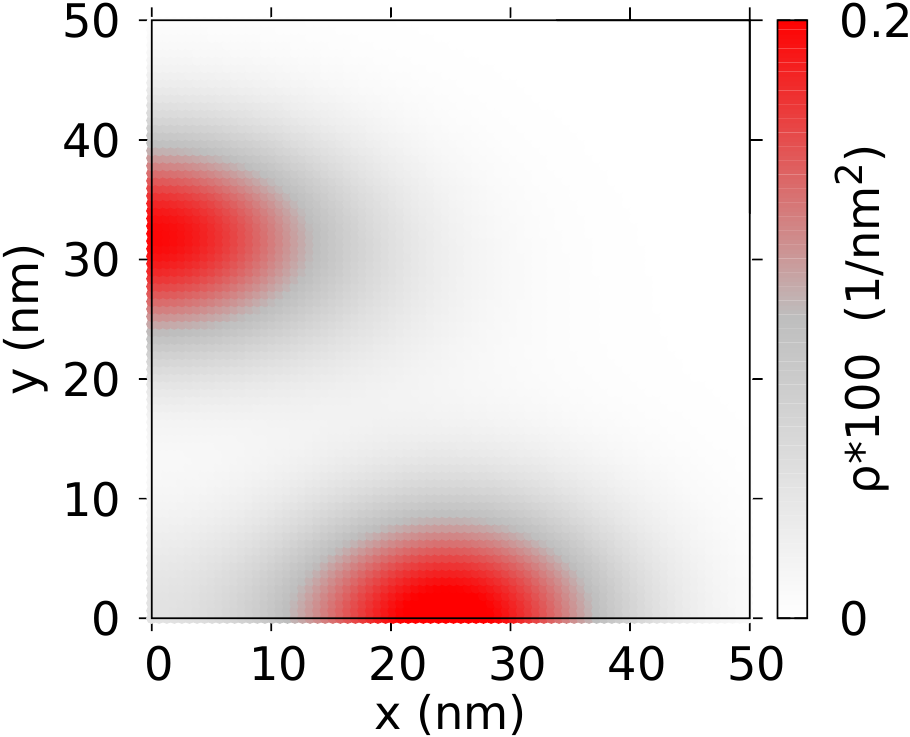} \put(-55,55){$H_0/60$   (i)}&  \includegraphics[width=0.34\columnwidth]{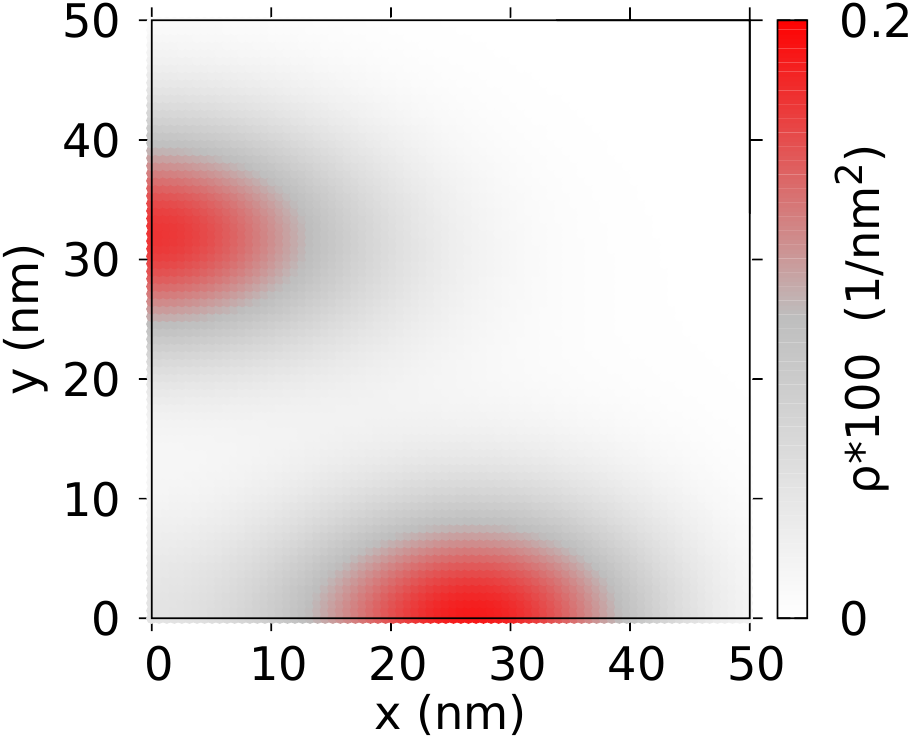}\put(-57,55){$H_0'/60$  (j)} &  \includegraphics[width=0.34\columnwidth]{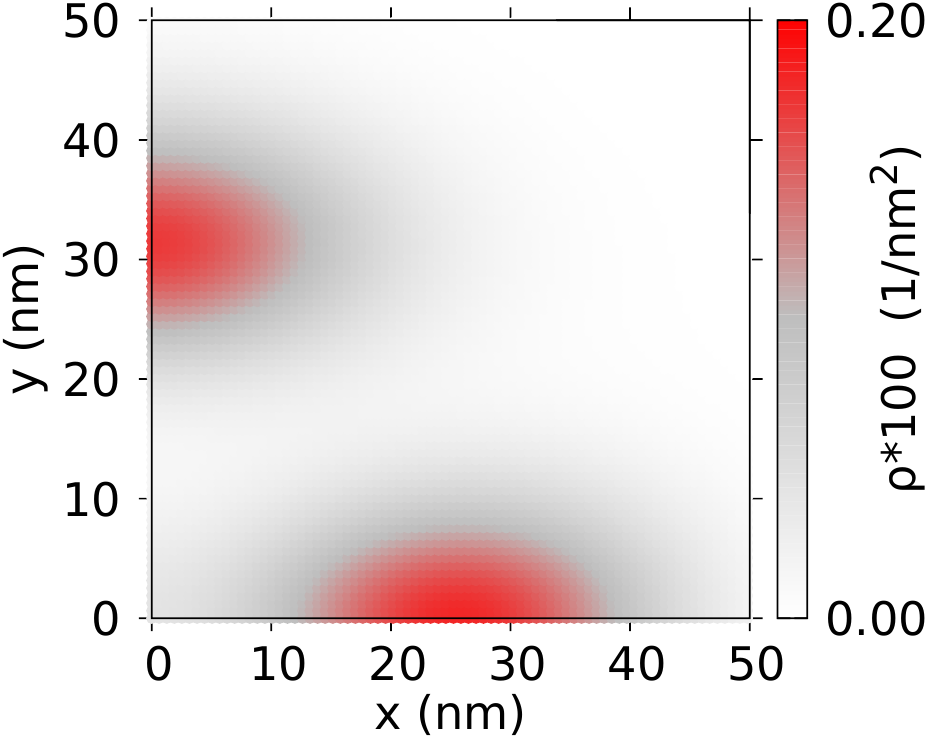}\put(-60,55){$H_0''/60$ (k)}
  \end{tabular}
\caption{
(a) Ground-state energy for 4 electrons confined in circular quantum dot as a function of the number
of single-electron eigenstates used for construction of the basis for $B=0$. For the black (red)
line the eigenstates of the single-electron Hamiltonian $H_0$ (Hamiltonian $H_0'$) with the bare confinement (confinement with a central Gaussian with $V_0=8$ meV and $\sigma=35$ nm the were used. 
(b) Energy shift that appears when the states with triple or quadruple excitations above the single-electron
$32$-nd energy level are excluded from the basis constructed using $H_0$ (black) and $H_0'$ (red) eigenstates.
(c-k) The ground-state charge density in the $x>0, y>0$ quarter of the QD as obtained
using the Slater determinants basis constructed with $H_0$ (c,f,i), $H_0'$ (d,g,j) and
$H_0''$ (e,h,k) Hamiltonians using $\nu=24$ (c-e), $\nu=40$ (f-h) and $\nu=60$ (i-k)
lowest-energy single electron wave functions.
}
 \label{conv}
\end{figure}
\subsection{Diagonalization of the $N$-electron Hamiltonian}
The system of $N$-confined electrons is described with the Hamiltonian
\begin{equation}
H_N=\sum_{i=1}^N H_0(i)+\sum_{j>i}^N \frac{e^2}{4\pi\epsilon_0\epsilon \, r_{ij}} \, . \end{equation} 
We take the dielectric constant $\epsilon=9$ assuming that the phosphorene is embedded in Al$_2$O$_3$.

In the standard configuration-interaction method \cite{ci,ci2,ci3,ci4,ci5}
the $N$-electron Hamiltonian is diagonalized on the basis of Slater determinants constructed with the single-electron Hamiltonian $H_0$ eigenstates.
Each of the Slater determinants defines a configuration, e.g. a distribution of electrons over the single-electron states. The number of Slater determinants to be used in the calculation is established by a study of the
convergence of the energy estimates. Reaching convergence in the present calculation is challenging because of the strong electron-electron
interaction in phosphorene \cite{fyang}.
The energy of the ground-state estimated for $N=4$ at $B=0$ 
in the circular potential of Fig. \ref{sys}(c) is plotted with the black line in Fig. \ref{conv}(a) as a function of the number of the lowest-energy single-electron states $\nu$
that span the Slater determinant basis.
 The Hamiltonian $H_N$ commutes with the operator
of the $z$ component of the total spin and also with the parity operator due to the point symmetry of the potential.
The symmetries allow for a few-fold reduction of the number of basis elements.
In  Fig. \ref{conv} the four-electron ground state at $B=0$
is the spin singlet $S_z=0$ of an even spatial parity. Only Slater determinants of these symmetries
contribute to the ground-state wave function. For $\nu=60$ 
 the Slater determinants set  counts $\binom{60}{4}=$487 635 elements,
 of which only about 94 500 determinants correspond to $S_z=0$ and either even or odd parity.
The right vertical axis in Fig. 2(a) shows the number of Slater determinants of the spin-parity symmetry 
that are compatible with and contribute to the ground state. 

The convergence of the CI method using the $H_0$ single-electron eigenstates (black line in Fig. \ref{conv}(a)) is slow. The electron-electron interaction has a pronounced effect on the electron localization, since the electrons in phosphorene are quite heavy as compared to those in e.g. GaAs, and the deformation of the charge
density in terms of the single-electron energy is cheap. This in turn results
in a high numerical cost of the convergent calculations that require a large number
of single-electron states to be included in the convergent basis.
Therefore, due to the strong electron-electron interaction, the set of $H_0$ eigenstates is not the best starting point
for a convergent CI calculation. The literature  indicates
a number of methods to speed-up the convergence, including 
the HF+CI method \cite{ci6,hfci,hfci0} where the basis
for the CI method is based on the Hartree-Fock single-electron spin-orbitals. 
In the HF+CI approach, the mean-field effects of the electron-electron interaction are accounted for 
already in the single-electron basis and the CI is responsible only for description of the electron-electron correlation effects that evade the mean-field treatment.
The HF charge density in the unrestricted version of the method breaks the symmetry of the confinement potential and its restoration is challenging \cite{chal1,chal2} on its own at the CI stage.  
Convergence speed-up by the choice of the single-electron basis is achieved \cite{no1,no2} with the natural orbitals \cite{no,uzi2} introduced by L\"owdin \cite{no0}.

In this work, we apply a a simple approach that allows for the convergence speed-up by
replacing the potential $W$ in $H_0$ by another potential that produces single-electron wave
functions of the low-energy spectrum that cover a larger area than the ones for the bare
potential $W$.  For preparation of the single-electron basis we diagonalize the single-electron Hamiltonian $H_0'$ with the potential 
\begin{equation} W'=W+V_0\exp(-(x^2+y^2)/d^2).\end{equation}

The $H_0'$, eigenfunctions are used for the Slater determinants to diagonalize the Hamiltonian $H_N$.
 $V_0$ and $d$ of Eq. (4) are used as variational parameters in terms of the $N$-electron energy \cite{dostajemy}.
The results for 4 electrons and the basis of the $H_0'$ eigenstates
for optimized Gaussian parameters given by the red line in Fig. \ref{conv}(a) 
exhibit a substantial convergence speed-up with respect to $H_0$ basis. 
In particular, the basis of $H_0'$ eigenstates with $\nu=36$ and about 12 thousand
elements produce a similar ground-state energy estimate as the $H_0$ basis with $\nu=52$
and as much as about 53 thousand Slater determinants.

Figure \ref{conv}(a) contains also the results obtained with eigenfunctions
of Hamiltonian $H_0''$ (green line) using  potential
$ W''(x,y)=W(x)/s $, where $s$ is the scaling factor of the bare potential
with its variationally optimal value of $s=2.13$. 
The scaling enlarges the area covered by the low-energy single-electron wave
functions in a manner that becomes equivalent in terms of the 4-electron ground-state energy for the one using $W'$ potential
for $\nu>32$. The low-energy single-electron wave functions for $W$, $W'$ and $W''$ potentials
are given in the Appendix.

Figure \ref{conv}(c-k) shows the ground-state charge density 
in the $x>0$, $y>0$ quarter of the QD as obtained
for $\nu=24$ (first row of plots), $\nu=40$ (second row of plots) and
$\nu=60$ (third row of plots) with the $H_0$ (left column), $H_0'$ (central column) 
and $H_0''$ (right column) eigenfunctions. 
For $\nu=60$ the results are similar for all the three bases. 
For lower $\nu$ the results for $H_0$ (Fig. \ref{conv}(c,f)) and $H_0''$ (Fig. \ref{conv}(e))
the islands appear closer to the origin than in the convergent result.


In order to illustrate the role of the modified potential in the description of the electron-electron interaction,
we plotted in Fig. \ref{conv}(b)  the energy overestimate that is obtained once
 the basis of Slater determinants is reduced by exclusion of all
 configurations with more than two electrons \cite{ci6} above the $32$-nd single-electron energy level.
For the Hamiltonian $H_0$ the cost of the limited basis is much larger than for $H_0'$ and grows fast with $\nu$.  For $H_0'$ the overestimate is much lower. The single-electron effects due to $W'$ potential are included in the basis, and the double excitations that stay in the basis cover most of the electron-electron correlation effects. A similar result is obtained in the HF+CI method \cite{hfci,hfci0}.

\subsection{Charge density and pair correlation function}
For analysis of the electron localization, we extract the charge density  and the pair correlation function from the $N$-electron wave 
function $\Psi$. The charge density is obtained as \begin{equation} \rho({\bf r})=\langle \Psi| \sum_{i=1}^N \delta({\bf r}_i-{\bf r})|\Psi \rangle \,. \end{equation}
The pair correlation function extracts the relative localization of the electrons with one of the carrier positions
fixed \begin{equation} \rho_{12}({\bf r},{\bf r}_f)=\langle \Psi| \sum_{i,j=1}^N \delta({\bf r}_i-{\bf r})\delta({\bf r}_j-{\bf r}_f)|\Psi \rangle \,.\end{equation}
In the following we fix the position ${\bf r}_f$ of one of the electrons near the local charge density maximum for a discussion of $\rho_{12}$ plots.
The spin density can be calculated as \begin{equation} \rho^{\alpha_i}({\bf r})=\langle \Psi| \sum_{i=1}^N \delta({\bf r}_i-{\bf r})|\alpha_i\rangle\langle \alpha_i||\Psi \rangle \,. \end{equation}
where the projection on the spin eigenstates uses $|\alpha_i\rangle $ that stands for the single-electron spin eigenstate for $i$-th electron.
Similarly, the relative localization including the spin configuration of the electron pair can be included in the $\rho_{12}$ pair-correlation function. In particular, opposite spin distribution can be obtained
using spin projections
\begin{eqnarray} \rho_{12}^{\uparrow\downarrow}({\bf r},{\bf r}_f)&=&\langle \Psi| \sum_{i,j=1}^N \delta({\bf r}_i-{\bf r})\delta({\bf r}_j-{\bf r}_f)\times\nonumber\\&\times& (|\alpha_i\beta_f \rangle\langle \alpha_i \beta_f|
+|\beta_i\alpha_f \rangle\langle \beta_i \alpha_f|)\Psi \rangle \,,\end{eqnarray}
with the spin eigenstates $|\alpha\rangle\neq|\beta\rangle$ .

\section{Results}
In this section we first (Section III.A) discuss case for the circular external potential 
where the formation of Wigner molecules in the laboratory frame occurs due to anisotropy of the effective mass.
Section III.B contains the results for an elongated confinement potential near the quasi 1D confinement limit.
 \begin{figure}
 \begin{tabular}{l}
  \includegraphics[width=0.7\columnwidth]{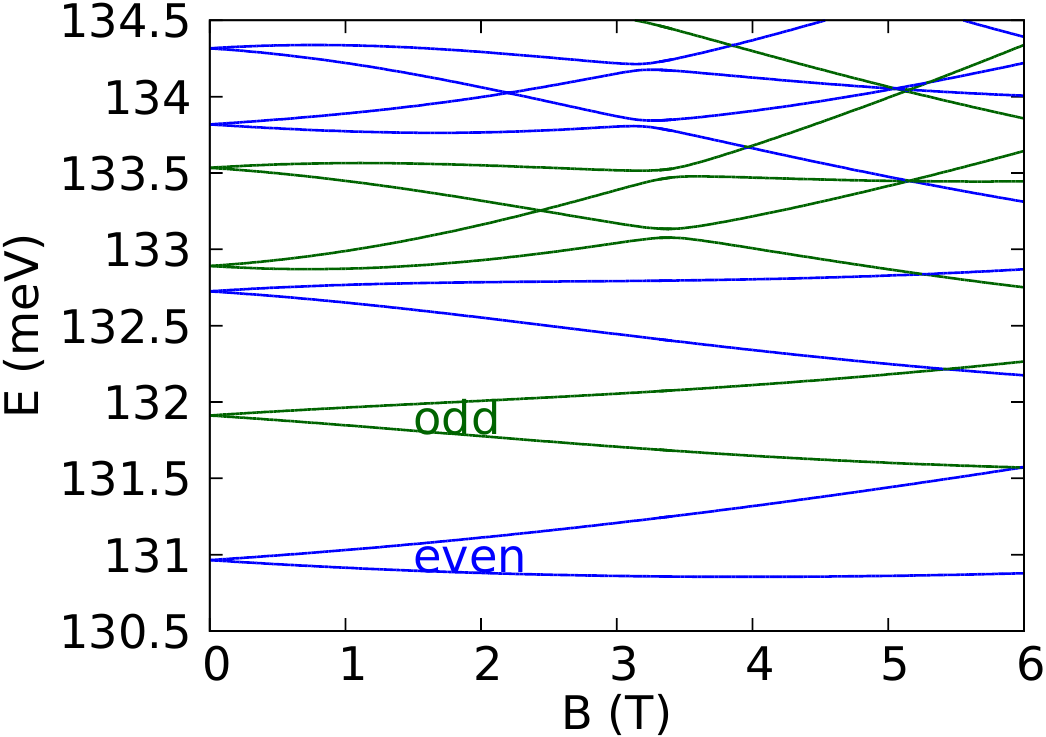} \put(-20,30){(a)} \put(-30,51){N=1}\\
 \includegraphics[width=0.75\columnwidth]{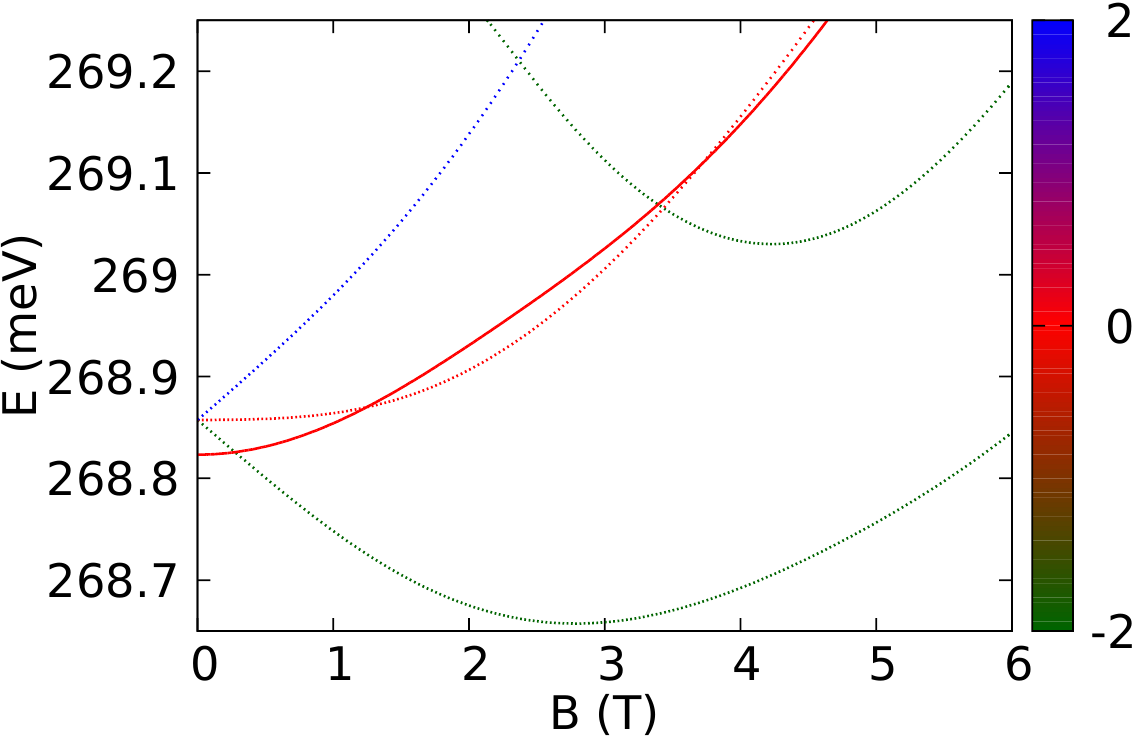} \put(-33,25){(b)}\put(-42,55){N=2} \\
 \includegraphics[width=0.75\columnwidth]{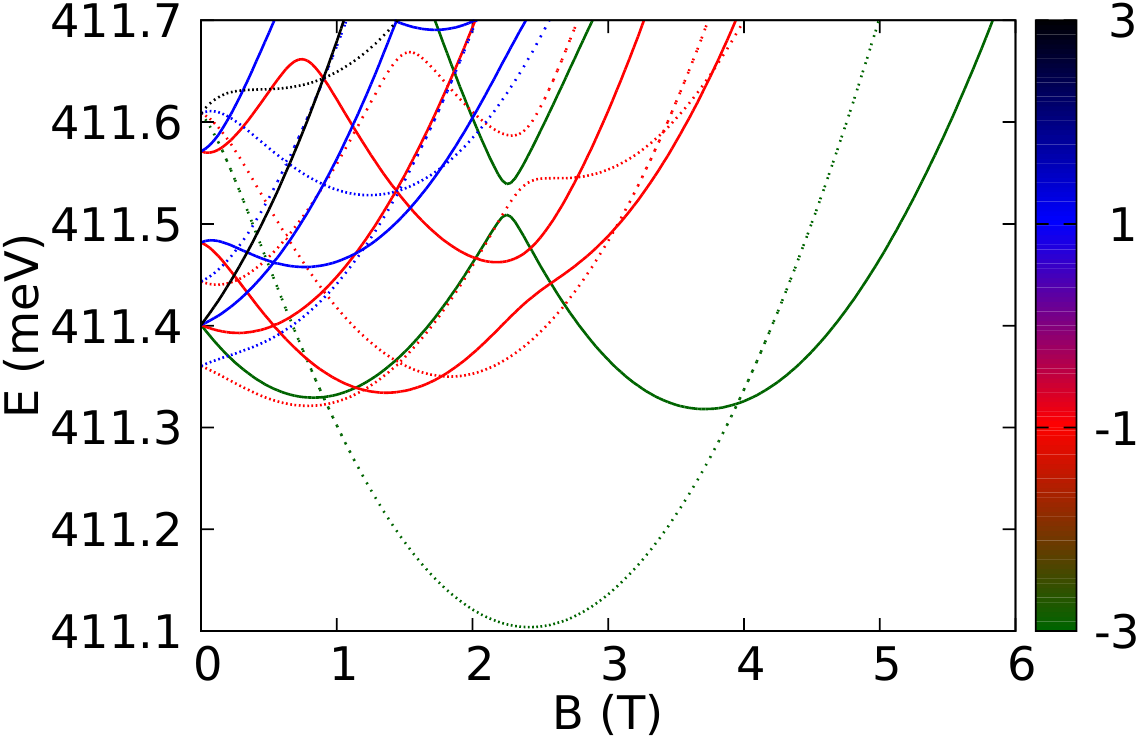} \put(-33,25){(c)}\put(-42,45){N=3} \\ 
 \includegraphics[width=0.75\columnwidth]{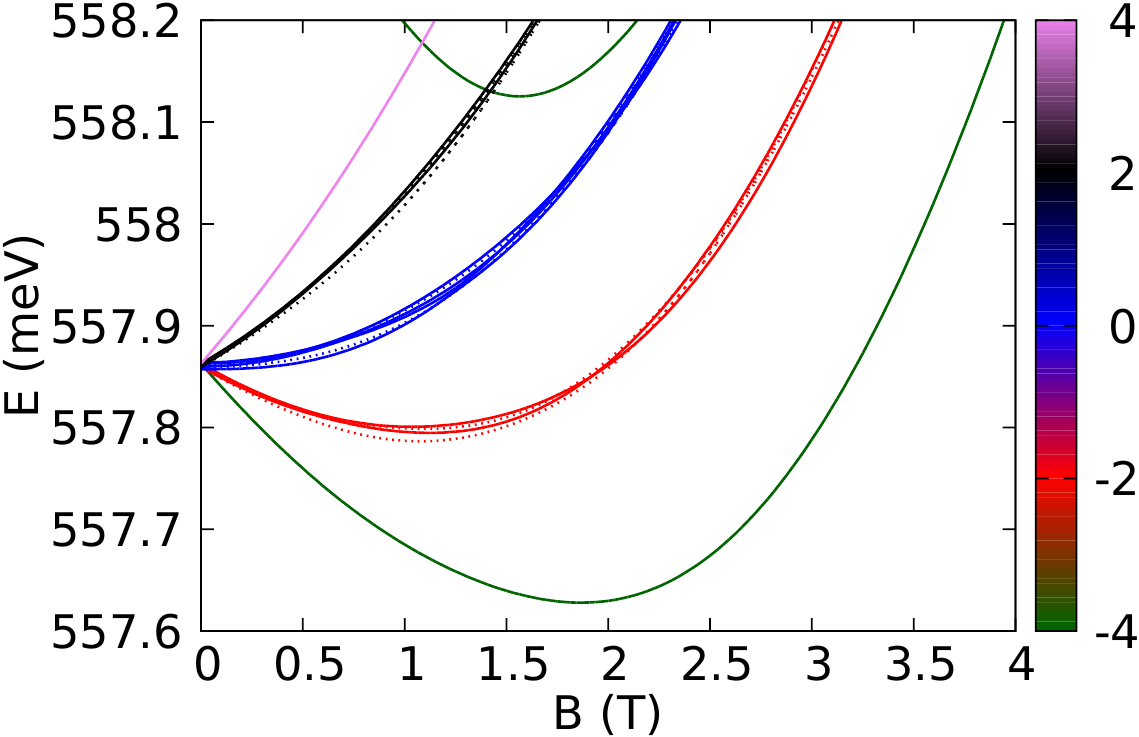}  \put(-33,25){(d)} \put(-42,55){N=4}\\
  \includegraphics[width=0.75\columnwidth]{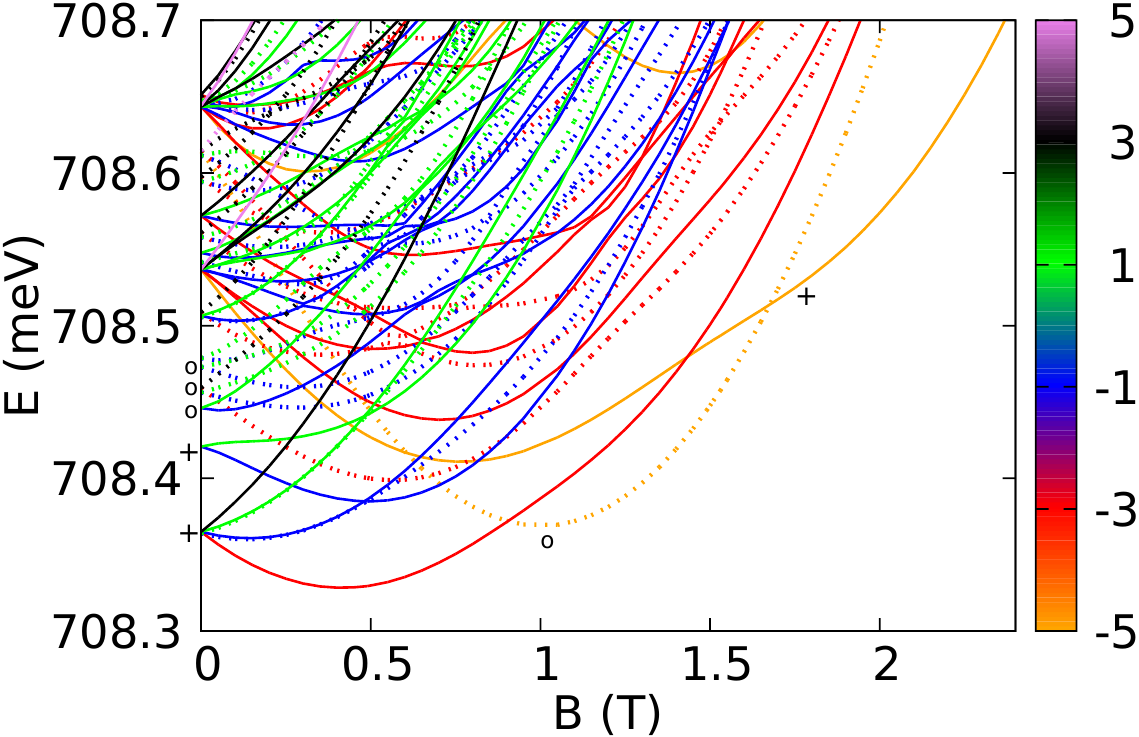} \put(-33,25){(e)} \put(-42,55){N=5}\\
 \end{tabular}\caption{Spectra for circular potential for $N=1$ (a), 2 (b), 3 (c), 4 (d), and 5 (e) electrons. 
In (a) even and odd parity energy levels are plotted with blue and green lines, respectively.
In (b-e) the even parity energy levels are plotted with the solid lines and the odd parity
energy levels with the dotted lines. In (e) the symbols of '+' and 'o' mark the states with (1,4) and (0,4) charge
configurations, respectively (see text).
The color scale is separate for each figure and given to the right of the plot.
}
 \label{scirc}
\end{figure}
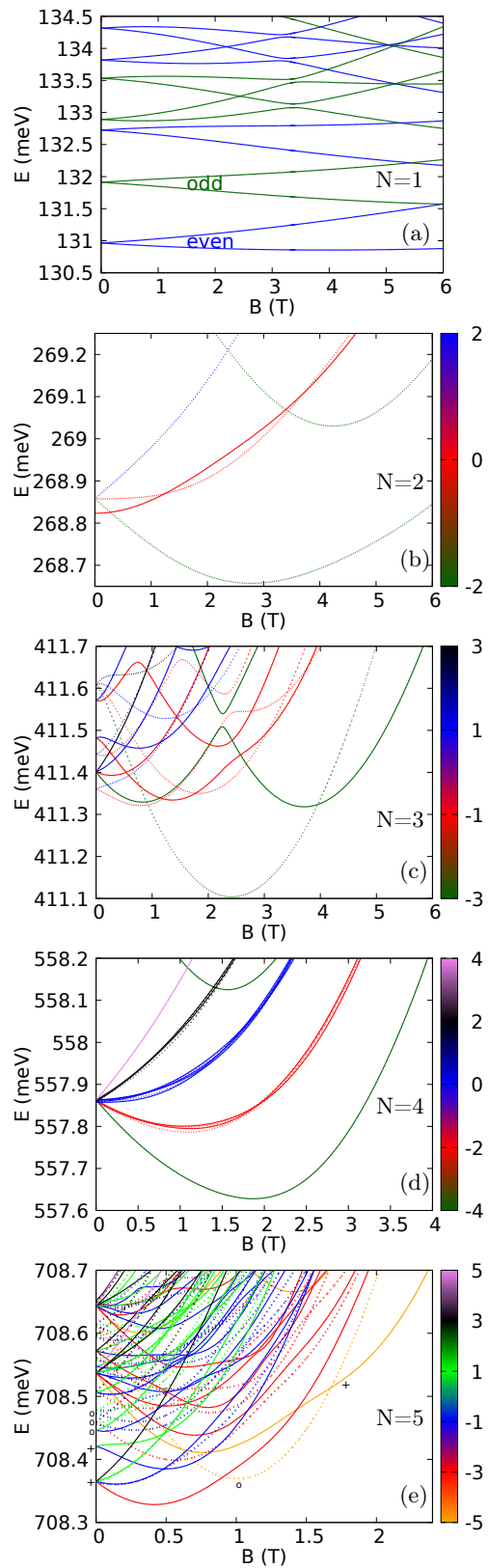%
\subsection{Circular potential}
We discuss first the relatively simple case for $N\leq 4$ (Section III.A.1) where in the low energy states the Wigner molecule
formation in the laboratory frame is present $(N=2,N=4)$ or absent $N=3$. Section III.A.2 covers the case for $N=5$ where
states of both types are present in the low-energy part of the spectrum.

 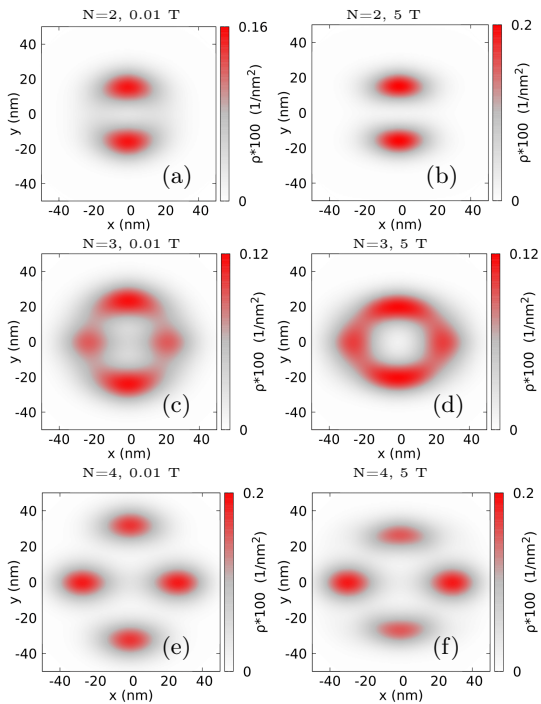
\begin{figure}
 \begin{tabular}{ll}
  \includegraphics[width=0.4\columnwidth]{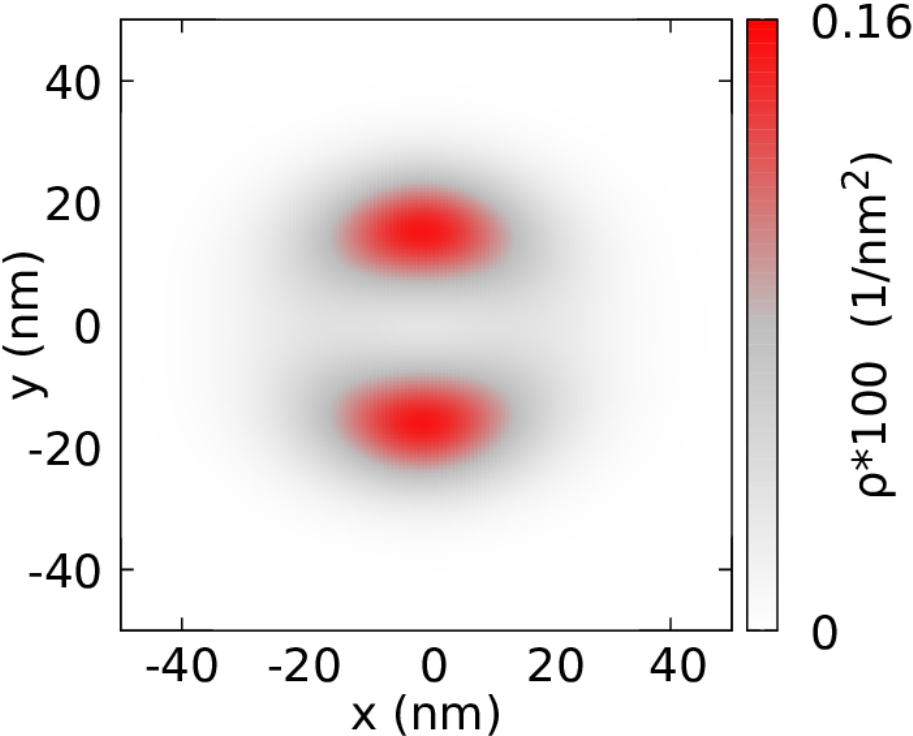} \put(-40,18){(a)} \put(-70,80){\tiny N=2, 0.01 T}&
    \includegraphics[width=0.39\columnwidth]{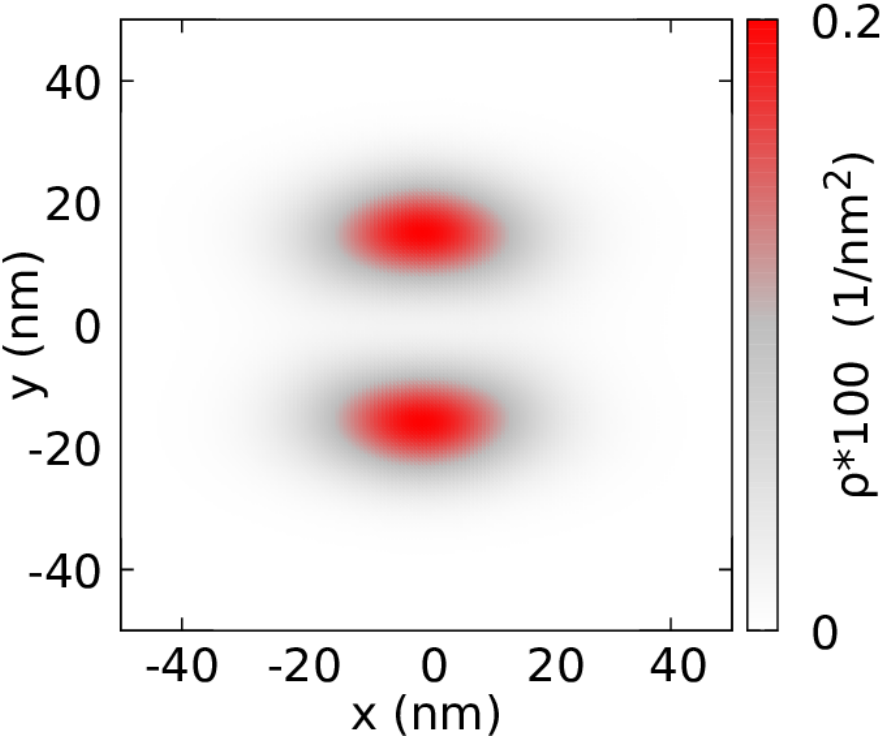} \put(-40,18){(b)} \put(-70,80){\tiny N=2, 5 T}\\
      \includegraphics[width=0.4\columnwidth]{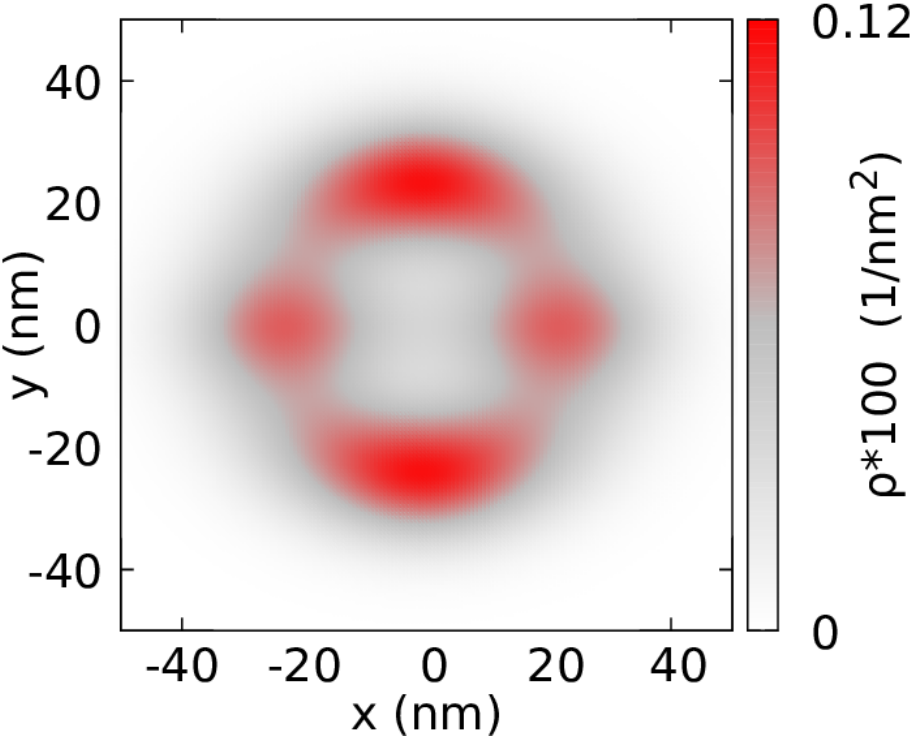} \put(-40,18){(c)} \put(-70,80){\tiny N=3, 0.01 T}&
    \includegraphics[width=0.4\columnwidth]{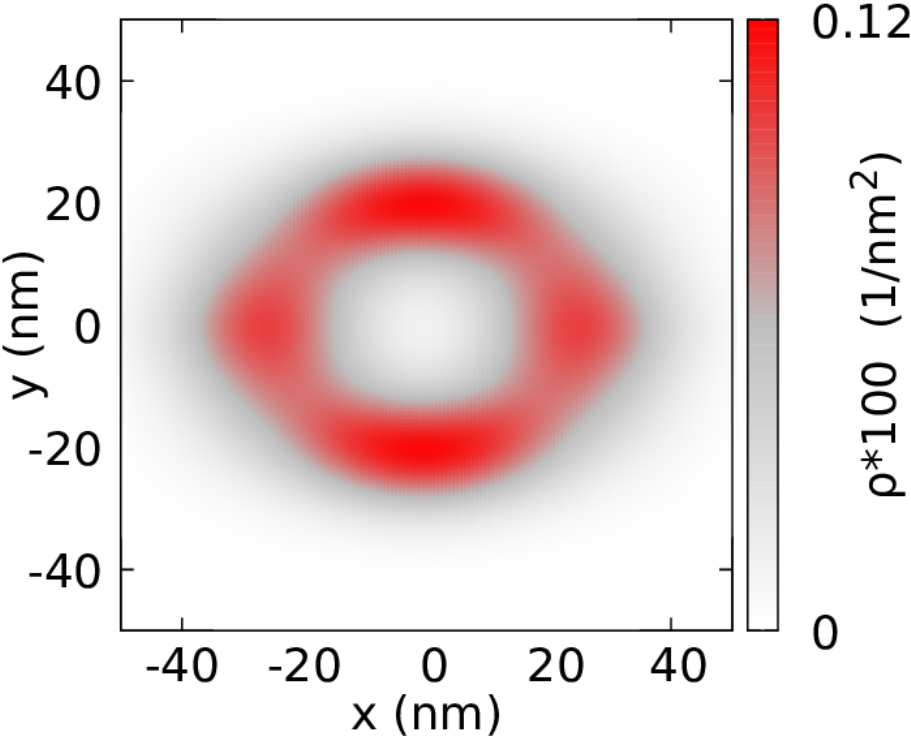} \put(-40,18){(d)} \put(-70,80){\tiny N=3, 5 T}\\
        \includegraphics[width=0.4\columnwidth]{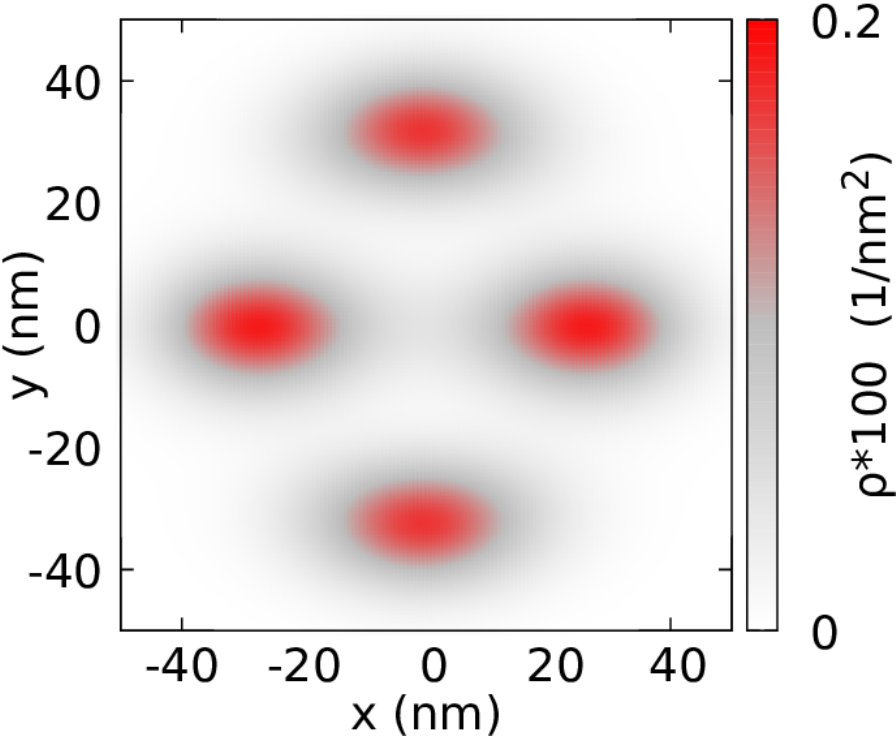} \put(-40,18){(e)} \put(-70,85){\tiny N=4, 0.01 T}
&
    \includegraphics[width=0.4\columnwidth]{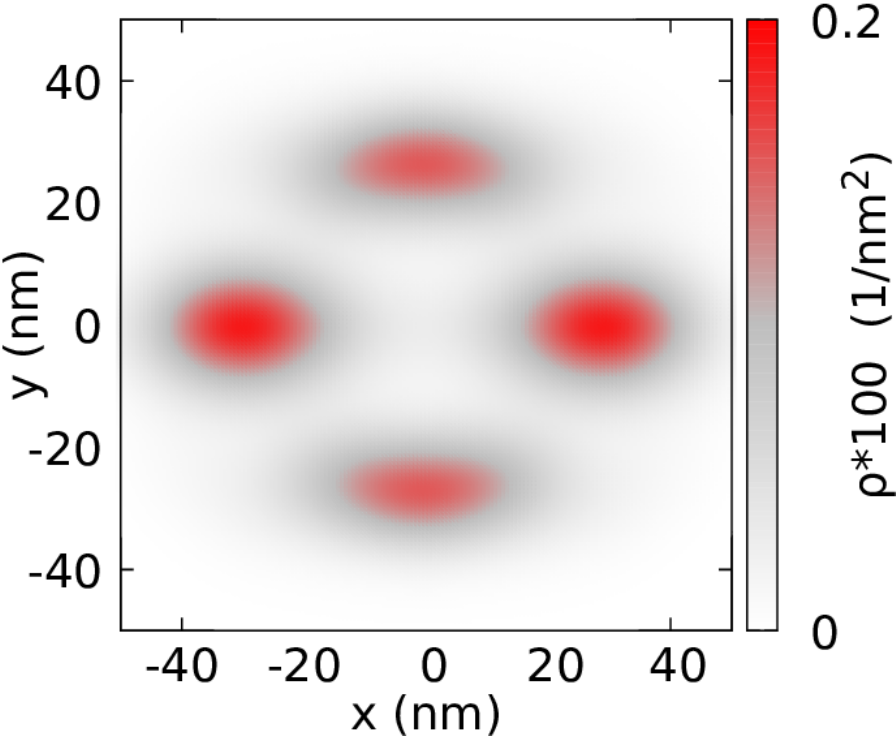} \put(-40,18){(f)} \put(-70,85){\tiny N=4, 5 T}\\

 \end{tabular}
\caption{
The ground-state charge density in the circular confinement for $N=2$(a,b), $3$(c,d) and $4$(e,f) 
for $B=0.01$ T (left column) and $B=5$ T (right column).
}
 \label{sd}
\end{figure}

\subsubsection{Results for $N\leq 4$ electrons}
The energy spectra for  the circular potential [see Fig. \ref{sys}(c)] are
plotted in Fig. 3. For $N=1$ [Fig. \ref{scirc}(a)] the energy levels for $B=0$ are degenerate
with respect to spin. For $N=2$ [Fig. \ref{scirc}(b)] the ground state is an even parity spin singlet
which is replaced by an odd-parity spin-polarized triplet for $B\simeq 0.25$ T. 
The energy spacing between the lowest-energy states in Fig. \ref{scirc}(b-f) are much smaller than for the single-electron
spectra, which is a signature of strong electron-electron interaction \cite{strongsup}. The spin triplets that we find correspond to odd spatial parity which is characteristic
to the two-electron system \cite{jose}.

For $N=3$ electrons [Fig. \ref{scirc}(c)] three different symmetry states  appear in the ground state starting from 
an odd-parity spin doublet for $B=0$. Above 1 T the ground state becomes spin-polarized
first in the odd parity and next, above $\simeq 4$ T, in the even parity state.

 For $N=4$ [Fig. \ref{scirc}(d)] the ground state at $B=0$ is nearly degenerate with respect
to the parity and the spin. The ground state at $B=0$ is an even parity singlet.
The spin polarization in the ground state appears already for $B\simeq 0.03$T.

The results for $N\in[2,4]$ can be summarized in the following manner:

(i) For even $N$ the spectrum at $B=0$ contains a few nearly degenerate energy levels near the  ground state. Already a low magnetic field 
of a fraction of tesla leads to a complete spin polarization in the ground state.t
The systems with $N=2$ and $N=4$ are also similar from the point of view of the Wigner molecular charge
density, which contains separate $N$ single-electron
islands (Fig. \ref{sd}(a,b) and Fig. \ref{sd}(e,f)).

(ii) For $N=3$ the ground state at $B=0$ is exactly two-fold degenerate
with respect to the spin, and a few ground state transitions appear 
in the field of the order of a few tesla before the high field symmetry is established.
The charge density exhibits 4 local charge maxima Fig. \ref{sd}(c,d).
These are not the single-electron islands. Furthermore, the charge density  is smeared over the area between the maxima [Fig. \ref{sd}(c,d)
and does not vanish as effectively as for even $N$ [Fig. \ref{sd}(a,b)]
The three-electron charge density is not ordered in the Wigner-molecule form.

\begin{figure}
 \begin{tabular}{ll}
  \includegraphics[width=0.5\columnwidth]{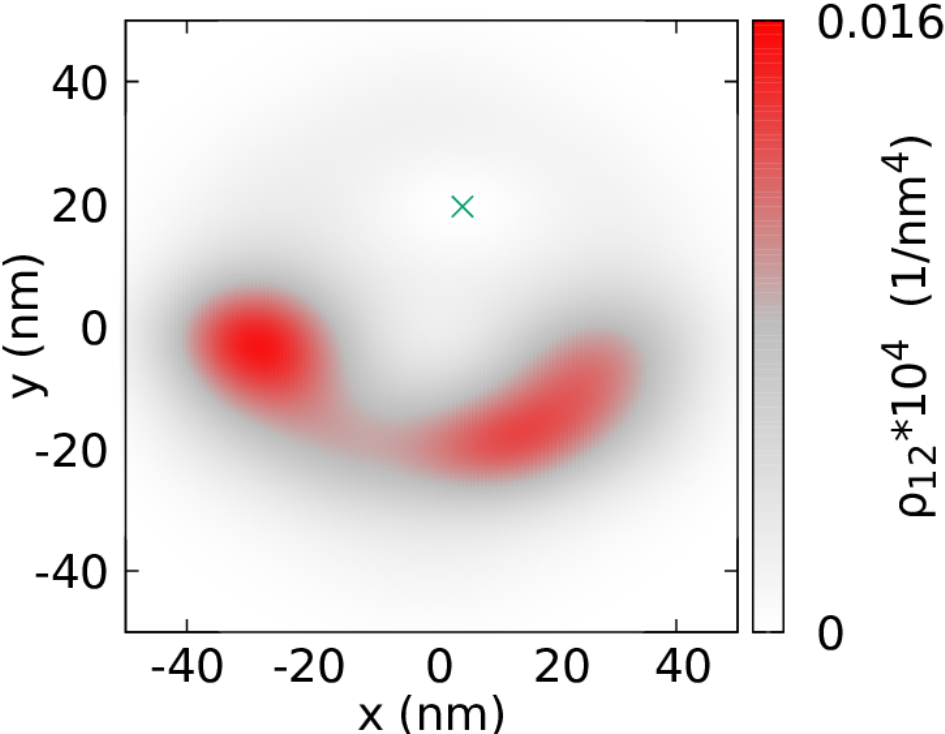} \put(-40,20){(a)}  \put(-60,85){N=3}& \includegraphics[width=0.5\columnwidth]{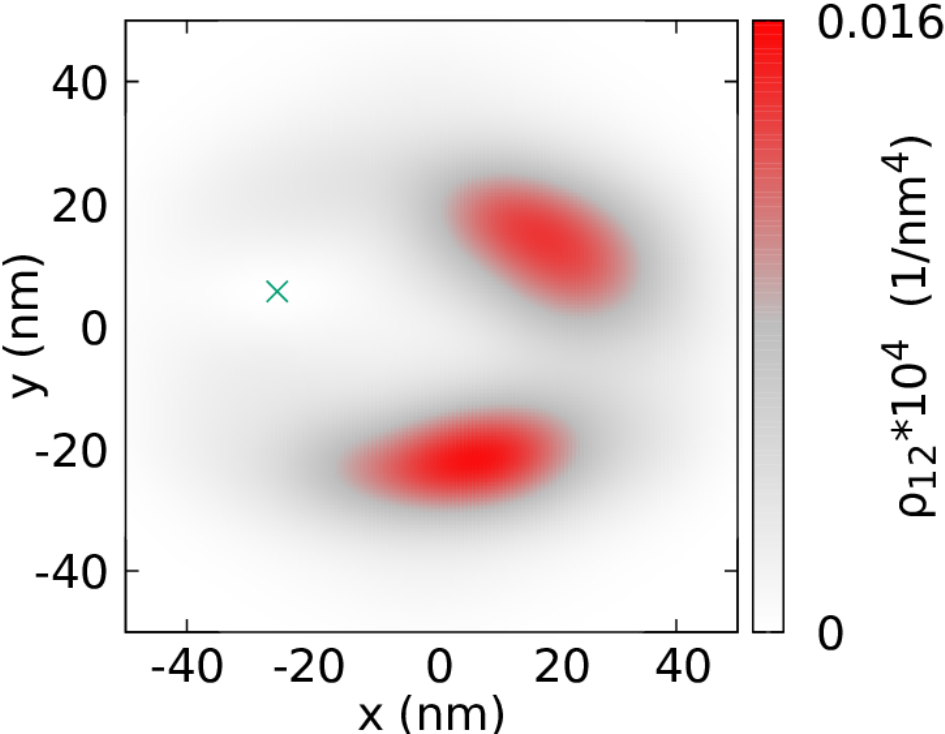} 
  \put(-40,20){(b)}  \put(-60,85){N=3}\\
 \includegraphics[width=0.5\columnwidth]{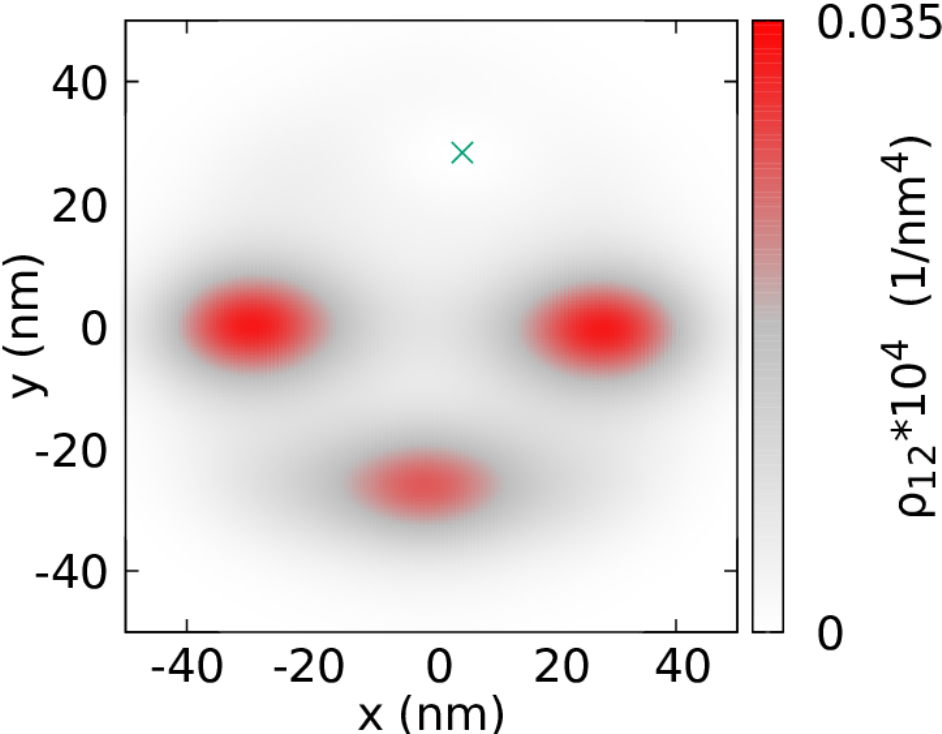} 
 \put(-40,20){(c)}  \put(-60,85){N=4}&
 \includegraphics[width=0.5\columnwidth]{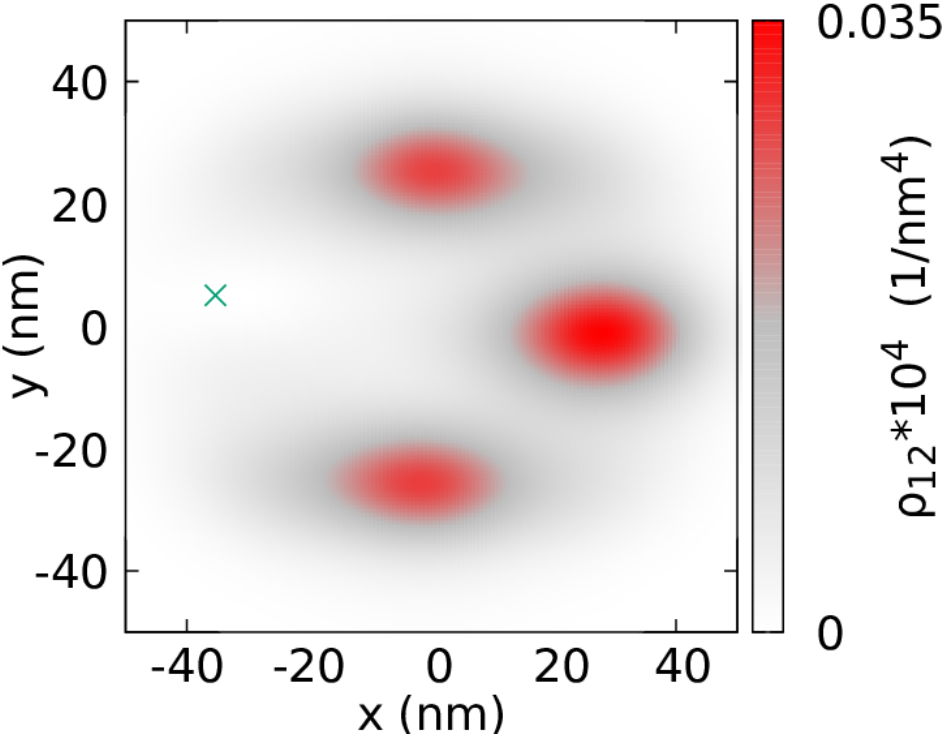}   \put(-40,20){(d)}  \put(-60,85){N=4}
   \\

  \end{tabular}
\caption{Pair correlation function plots for the circular confinement potential. The spin-polarized
ground state at $B=5$T is considered. The crosses indicate the position of the fixed electron (see Eq.(6)).
}
 \label{pcf}
\end{figure}
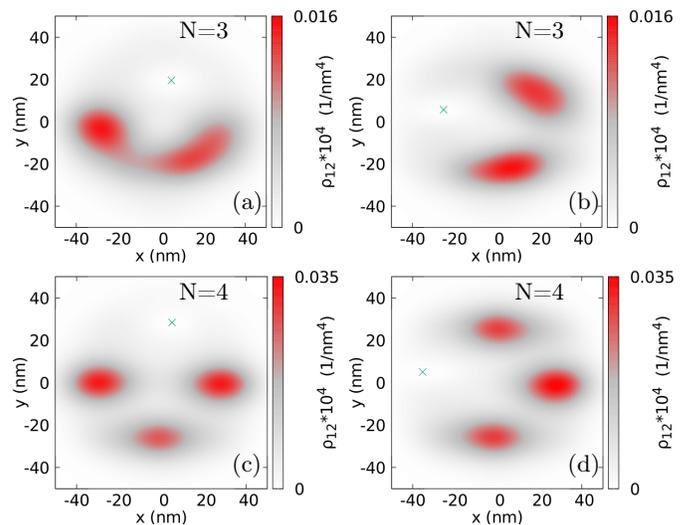

The effects of the electron-electron correlation in the localization of the carriers can be observed in 
the pair-correlation function plots given in Fig. \ref{pcf}.
 For these plots we fix the position of one of the electrons (see ${\bf r}_f$ in Eq. (6)) that
is marked by the cross in each panel of Fig. \ref{pcf}. 
For illustration of the system reaction to the electron position,
we fixed one of the electrons slightly off the local density maxima of Fig. \ref{pcf}
near the left (top) edge of the charge distribution in the left (right) column of Fig. \ref{pcf}.
For 3 electrons [Fig. \ref{pcf}(a,b)] the conditional probability exhibits two separate maxima.
The maxima move once the fixed electron position is changed [cf. Fig. \ref{pcf}(a,b)],
which corresponds to the ring-like charge distribution in Fig. 4(c,d).
On the other hand, for $N=4$ the probability maxima at the right and bottom edges 
of the quantum dot stay  in the same place when the fixed electron position is changed  [Fig. \ref{pcf}(c,d)].
Note that for $N=4$  the charge density produces the pronounced
single-electron maxima [Fig. 4(e,f)]. 

 \begin{figure}
 \begin{tabular}{l}
  \includegraphics[width=0.7\columnwidth]{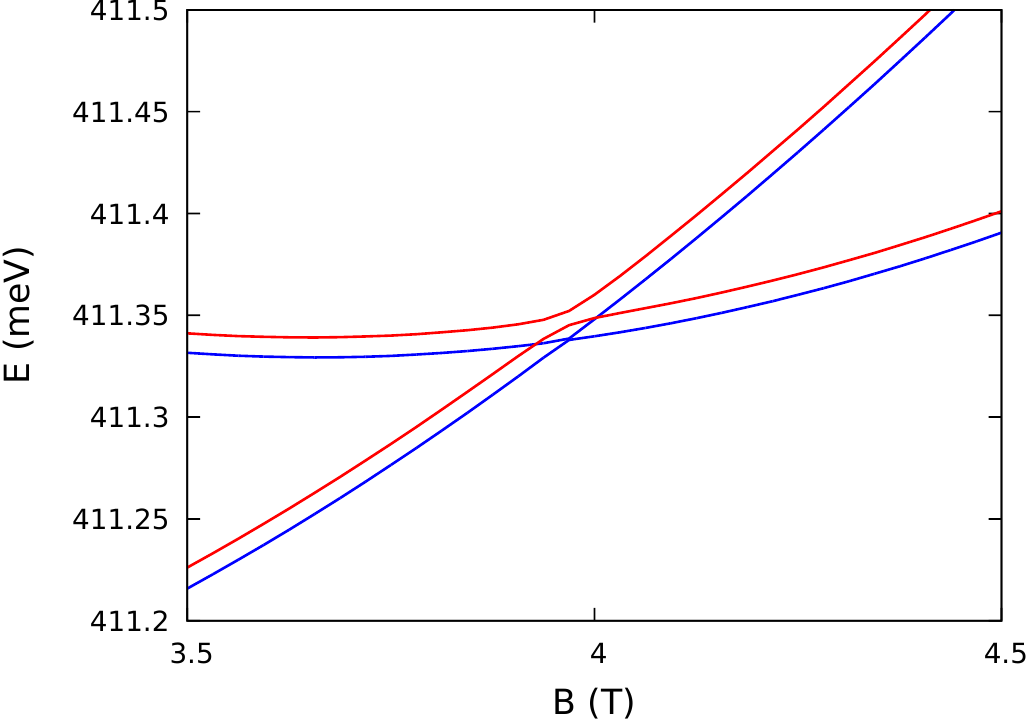} \put(-80,105){N=3}\\
    \end{tabular}
    \caption{
    The energy spectra with (red lines) and without (blue lines) a Gaussian impurity [Eq. (9)] that perturbs the
    circular symmetry of the confinement potential. The energy levels for the clean system (blue lines)
    are taken from Fig. 3(c). In Eq. (9) we apply $D=0.125$ meV, $y_0=20$ nm and $R_p=5$ nm.
 }
    \label{zabu}
    \end{figure}

 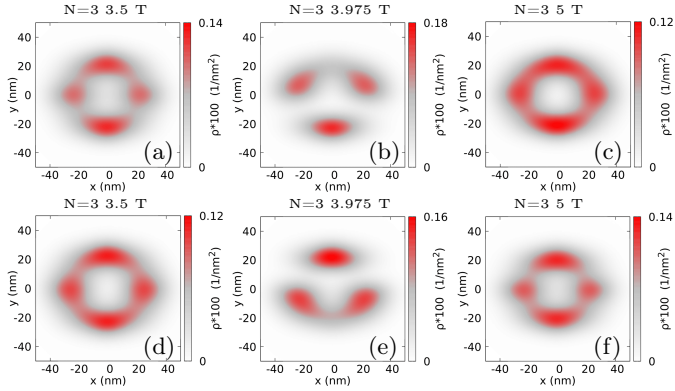
\begin{figure}
 \begin{tabular}{lll}
        \includegraphics[width=0.33\columnwidth]{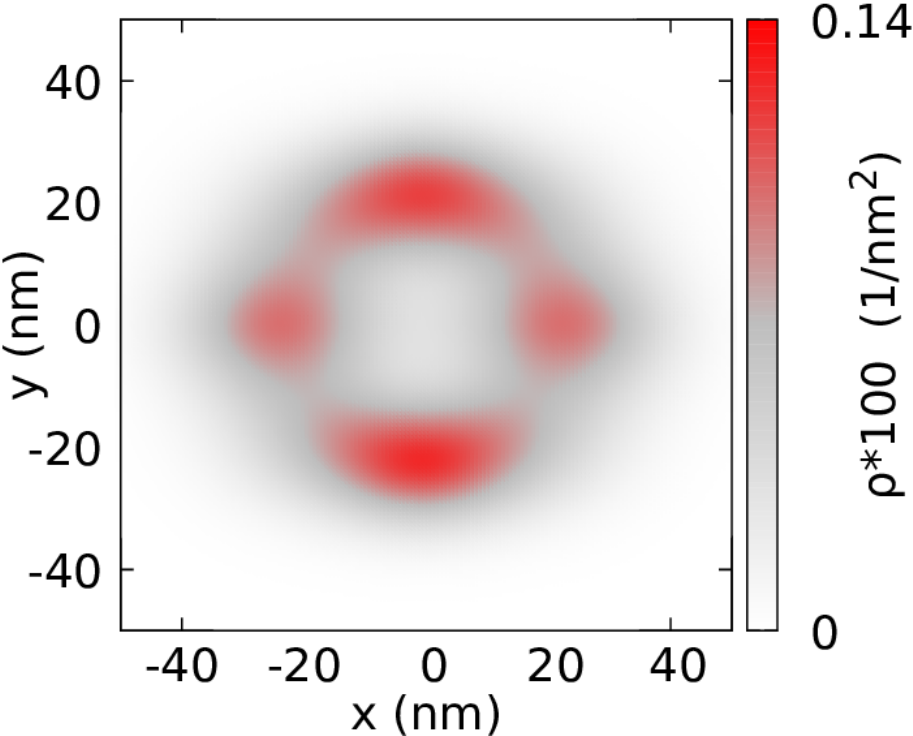} \put(-30,12){(a)} \put(-60,67){\tiny N=3 3.5 T}&
      \includegraphics[width=0.33\columnwidth]{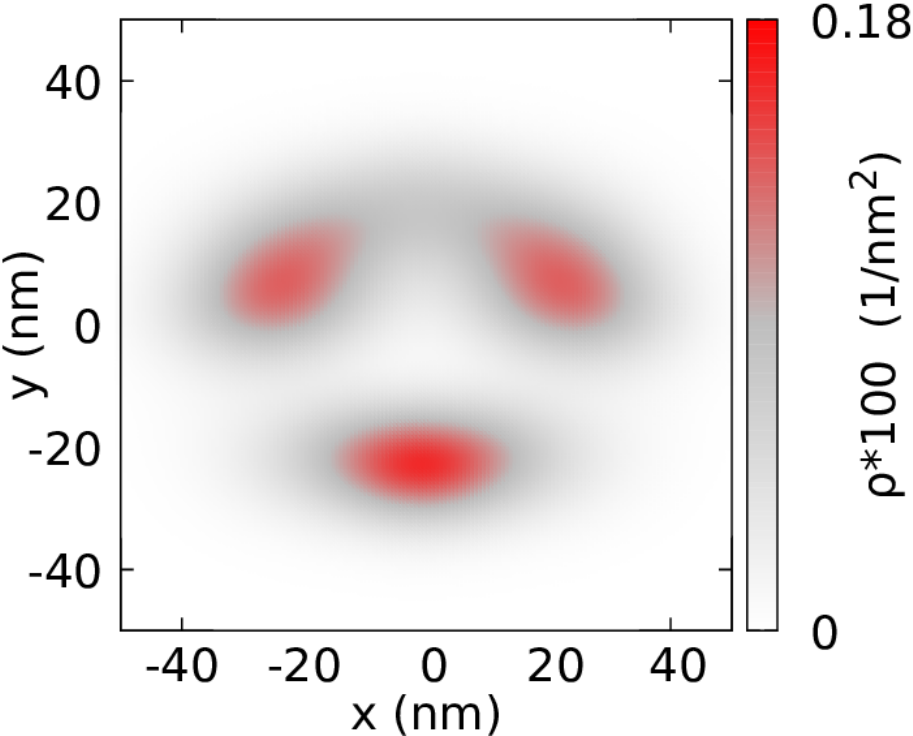} \put(-30,12){(b)} \put(-60,67){\tiny N=3 3.975 T}&
    \includegraphics[width=0.33\columnwidth]{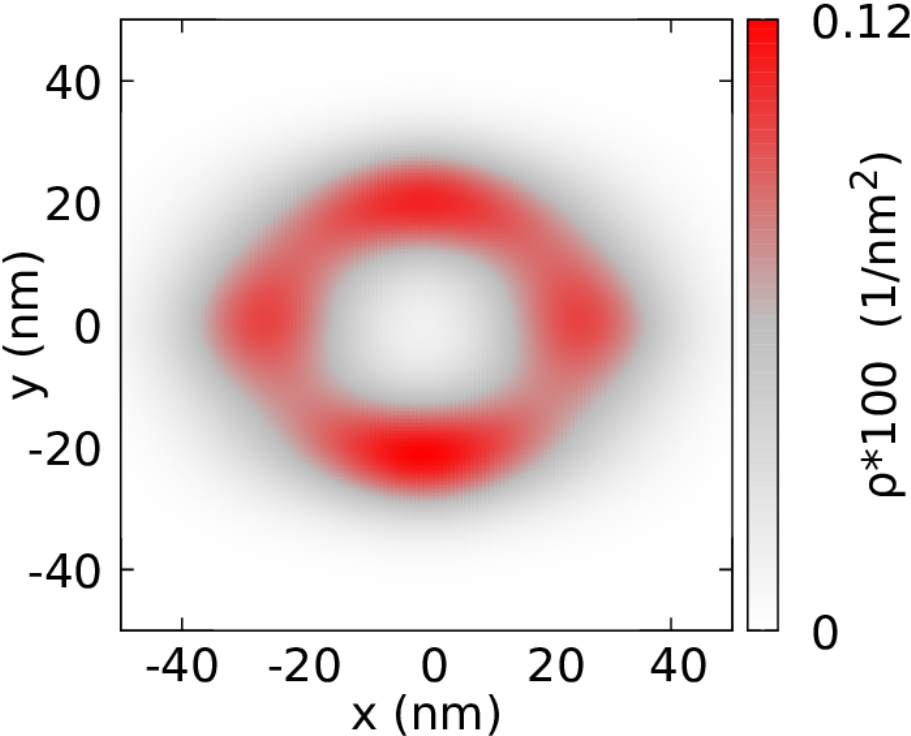} \put(-30,12){(c)} \put(-60,67){\tiny N=3 5 T}\\
        \includegraphics[width=0.33\columnwidth]{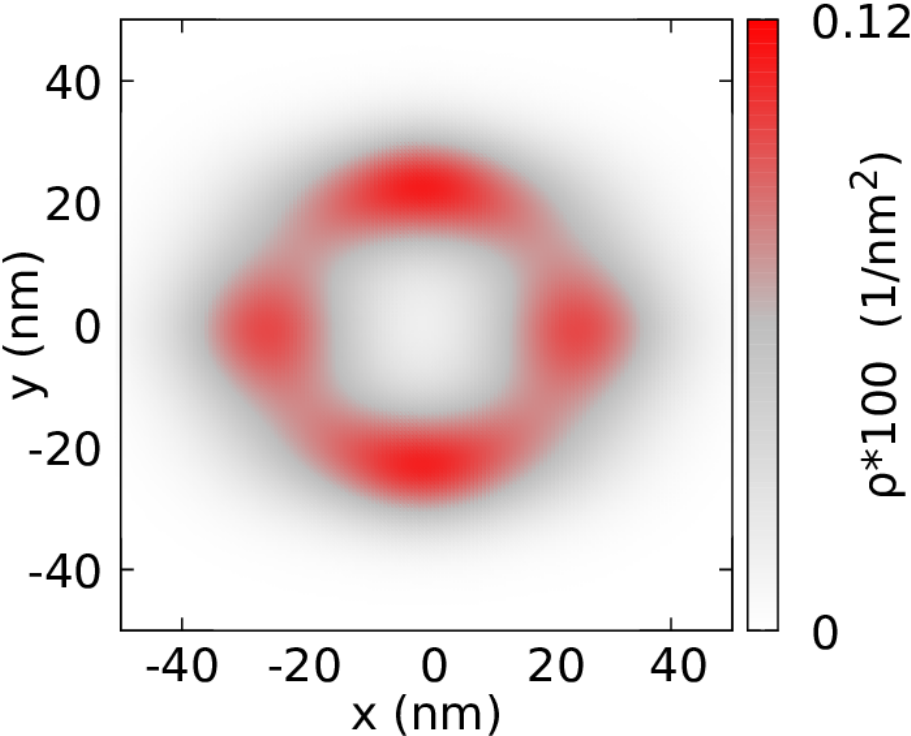} \put(-30,12){(d)} \put(-60,67){\tiny N=3 3.5 T}&
      \includegraphics[width=0.33\columnwidth]{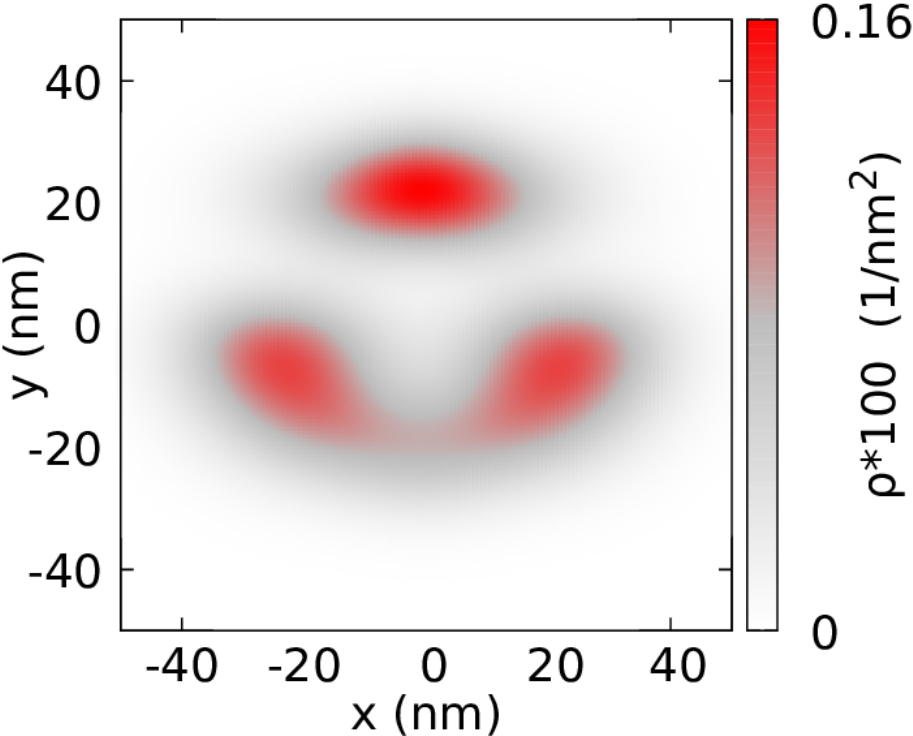} \put(-30,12){(e)} \put(-60,67){\tiny N=3 3.975 T}&
    \includegraphics[width=0.33\columnwidth]{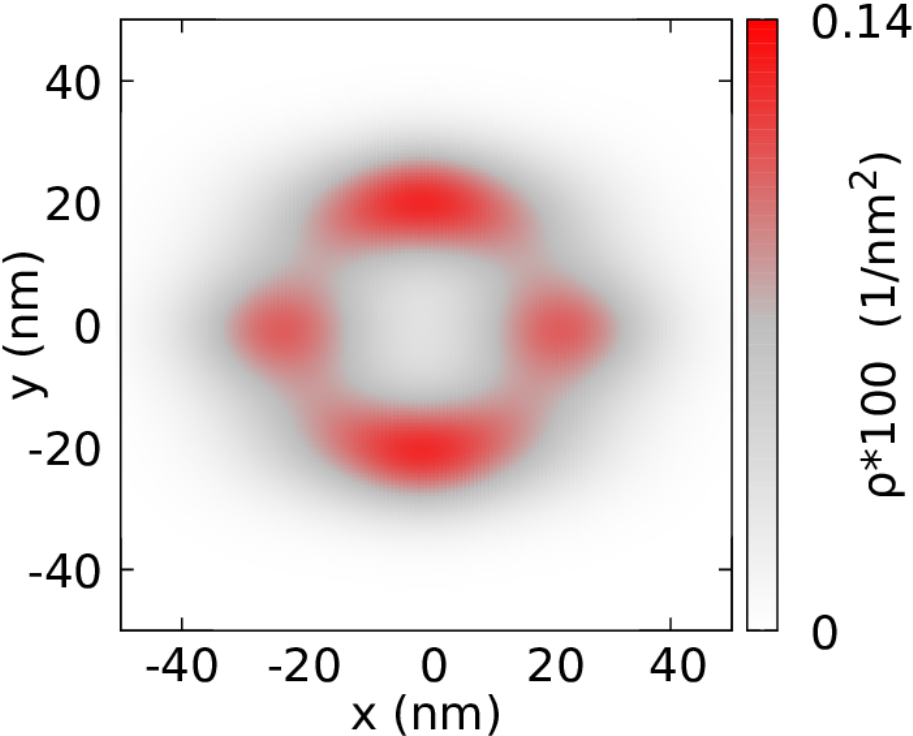} \put(-30,12){(f)} \put(-60,67){\tiny N=3 5 T}\\
 
 \end{tabular}
\caption{
The charge density for $N=3$ (a-f) 
levels marked in red in Fig. \ref{zabu} for the circular quantum dot with an off-center Gaussian impurity.
The first row (a-c) shows the data for $N=3$ ground state. The excited state for 3 electrons is shown
in the second row (d-f).  The central column (b,e) shows the results at the center
of the avoided crossing displayed in Fig. \ref{zabu}.
}
 \label{zaburo}
\end{figure}

For $N=3$ the single-electron islands forming a Wigner molecule in the real space
are not observed since the number of local charge maxima is not equal to $N$ [Fig.4(c,d)]. 
The single-electron charge maxima can appear in the real space
when the symmetry of the confinement potential is lowered \cite{wimabrok} by e.g. an off-center impurity.
We use a weak   Gaussian perturbation
introduced to the confinement potential used in $H_0$, \begin{equation} V_p(x,y)=D\exp(-(x^2+(y-y_0)^2)/R_p^2)\,,\end{equation}
with parameters  $D=0.125$ meV, $R_p=5$ nm, and $y_0=20$ nm.

 We considered the spin-polarized states at the magnetic field,  near the ground-state symmetry transition from even to odd parity, below 4 T for $N=3$ Fig. \ref{scirc}(c).
The blue lines in Fig. \ref{zabu} present the energy levels of a clean
system adopted from Fig. \ref{scirc}(c). The red lines show the energy levels for
the perturbed system. The off-center Gaussian impurity opens avoided crossings between the energy levels [Fig. \ref{zabu}] which for the clean potential correspond to opposite parity.

The charge density for the ground (excited) state of $N=3$ is given in Fig. \ref{zaburo}(a-c) for the magnetic field that is swept across the avoided crossing. The charge density
outside the avoided crossing [Fig. \ref{zaburo}(a,c,d,f)] produces four maxima along the ring-like charge distribution,
which deviates from the point symmetry due to the $V_p$ potential. 
At the avoided crossing the wave functions from otherwise crossing levels mix.
In the ground state [Fig. \ref{zaburo}(b)] three well-separated single-electron islands appear and the
charge density is low at the center of the perturbation ($(x,y)=(0,y_0)$). 
The excited state [Fig. \ref{zaburo}(e)] at the avoided crossing produces a Wigner molecule that is rotated by the $\pi$ angle.
In this sense, the even and odd-parity $N=3$ states for a clean potential can be understood as superposition
of states producing Wigner crystallization in real space with two equivalent but rotated charge distribution.
This result for three electrons in circular potential but anisotropic effective mass is similar
to the ones found previously for isotropic mass but anisotropic confinement potential \cite{aniso1,aniso2}.

\subsubsection{5-electron system in the circular quantum dot}
\begin{figure}
 \begin{tabular}{ll}
 \includegraphics[width=0.43\columnwidth]{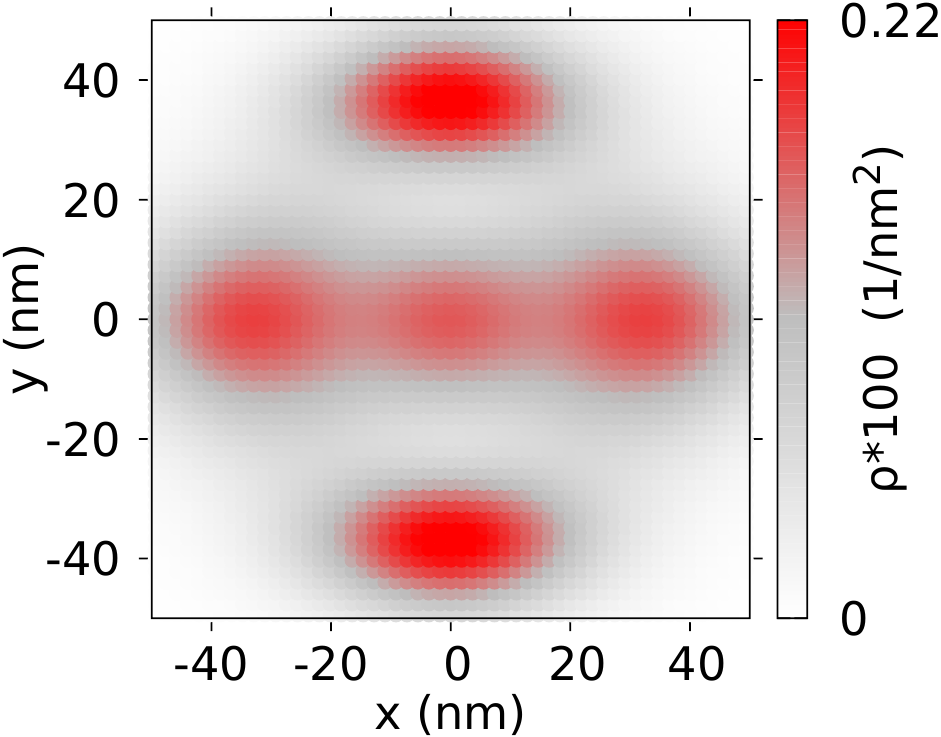} \put(-40,20){(a)} &
\includegraphics[width=0.43\columnwidth]{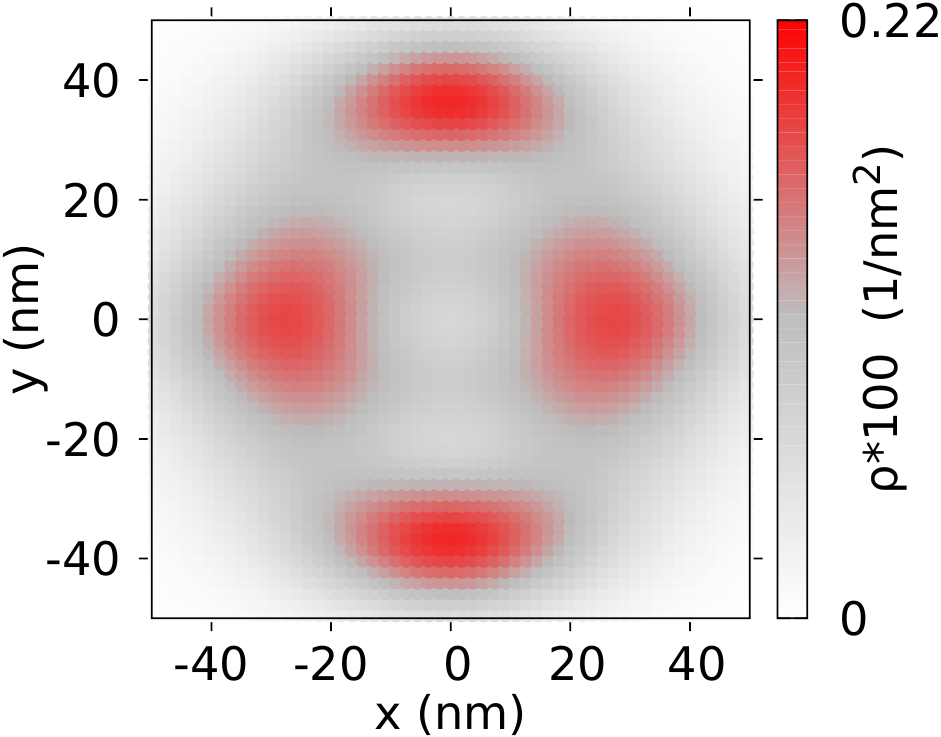} \put(-40,20){(b)}\\
\includegraphics[width=0.43\columnwidth]{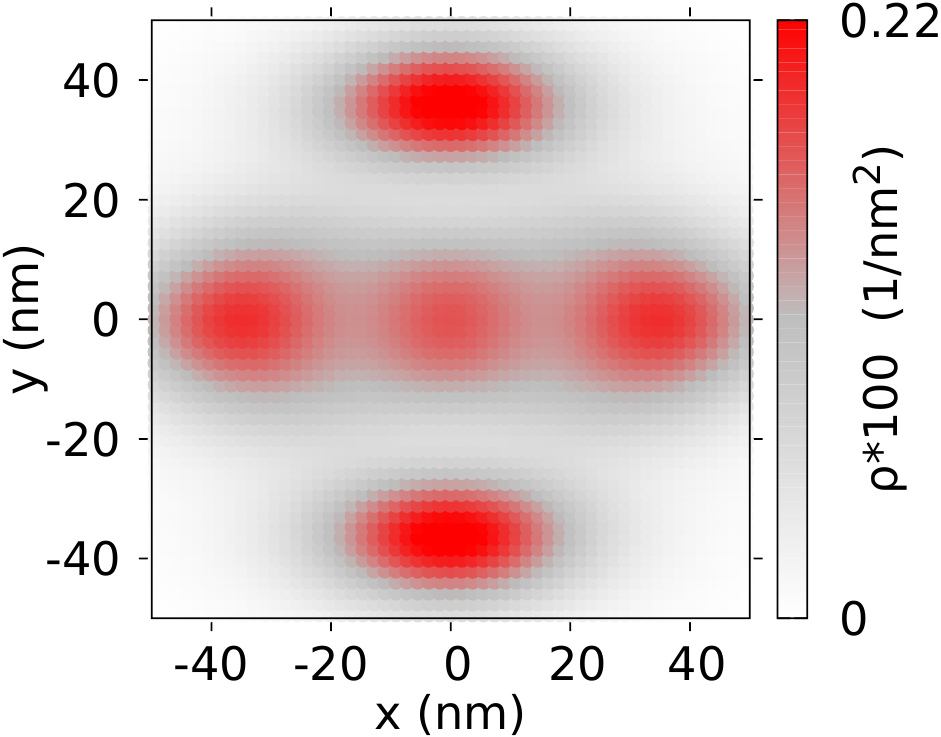} \put(-40,20){(c)}&
 \includegraphics[width=0.43\columnwidth]{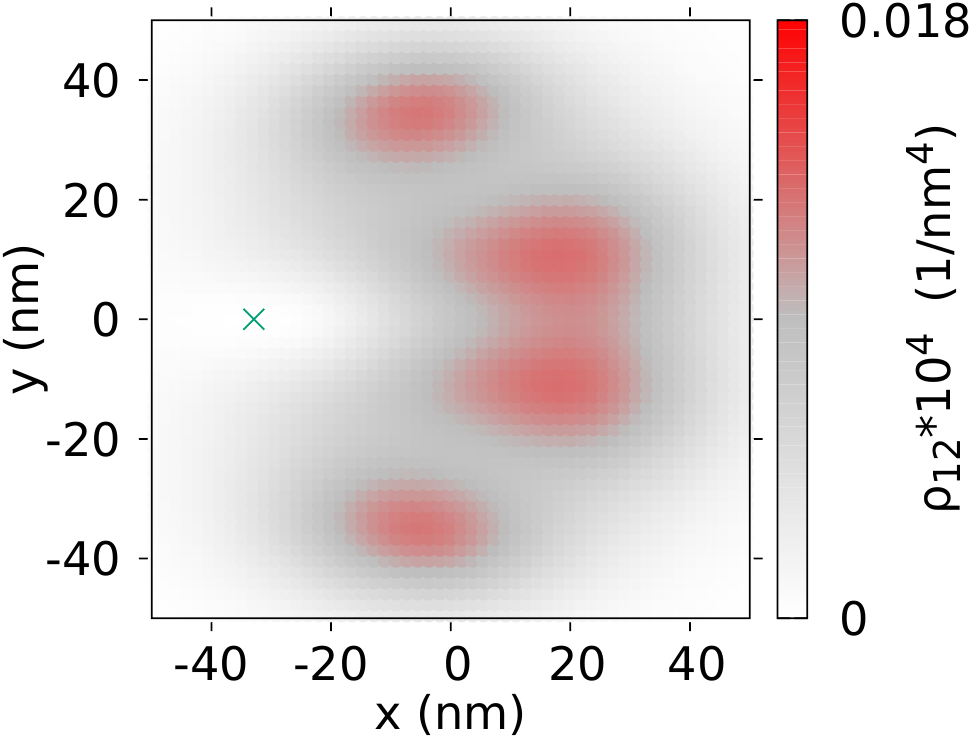} \put(-40,20){(d)} \\

 \end{tabular}
\caption{The ground-state charge density for $N=5$ electrons in $B=0.01$ T (a), $B=1$T (b) and $B=1.75$ T (c) -- see the spectrum 
in Fig. 3(e). In panel (d) we plot the pair correlation function for the state of panel (b) one of the electrons fixed at the position
marked by the cross.
}
 \label{6}
\end{figure}
The five-electron systems has a more complex structure than systems with two to four electrons
 due to two charge distributions that appear
in the low energy spectrum [see Fig. 3(e)]. In ground-state at $B=0$ we find 6 nearly degenerate energy levels:
a $S_z=\pm 1$ odd-parity doublet slightly below $S_z=-3,-1,1,3$  even parity quartet. 
The charge density in all these energy levels is organized in the Wigner molecule form
that is plotted in Fig. \ref{6}(a) with one central electron island and four others shifted off the center
of the QD forming a cross-like structure that we will denote as (1,4). The near degeneracy of the ground-state is a counterpart
of the four-electron ground-state for $N=4$, where the Wigner molecule is also found. 
For five electrons the first excited energy level at $B=0$ is a even parity $S_z=\pm 1$ doublet
also with (1,4) charge distribution.
In several higher excited energy levels the states at the energy of $\simeq 708.45$ meV 
the charge density is organized in a ring-like structure with four charge density maxima
without the single-electron islands. We will denote this structure as (0,4).
In higher excited levels the states with (1,4) and (0,4) structure interlace on the energy scale.
For $B>0.9$ T the ground-state becomes fully spin-polarized [Fig. 3(e)] and the (0,4) structure [Fig. \ref{6}(b)] appear
in the ground-state. Above 1.7 T the spatial parity of the ground-state change from odd to even
and the (1,4) structure [Fig. \ref{6}(c)] re-appears in the ground-state. 
In presence of both types of states near the ground-state -- with and without single-electron islands in the laboratory frame -- the spectrum contains the features of
both $N=2,4$ (ground-state near degeneracy at $B=0$) and symmetry transformations at higher field as for $N=3$.

The (0,4) state for five electrons has a similar character as the three-electron ground-state.
For three electrons the charge density is a superposition of two equivalent configurations
one being an inversion of the other.
The five-electron charge configurations with the (0,4) state correspond to the superposition one or two electrons
at the right/left ends of the charge density near $y=0$ line. 
One of the two equivalent structures for five electrons can be observed in the pair correlation
function plot of Fig. \ref{6}(d) for one of the electrons fixed at point (-29.4 nm,0). 

\subsection{A linear confinement}
We discuss two perpendicular orientations of the elongated quantum dot to study the interplay of the potential
and effective mass anisotropies. 

\begin{figure}
 \begin{tabular}{l}
\includegraphics[width=0.67\columnwidth]{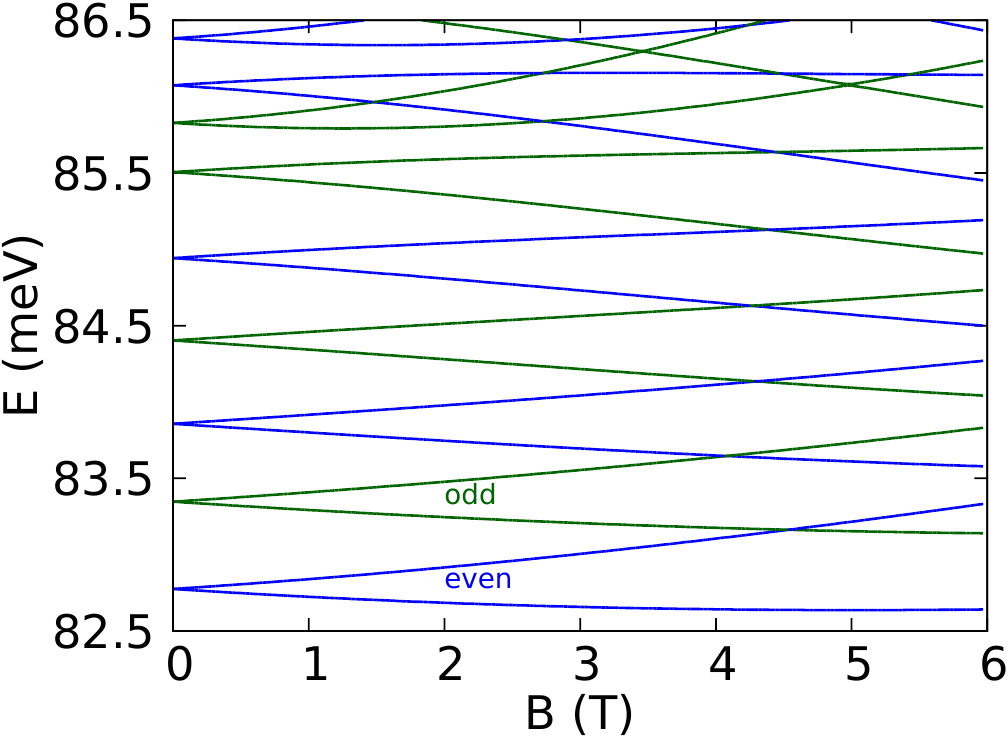} \put(-135,18){\tiny (0,0)}
\put(-135,34){\tiny (0,1)}\put(-135,45){\tiny (0,2)}\put(-135,60){\tiny (0,3)}
\put(-135,74){\tiny (0,4)}\put(-135,87){\tiny (0,5)}\put(-135,96){\tiny (1,0)} \put(-20,23){(a)}
  \\
 \includegraphics[width=0.7\columnwidth]{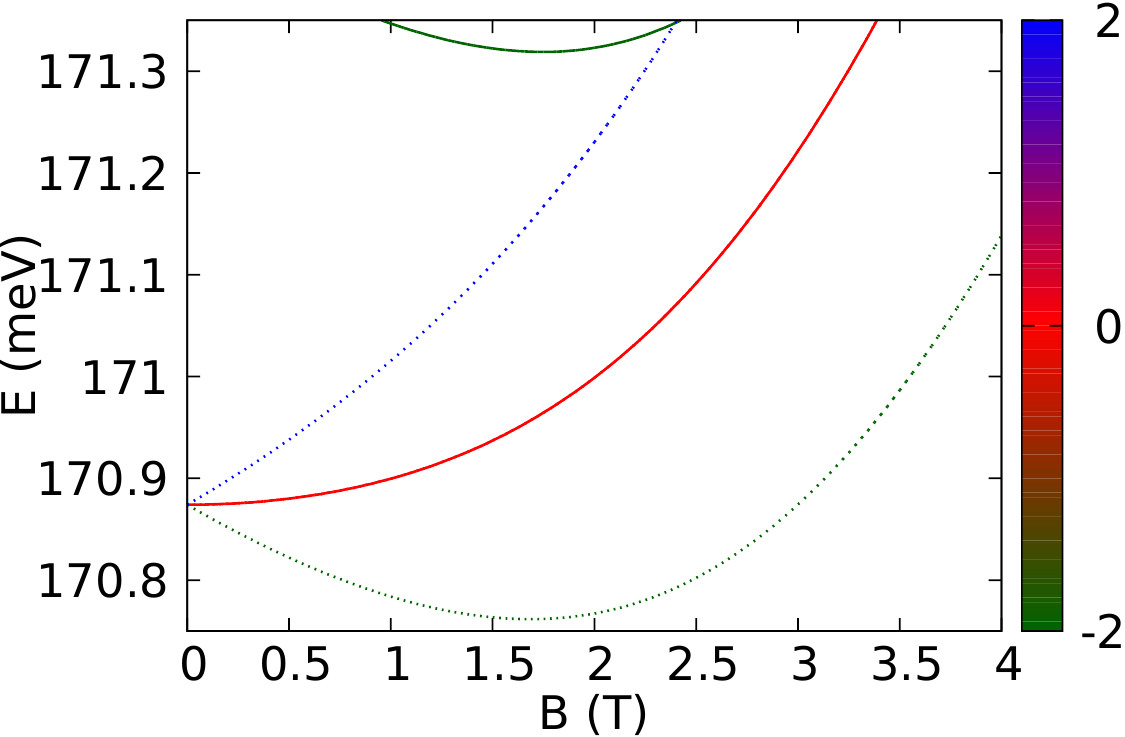}\put(-32,19){(b)}\\
\includegraphics[width=0.7\columnwidth]{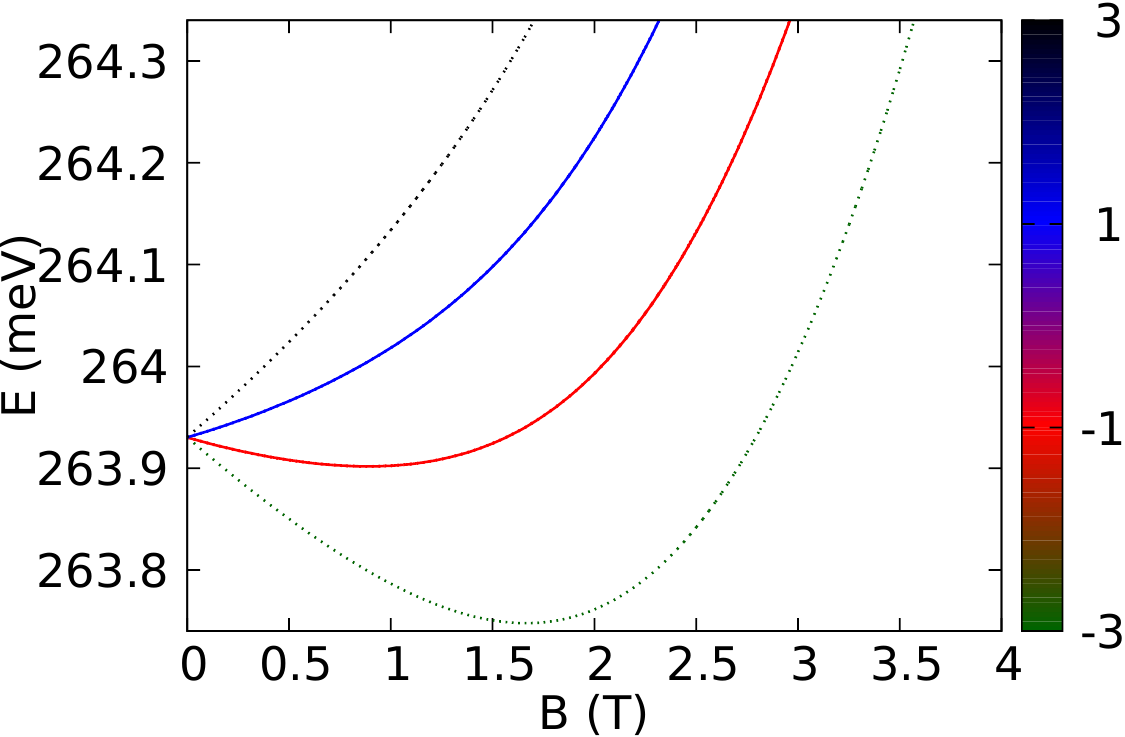} \put(-32,19){(c)}\\
\includegraphics[width=0.7\columnwidth]{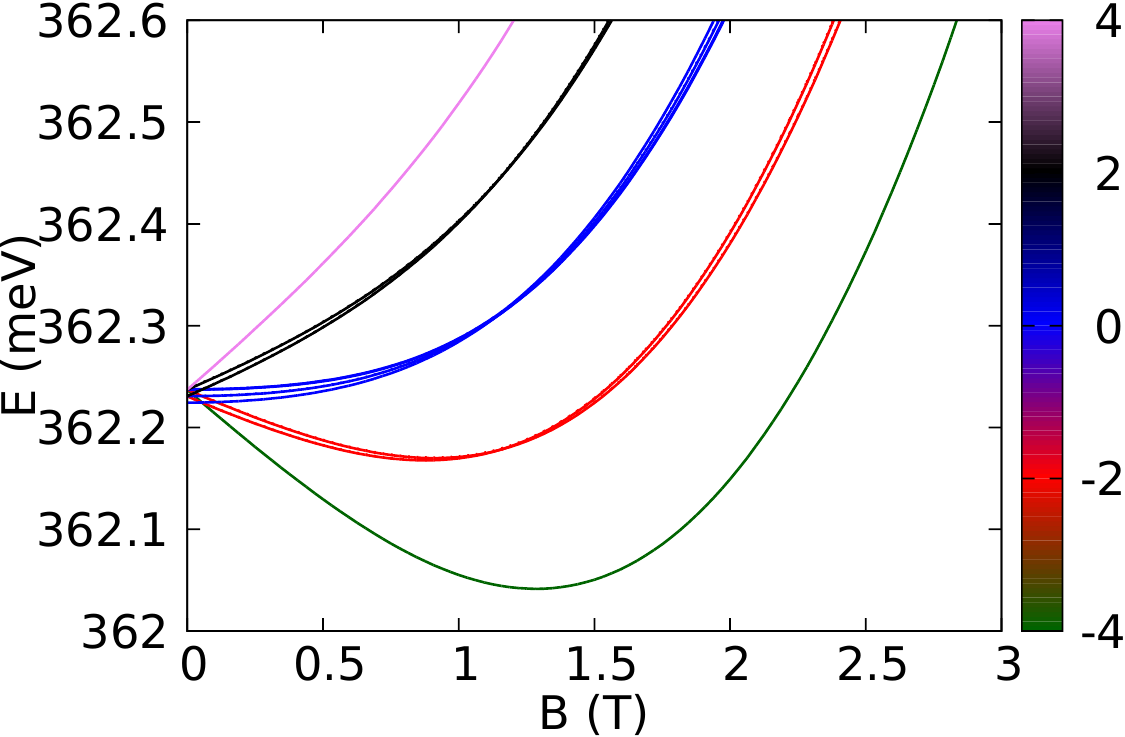}\put(-32,19){(d)}\\
 \end{tabular}
\caption{ 
Spectra for the potential of Fig. \ref{sys}(d) elongated along the zigzag crystal direction ($y$) for $N=1$ (a), 2 (b), 3(c), 4(d). 
In (a) even and odd parity levels are plotted with blue and green lines, respectively.
Notation $(n_x,n_y)$ close to $B=0$  shows the number of excitations (sign changes) in 
the wave function at 0T.
In (b-e) the even parity levels are plotted with the solid lines and the odd parity
 levels with the dotted lines. The color of the lines stands for $S_z$.
The color of the lines in (b-d) shows the $S_z$ value.
}
 \label{1ewy}
\end{figure}

 \begin{figure}
 \begin{tabular}{lll}
 \includegraphics[width=0.33\columnwidth]{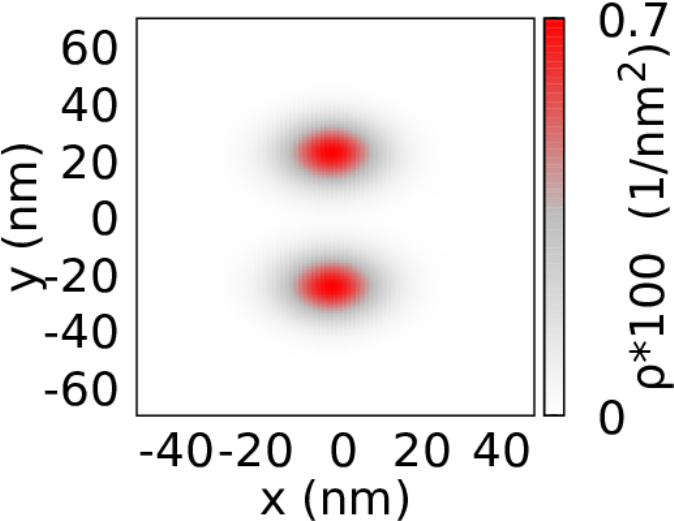} \put(-30,20){(a)} &
\includegraphics[width=0.33\columnwidth]{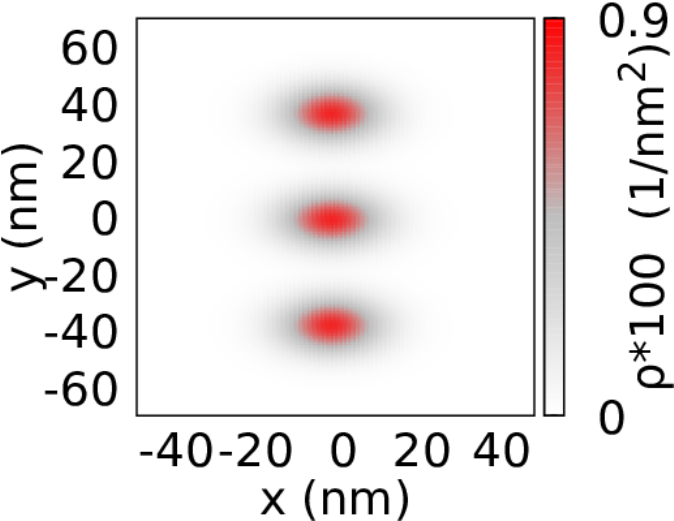} \put(-30,20){(b)}&
\includegraphics[width=0.33\columnwidth]{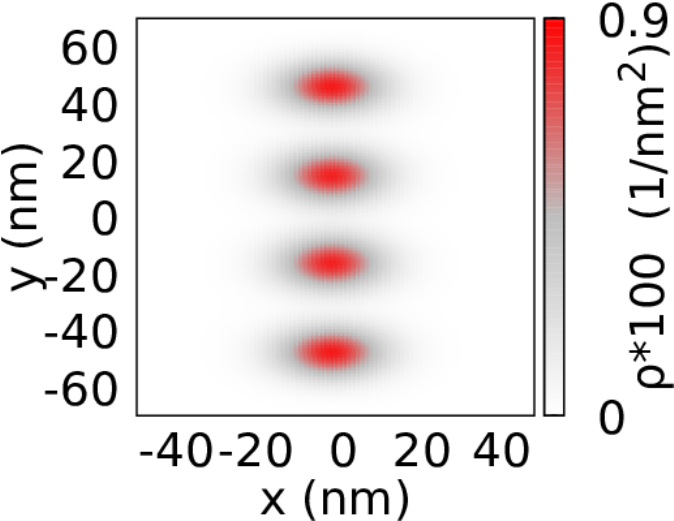} \put(-30,20){(c)}\\
 \includegraphics[width=0.33\columnwidth]{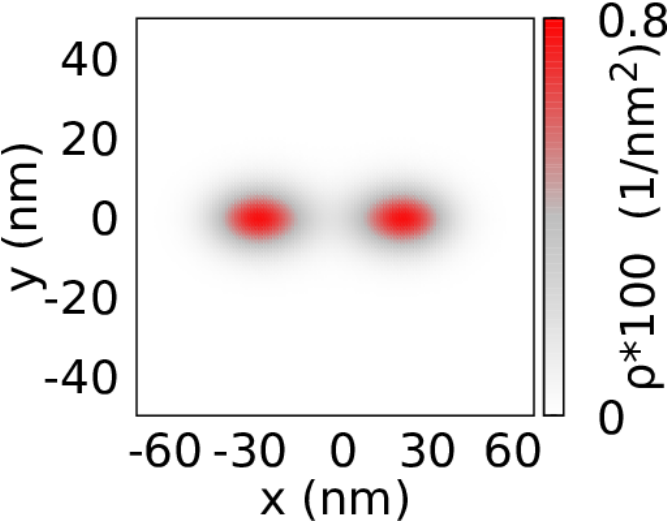} \put(-30,20){(d)} &
\includegraphics[width=0.33\columnwidth]{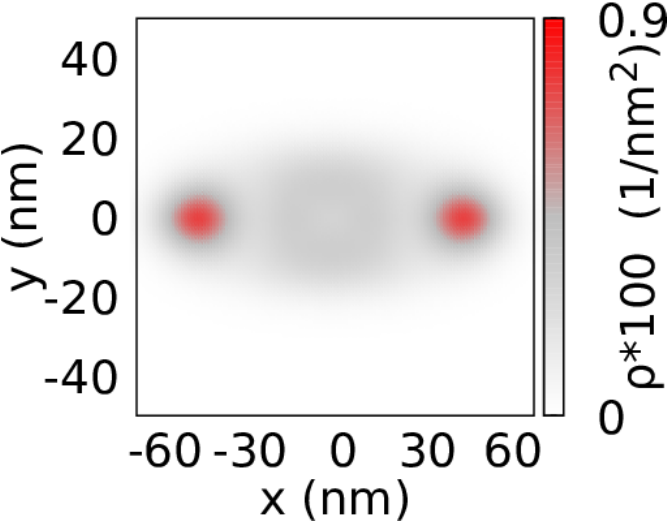} \put(-30,20){(e)}&
\includegraphics[width=0.33\columnwidth]{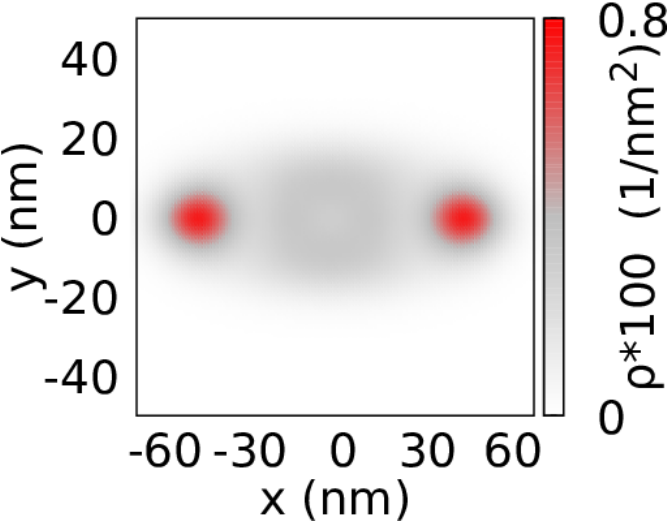} \put(-30,20){(f)}\\
\includegraphics[width=0.33\columnwidth]{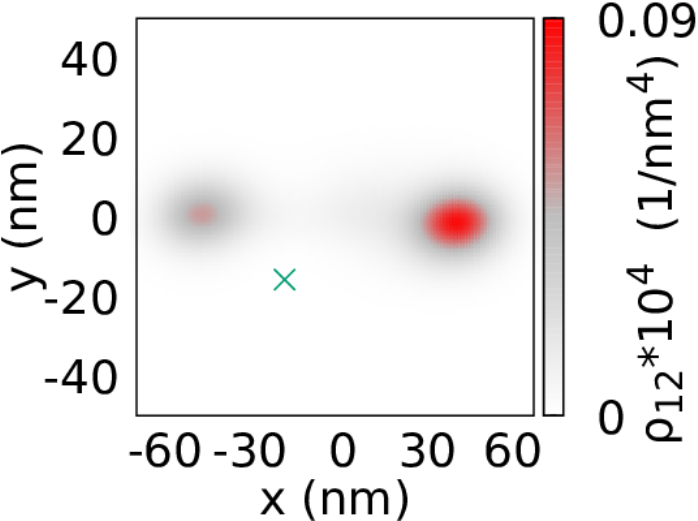} \put(-32,20){(g)} &
\includegraphics[width=0.33\columnwidth]{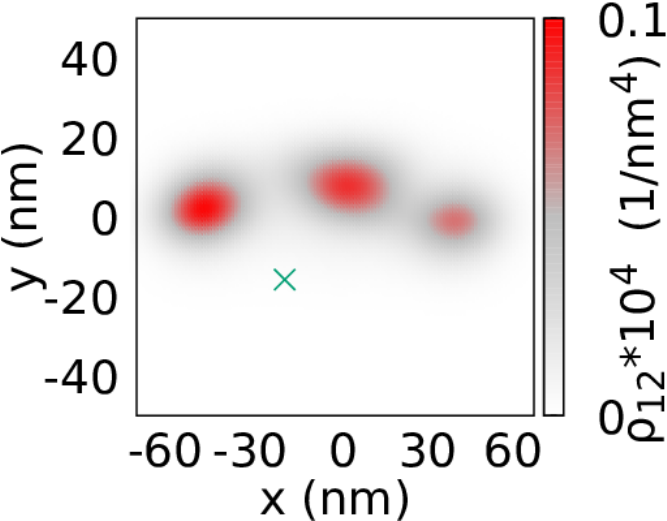} \put(-32,20){(h)} &
\includegraphics[width=0.33\columnwidth]{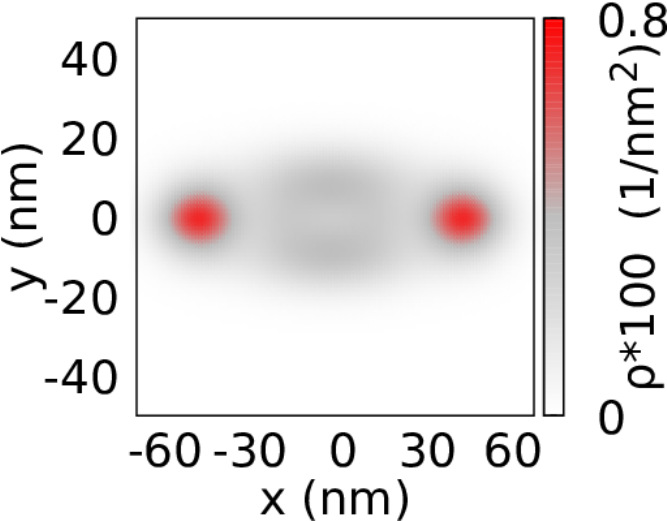} \put(-30,20){(i)} \\

 \end{tabular}
\caption{Charge density for the $B=0.01$ T ground state of the quantum dot elongated
along the $y$ axis (zigzag direction, spectra given in Fig. \ref{1ewy}) for $N=2$ (a), 3(b), and 4 electrons (c).
The second row of the plots indicates the results for the dot elongated along the $x$ direction (armchair direction, the spectra given in Fig. \ref{1ewx})
for $N=2$ (d), 3 (e) and 4 electrons (f). Panel (i) corresponds to $N=4$ and $B=5$T.
Panels (g) and (h) show the pair correlation function for $N=4$ and the dot elongated along the $x$ direction.
The fixed position of one of the electrons is marked by the cross. Panel (g) shows the pair correlation function for 
the other electrons with the same spin as the fixed one. In panel (h) the spin of the other electrons
is opposite to the one of the fixed one.
}
 \label{dens1}
\end{figure}

\subsubsection{Confinement along the zigzag direction}

The spectra for the potential of Fig. \ref{sys}(d) with the confinement potential elongated
along the $y$ axis, i.e. in the direction where the effective mass is larger, are given in Fig. \ref{1ewy}.
The numbers $(n_x,n_y)$ given in Fig. \ref{1ewy}(a) near the energy levels at $B=0$
(for $B=0$ and $N=1$ the parity
with respect to inversion along the $x$ and $y$ axes of the $H_0$ eigenstates are definite).
The first excitation within $x$, the direction of thinner confinement, occurs 3 meV above the ground state (see the energy level
marked by (1,0) in Fig. \ref{1ewy}(a)) above 5 states excited in the $y$ direction.
For $N=2$ (Fig. \ref{1ewy}(b))  the singlet-triplet ground state the degeneracy is nearly perfect at $B=0$.
For $B>0$ the $S_z=0$ state remains two-fold degenerate with spin-singlet and spin-triplet 
energy levels that coincide in energy. The separation of the electron charges [Fig. \ref{dens1}(a)]
is complete, and the system is effectively equivalent to a pair of electrons in 
a double quantum dot with vanishing tunnelling between the dots, which produces a vanishing exchange energy \cite{jose,burkard,petta}.
The ground-state degeneracy at $B=0$  is also obtained for $N=3$ (Fig. \ref{1ewy}(c)). The Wigner crystallized
charge density
of the three-electron system at $B=0$ is shown in Fig. \ref{dens1}(b).	
Parity  has no significant
impact on energy once the electron charges are separated
and the $S_z=\pm 1$ energy levels
are two-fold degenerate with respect to parity. Low-energy spin-polarized states $S_z=\pm 3$
occur only in the odd parity.

For $N=4$ the ground state at $B=0$ is only close to the degeneracy (Fig. \ref{1ewy}(d))
with the even parity ground state at $B=0$.  The spin polarization $S_z=\pm 4$ occurs only in  even parity states.
 The structure of the low-energy spectrum is similar to the one found for the circular quantum
 dot [Fig. \ref{scirc}(e)] with even-odd parity splitting of energy levels reduced almost to zero.

The results of Fig. \ref{1ewy}(c,d) for $B>0$ indicate that for $N=3$ ($N=4$) the low-energy spin-polarized 
energy levels occur only with the odd (even) parity symmetry. 
This is in agreement with Ref. \cite{szafran1d} that indicated that in 1D Wigner molecules
for $N=2M$ or $N=2M+1$ with integer $M$ 
the low-energy spin-polarized state has the spatial parity that agrees with the number $M$.
Note that for $N=3$  in the circular potential [Fig. \ref{scirc}(c)], for which
the Wigner molecules were not formed in the real-space charge density,
 both odd and even parity spin-polarized states appear in the ground state for a range of magnetic field. 

\subsubsection{Confinement along the armchair direction}
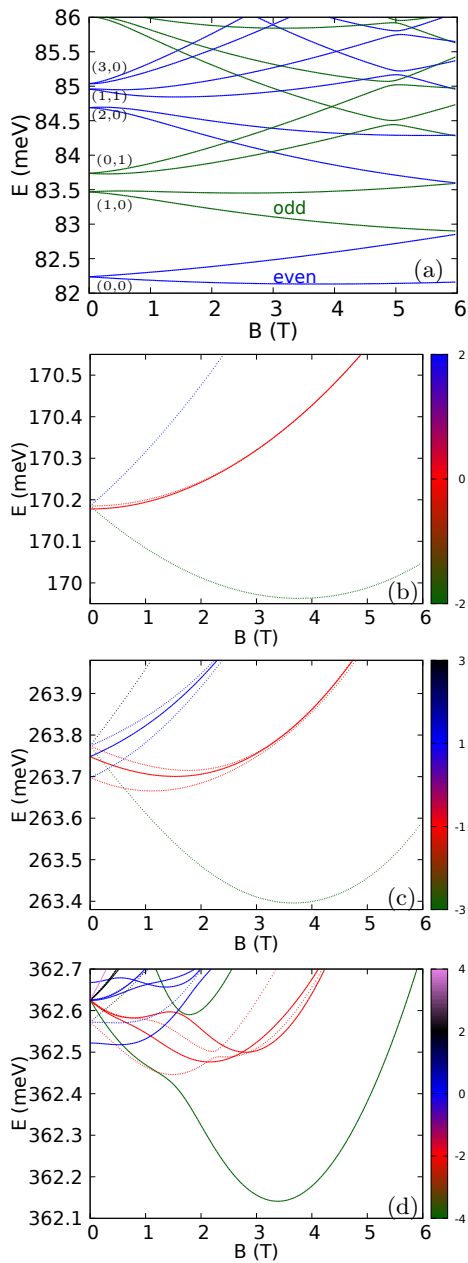
\begin{figure}
 \begin{tabular}{l}
\includegraphics[width=0.7\columnwidth]{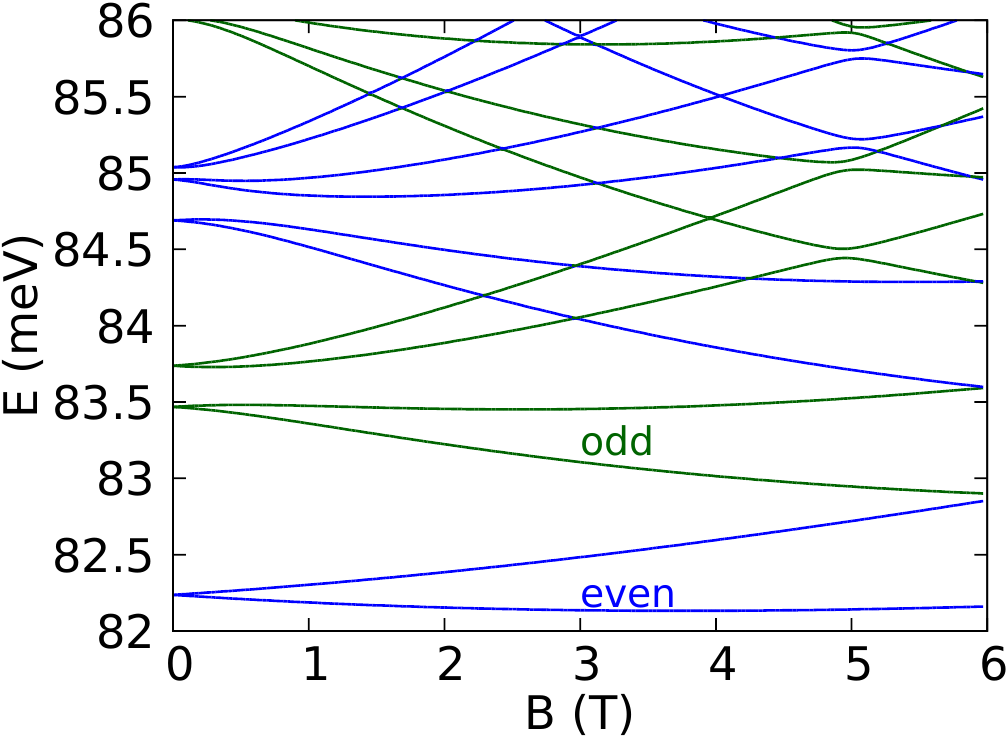} \put(-20,24){(a)}
\put(-140,19){\tiny (0,0)}
\put(-140,50){\tiny (1,0)}\put(-140,67){\tiny (0,1)}\put(-142,84){\tiny (2,0)}
\put(-142,91){\tiny (1,1)}\put(-142,102){\tiny (3,0)}
  \\
 \includegraphics[width=0.7\columnwidth]{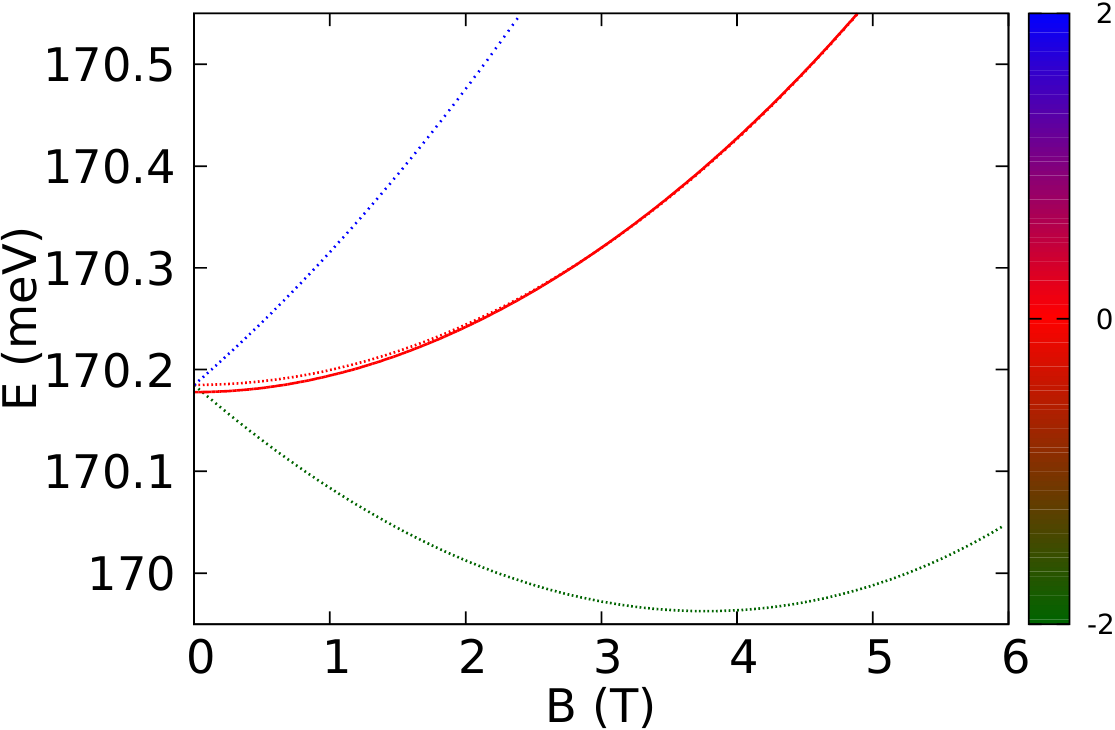} \put(-30,18){(b)}\\
\includegraphics[width=0.7\columnwidth]{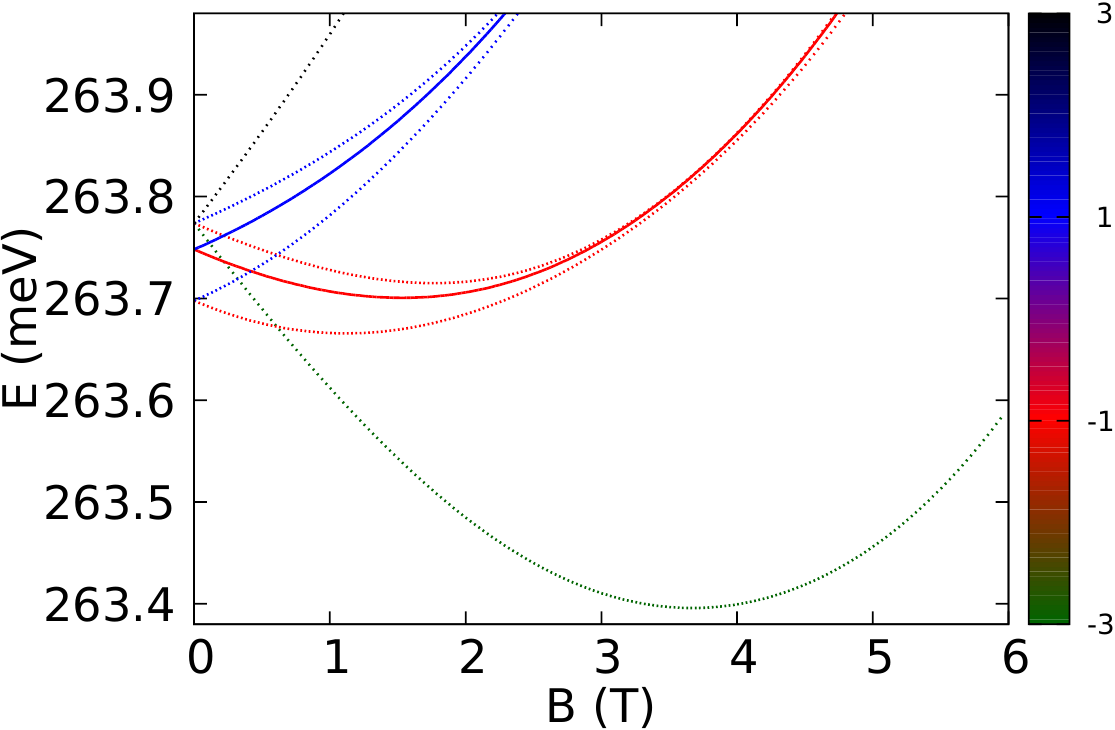}\put(-30,18){(c)}\\
\includegraphics[width=0.7\columnwidth]{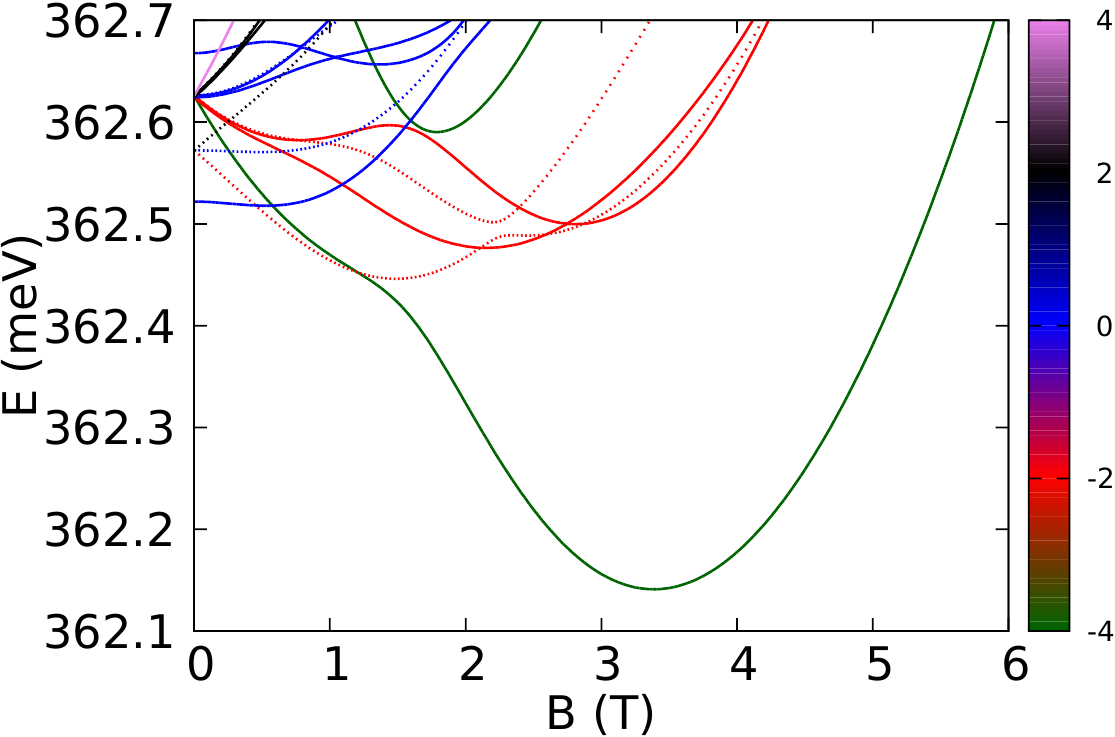}\put(-30,18){(d)}\\
 \end{tabular}
\caption{
Spectra for the potential of Fig. \ref{sys}(d) elongated along the armchair crystal direction ($x$) for $N=1$ (a),2 (b),3 (c),4 (d).
In (a) even and odd parity energy levels are plotted with blue and green lines, respectively.
The notation $(n_x,n_y)$ close to $B=0$ axis shows the number of excitations (sign changes) observed in 
the wave function at zero magnetic field.
The spin degeneracy is lifted for $B>0$ with the spin-down energy levels promoted by the Zeeman interaction.
In (b-d) the even parity energy levels are plotted with the solid lines and the odd parity
energy levels with the dotted lines. The color of the lines stands for the spin.
The color of the lines in (b-d) shows the $S_z$ value. The color scale is separate for each figure and given to the right of the plot.
}
 \label{1ewx}
\end{figure}
For the rectangular gate of Fig. \ref{sys}(a) with the longer side
oriented along the $x$ direction, 
the electron mass along the dot  is light and in the transverse direction the mass is heavy.
For the preceding subsection with confinement along the zigzag direction [Fig. \ref{1ewy}(a)], several lowest energy single-electron states
correspond to the same -- ground state -- energy level of the quantization in the direction
perpendicular to the quantum dot axis.
For the confinement along the armchair direction, the large $m_y$ mass allow
the states with excitations in the direction perpendicular to the confinement axis
to appear low in the energy spectrum [Fig. \ref{1ewx}(a)].
The first single-electron state excited in the transverse direction is the second excited energy level
in Fig. \ref{1ewx}(a) (see the energy level described by (0,1)).
 From this point of view, the system deviates  from the quasi 1D confinement, which should be characterized
 by a large number of excitations along the quantum dot below the energy when the first transverse excitation occurs. 
However, the charge density of the $N$-electron states is elongated along the $x$ direction
(see Fig. \ref{dens1}(d,e,f)). The lifting of the even-odd degeneracy is observed in the spectrum
for $N=2$ and $N=3$ [Fig. \ref{1ewx}(b,c)]. A lower value of $m_x$ allows for a non-negligible 
electron tunneling between the single-electron islands (cf. the charge density
in between the single-electron maxima for $N=2$ and $N=3$ in Fig. \ref{dens1}(a,d) and Fig. \ref{dens1}(b,e)),  thus lifting the degeneracy at $B=0$.
Near $B=0$, the spectrum for $N=4$ (Fig. \ref{1ewx}(d)) has a distinctly different character than for the zigzag confinement 
(Fig. \ref{1ewy}(d)) and for the circular confinement [Fig. \ref{scirc}(e)]. In both preceding cases
the electrons were separated in the single electron islands [Fig. \ref{sd}(e,f), Fig. \ref{dens1}(c)].
The ground state charge density for the armchair confinement possesses two single-electron
islands at the ends of the dot and a lower but more extended central maximum [Fig. \ref{dens1}(f) for $B=0.01$T and Fig. \ref{dens1}(i) for $B=5$T].
The charge densities of the two central electrons do not separate into  single-electron islands.
The effect can be attributed to both a large value of $m_y$ that allows the states
excited in the $y$ direction to contribute to the interacting states and a small value of $m_x$
which makes the formation of the single-electron islands along the $x$ direction more expensive
in terms of the kinetic energy.

Fig. \ref{dens1}(g,h) shows the spin-resolved pair correlation function for one of the electron positions
fixed in the point marked by the cross for $B=0.01$T. Fig. \ref{dens1}(g) (Fig. \ref{dens1}(h)) corresponds to the electron distribution
with the same (opposite) spin as the fixed electron spin. 
The electrons at the central maximum of the charge distribution [Fig. \ref{dens1}(f)]
possess opposite spin. 
Note that, with the electrons of the extreme ends of the dots, separated from the central
density island, the system at $B=0$ acquires a singlet / triplet spectral structure (Fig. \ref{1ewx}(d)) similar to the one for
the two-electron system in the circular quantum dot (Fig. \ref{scirc}(b)) -- the central two-electron system governs the form of the low energy
part of the spectrum. 

\section{Discussion}
The systems confined in quantum dots weakly coupled to  electron reservoirs are studied with the transport spectroscopy using the Coulomb blockade phenomenon \cite{leo,eqd}.
In the Coulomb blockade regime, the flow of the current across the dot is stopped 
when the chemical potential of the confined $N$-electron is outside the transport window 
defined by the Fermi levels of the source and drain. For a small voltage drop between the reservoirs,
the position of the chemical potential can be very precisely determined. 
The chemical potential $\mu_N=E_N-E_{N-1}$ is defined by the ground state energies of systems with $N$ and $N-1$ electrons  \cite{leo,eqd}. The ground-state energy crossing in the $N$ electron system produces $\Lambda$-shaped cusps
in the charging line as a function of the external magnetic field for the $N$-th electron added to the confined system, while the ground-state transitions for the $N-1$ system produces $V$-shaped cusps.
Transport spectroscopy allows reconstruction of the energy spectra with
a precision of the order of a few $\mu$eV \cite{fuhrer}.
The results presented above indicate that the formation of the Wigner molecule in the laboratory frame
leads to a near degeneracy of the ground state near $B=0$. The larger the electron separation
in the single-electron charge islands, the closer the degeneracy at $B=0$.
 The ground state becomes spin-polarized
at low magnetic field and no further ground-state transitions are observed. 
The systems without the Wigner molecular charge density undergo a number
of ground-state transitions that also appear  at higher magnetic field. 
The transport spectroscopy technique can also be used for detection of the
excited part of the spectra when the corresponding energy level enters the transport window \cite{leo,fuhrer}.
Detection of a dense set of levels near the ground state can be used as a signature of Wigner molecule formation \cite{corrigan,pecker}.

\section{Summary and conclusions}
We have studied the system of a few electrons in an electrostatic quantum dot confinement induced
within a phosphorene layer. A version of the configuration interaction approach dealing with
the strong electron-electron interaction effects has been developed. 
We indicated formation of Wigner molecules with single-electron islands separated in the real
space in a system of realistic yet small size. In circular quantum dots, Wigner crystallization
in the laboratory frame occurs due to the lowered Hamiltonian symmetry with the anisotropy
of the electron effective mass. The Wigner molecules appear in the laboratory frame when
the distribution of the single-electron islands is consistent with the Hamiltonian symmetry.
For circular quantum dots we found Wigner molecules for 2 and 4 electrons but not for 3 
electrons i.e. the conditions for the Wigner crystallization are not directly determined by the 
mean electron-electron distance. For 5 electrons in the circular confinement two forms of the charge density with or without
single-electron islands appear in the ground-state depending on the value of the external magnetic field. 
Formation of the Wigner molecules in a quasi-1D confinement
calls for orientation of the confinement potential with longer axis along the zigzag direction
since larger effective mass promotes the single-electron islands localization and the smaller 
armchair mass supports the quasi 1D confinement.
We studied the spectra for both circular and elongated quantum dots to find  the signatures
of the Wigner crystallization in  real space. The systems with single-electron islands
forming Wigner molecules in the laboratory frame are characterized by near degeneracy of the 
ground state at $B=0$ followed by spin polarization in low magnetic field and an energy gap
between the nearly degenerate ground state and excited states. These features are missing
for systems that do not form single-electron islands. Resolution of these signatures
is within the reach of transport spectroscopy techniques.

\section*{Acknowledgments}
This work was supported by the National Science Centre
(NCN) according to decision DEC-2019/35/O/ST3/00097.
Calculations were performed on the PLGrid infrastructure.

\section*{Appendix}

\renewcommand{\thefigure}{A\arabic{figure}}
\subsection*{Single-electron eigen-states with $H_0$, $H_0'$ and $H_0''$}
 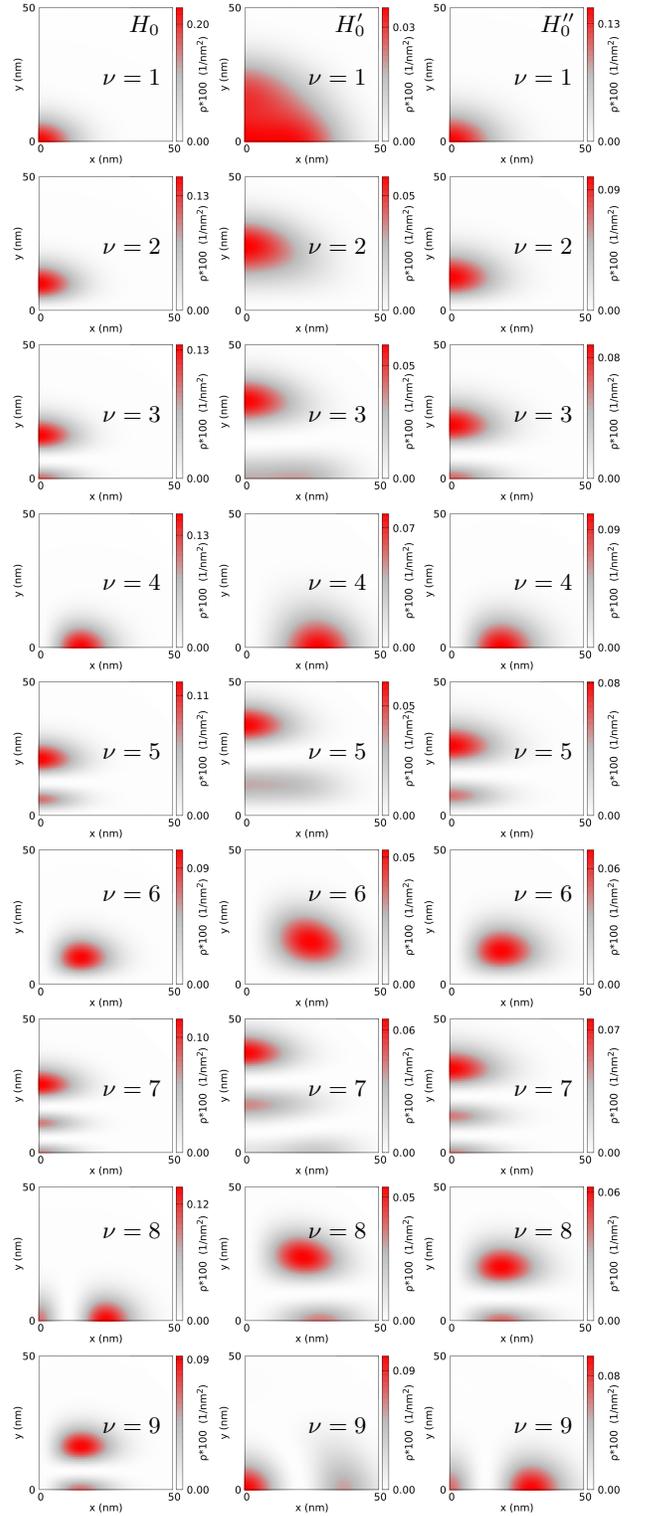
\begin{figure}
  \begin{tabular}{lll}
   \includegraphics[width=0.3\columnwidth]{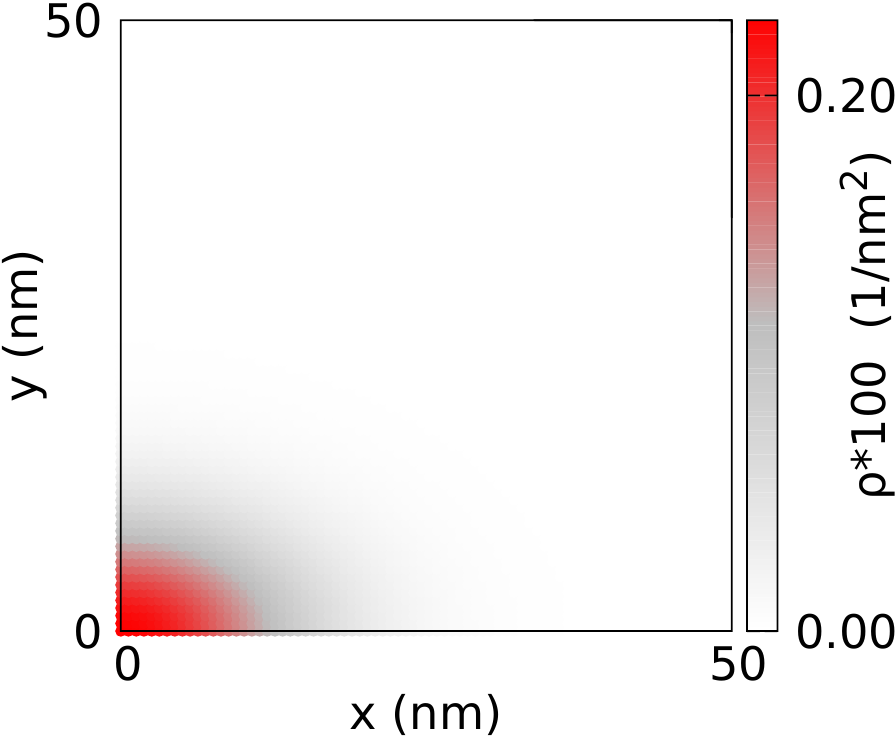}\put(-30,50) {$H_0$}\put(-40,30) {$\nu=1$}&  \includegraphics[width=0.3\columnwidth]{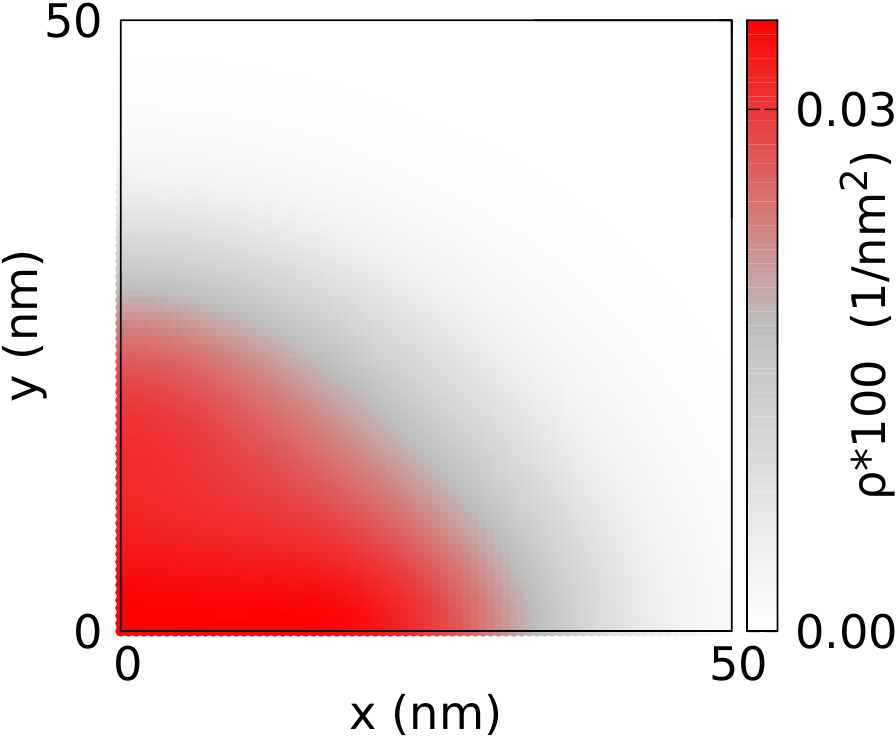} \put(-30,50) {$H_0'$}\put(-40,30) {$\nu=1$}&    \includegraphics[width=0.3\columnwidth]{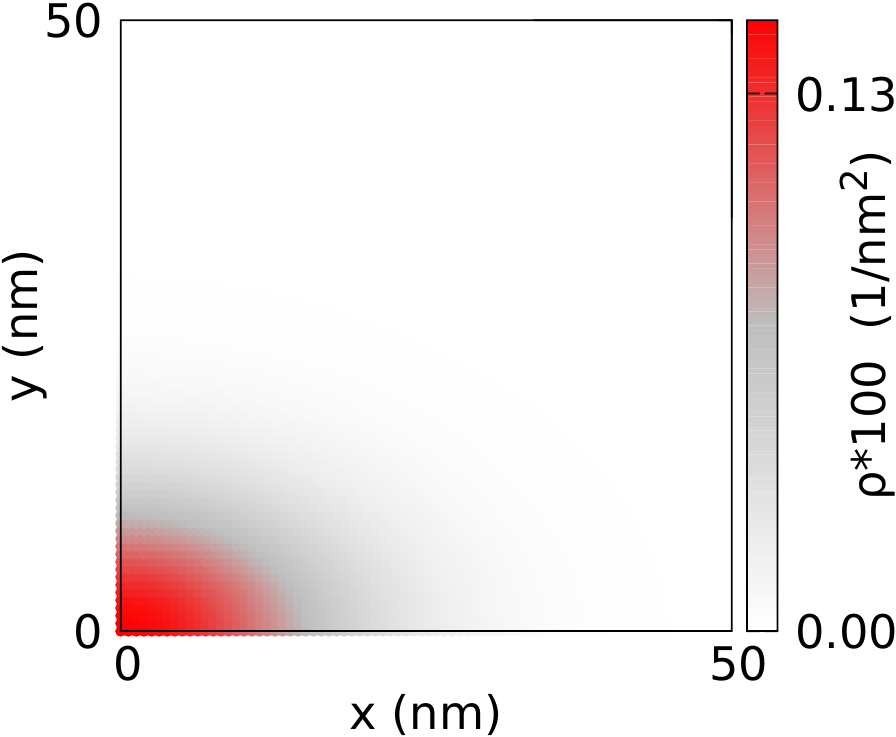}\put(-30,50) {$H_0''$}\put(-40,30) {$\nu=1$}\\
      \includegraphics[width=0.3\columnwidth]{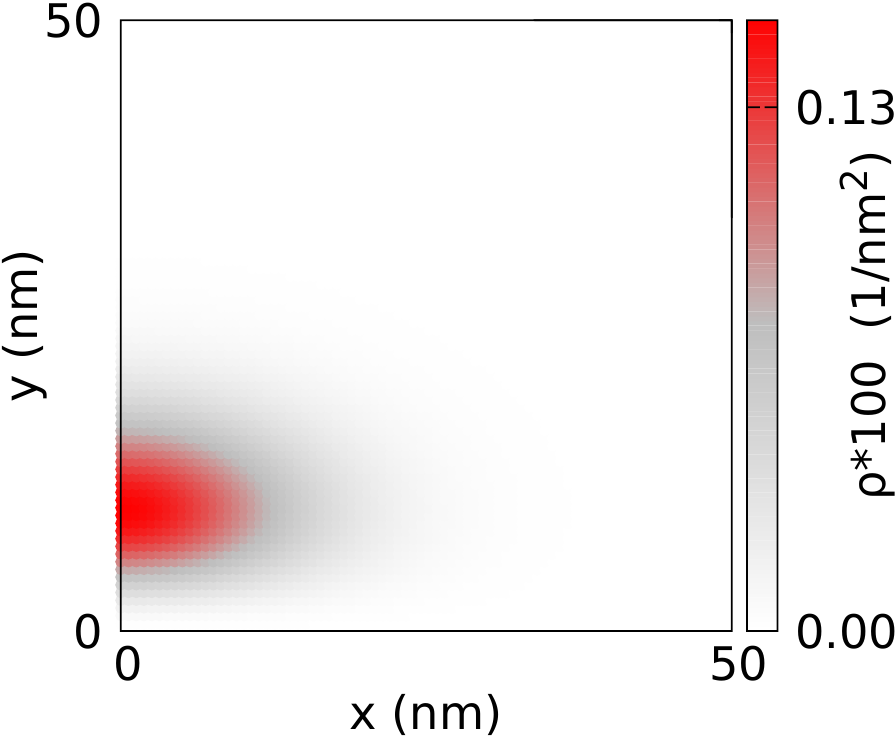}\put(-40,30) {$\nu=2$}  &\includegraphics[width=0.3\columnwidth]{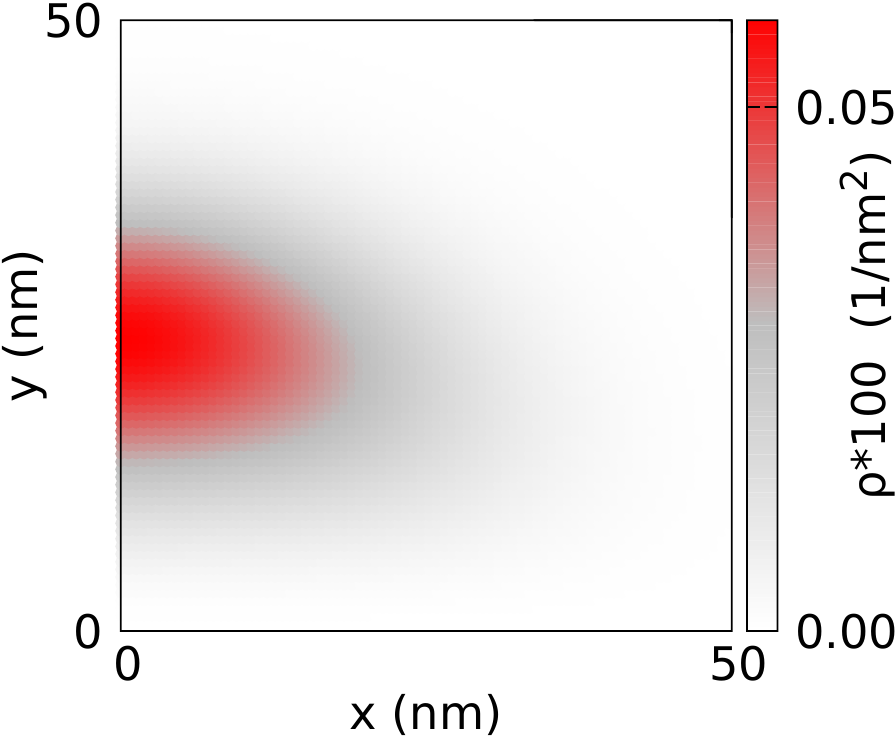} \put(-40,30) {$\nu=2$}&    \includegraphics[width=0.3\columnwidth]{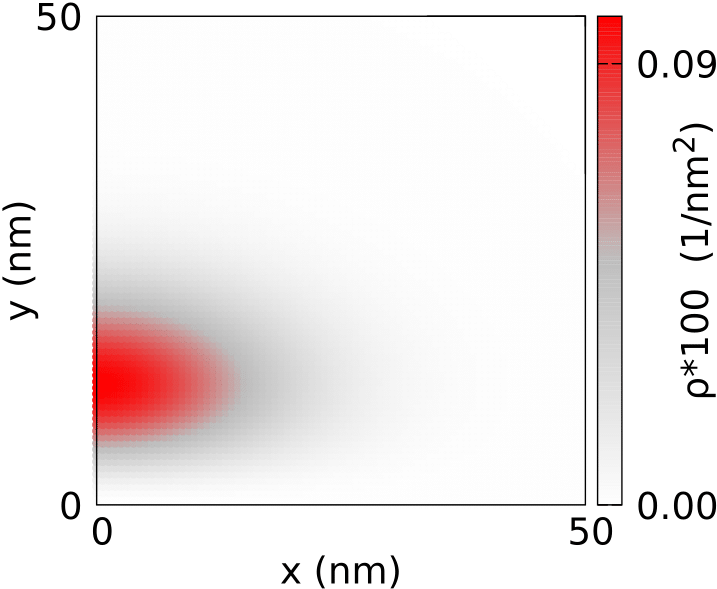}\put(-40,30) {$\nu=2$}\\
      \includegraphics[width=0.3\columnwidth]{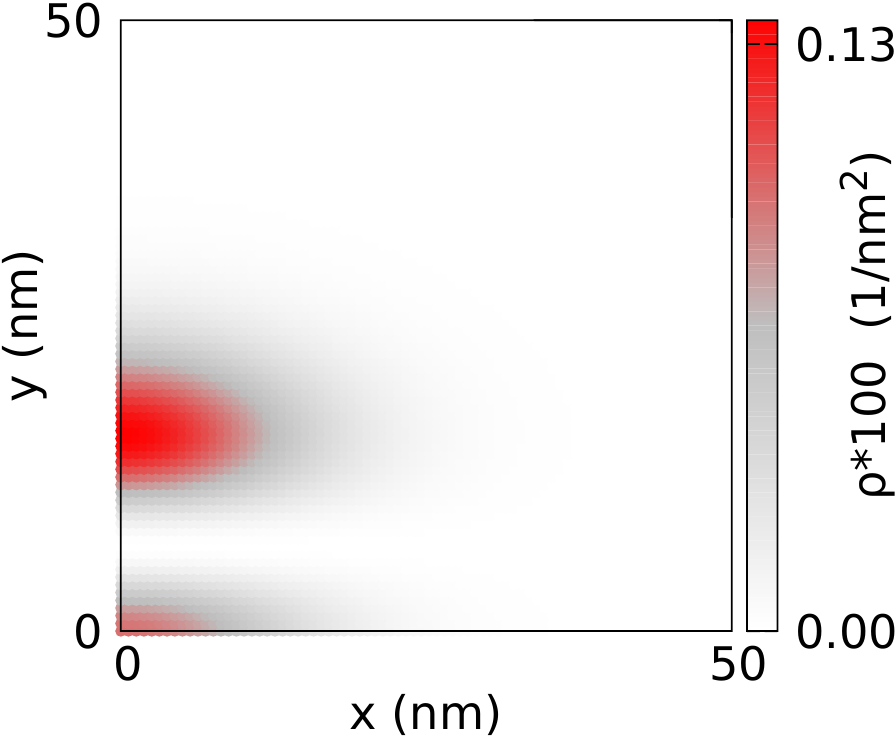}\put(-40,30) {$\nu=3$}&  \includegraphics[width=0.3\columnwidth]{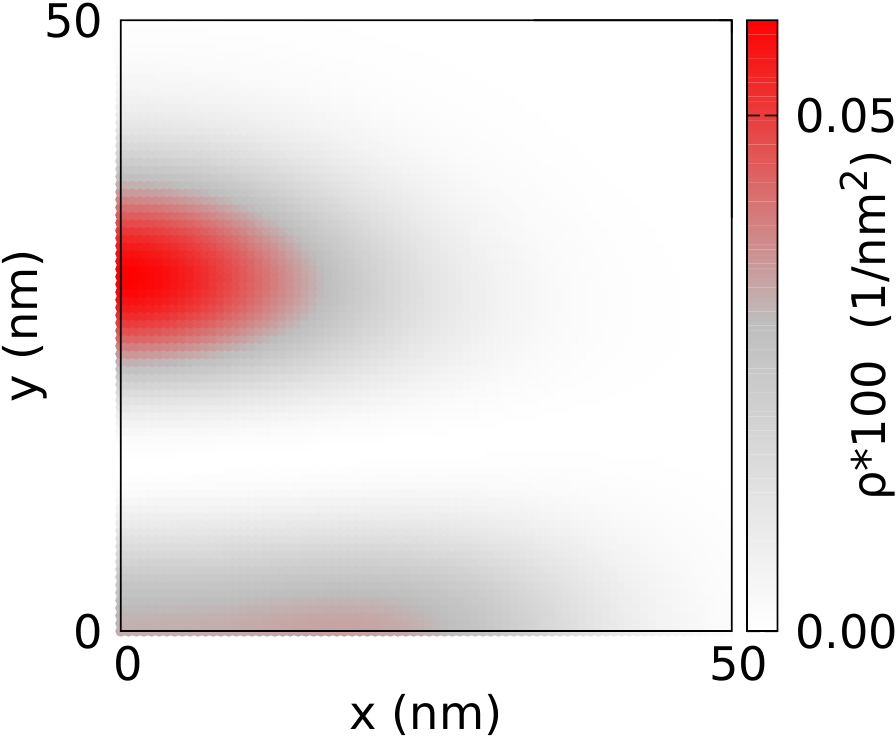} \put(-40,30) {$\nu=3$}&    \includegraphics[width=0.3\columnwidth]{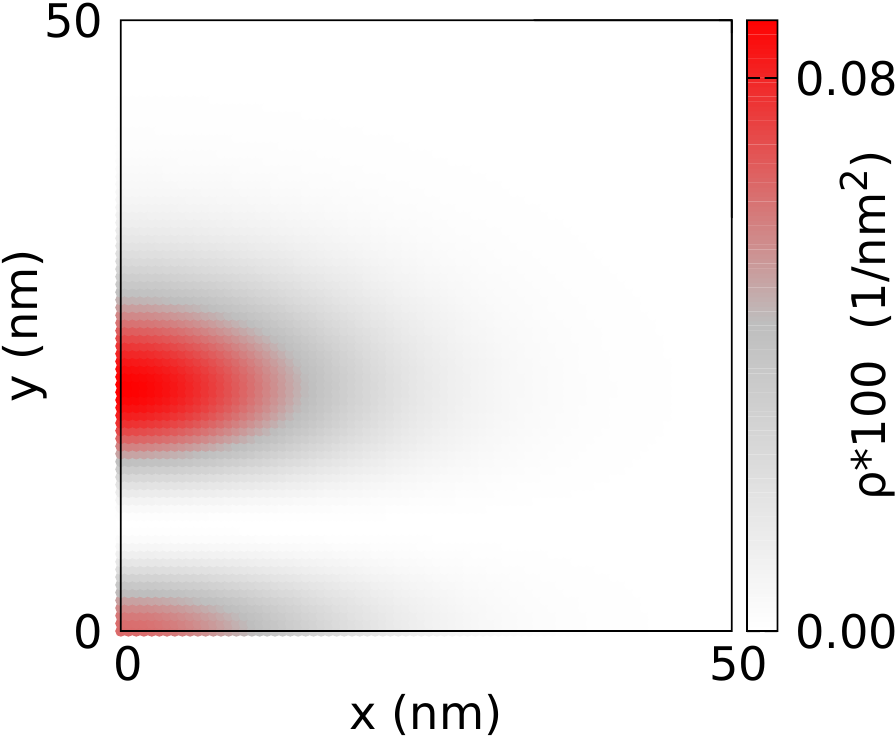}\put(-40,30) {$\nu=3$}\\
      \includegraphics[width=0.3\columnwidth]{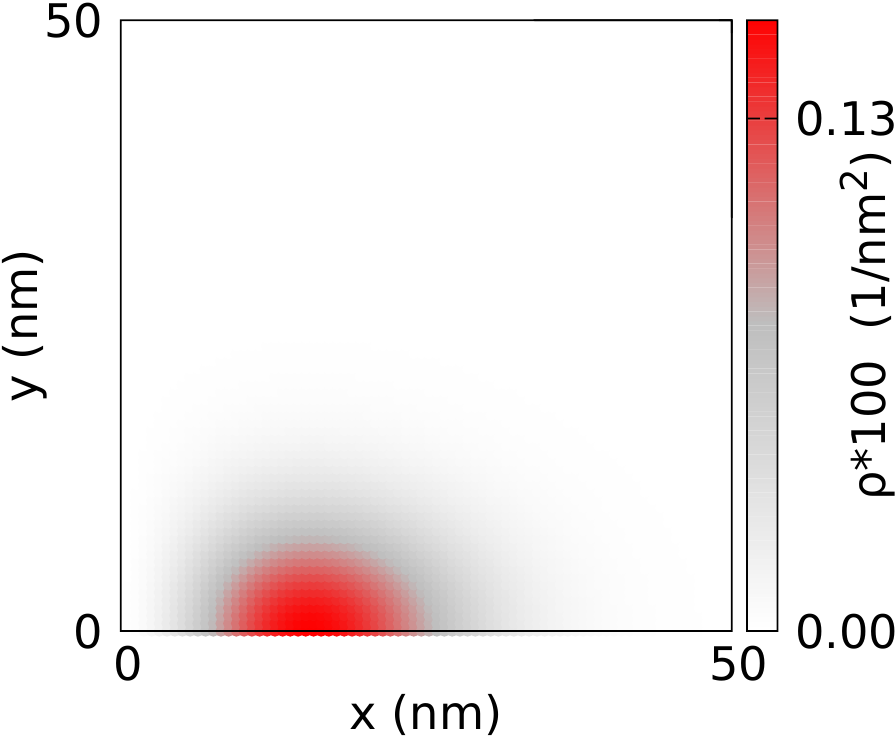}\put(-40,30) {$\nu=4$}&  \includegraphics[width=0.3\columnwidth]{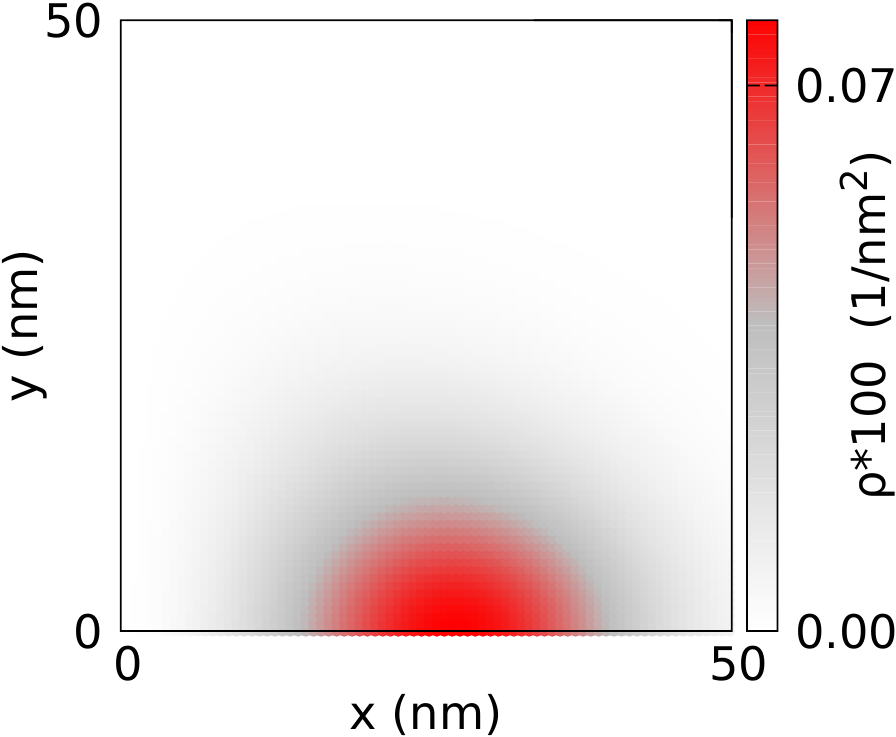} \put(-40,30) {$\nu=4$}&    \includegraphics[width=0.3\columnwidth]{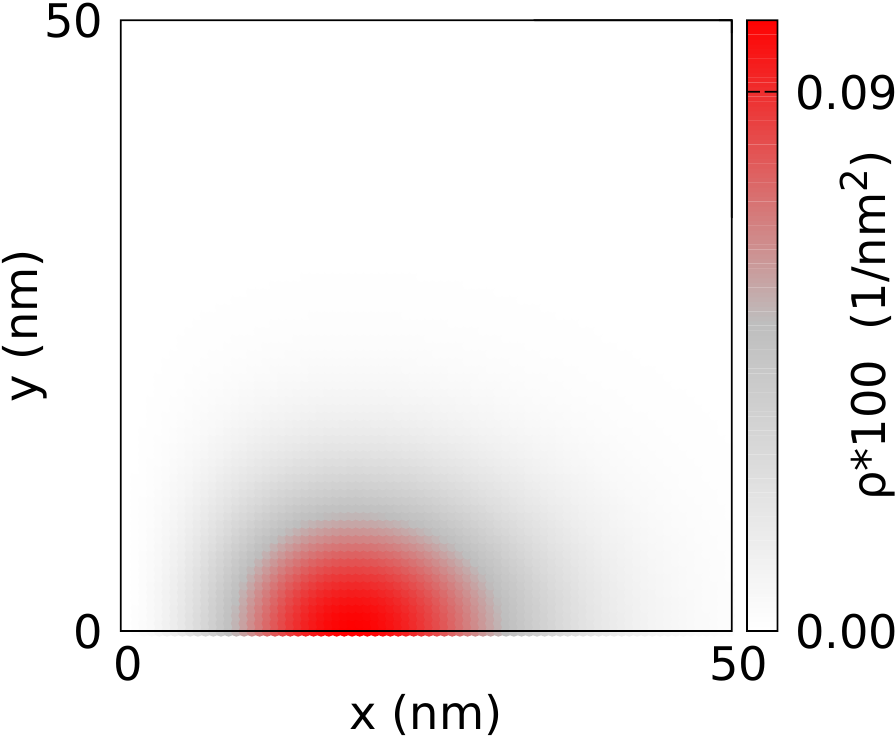}\put(-40,30) {$\nu=4$}\\
            \includegraphics[width=0.3\columnwidth]{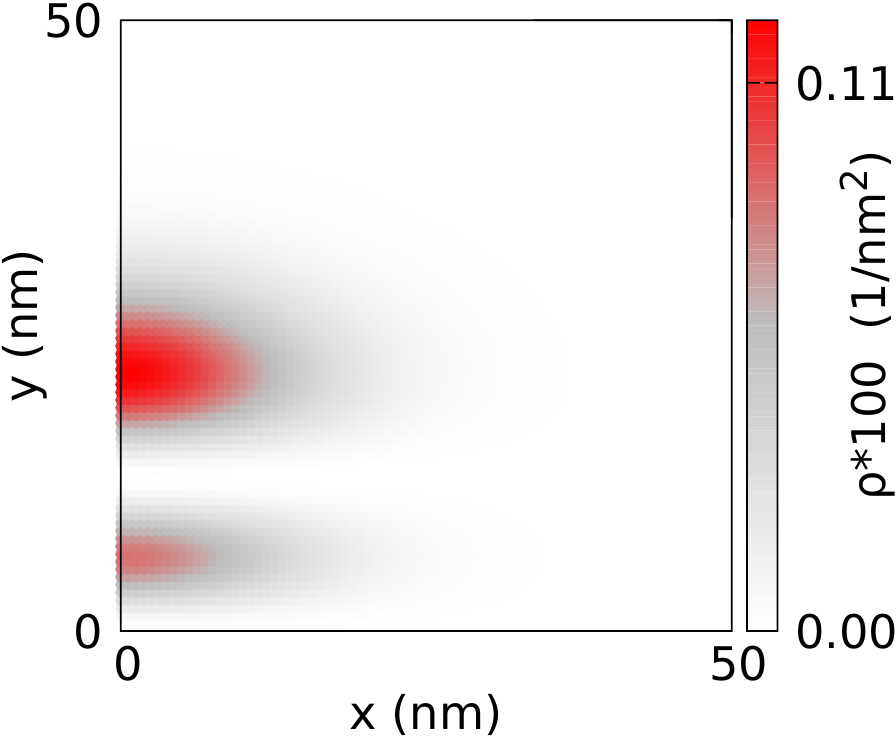}\put(-40,30) {$\nu=5$}&  \includegraphics[width=0.3\columnwidth]{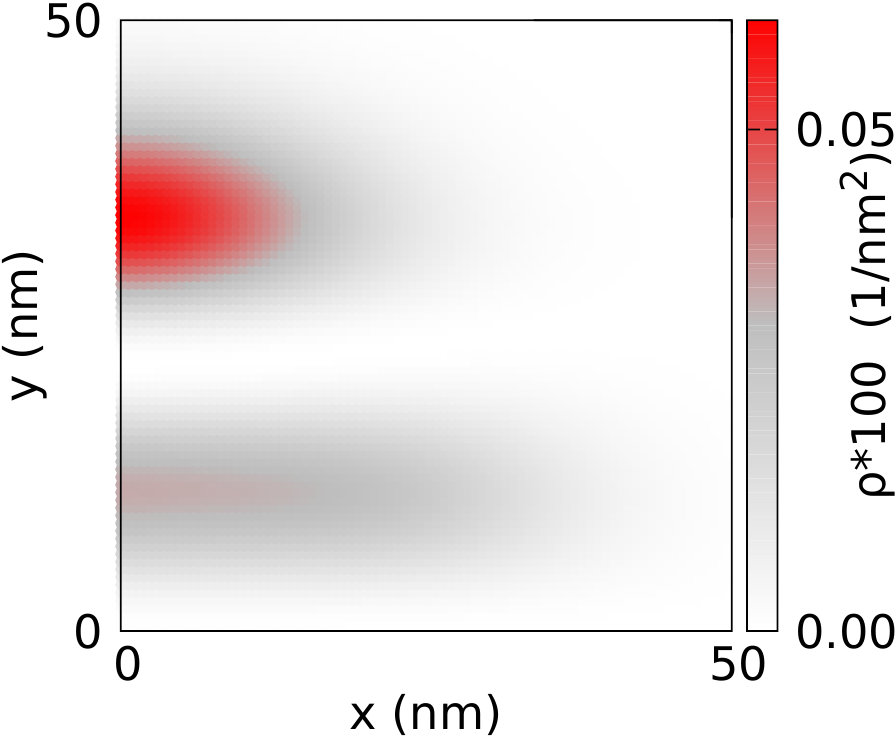} \put(-40,30) {$\nu=5$}&    \includegraphics[width=0.3\columnwidth]{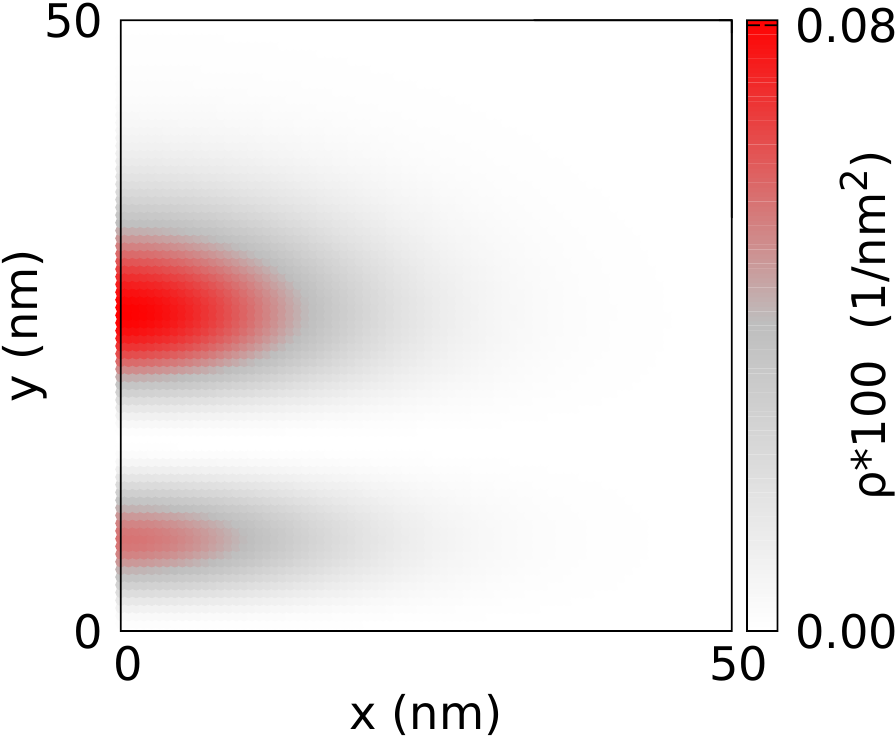}\put(-40,30) {$\nu=5$}\\
            \includegraphics[width=0.3\columnwidth]{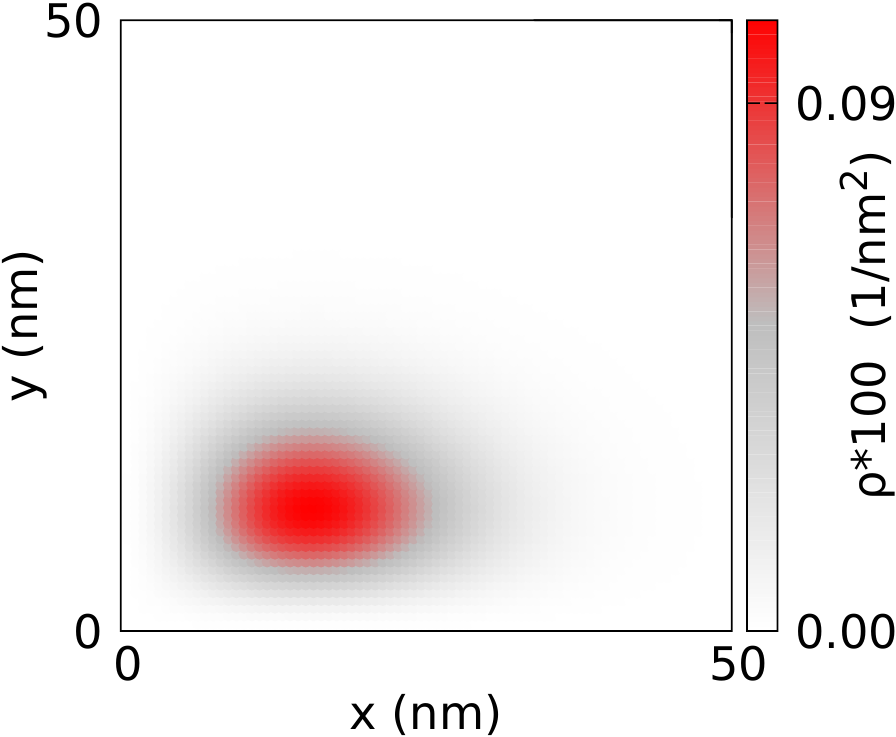}\put(-40,40) {$\nu=6$}&  \includegraphics[width=0.3\columnwidth]{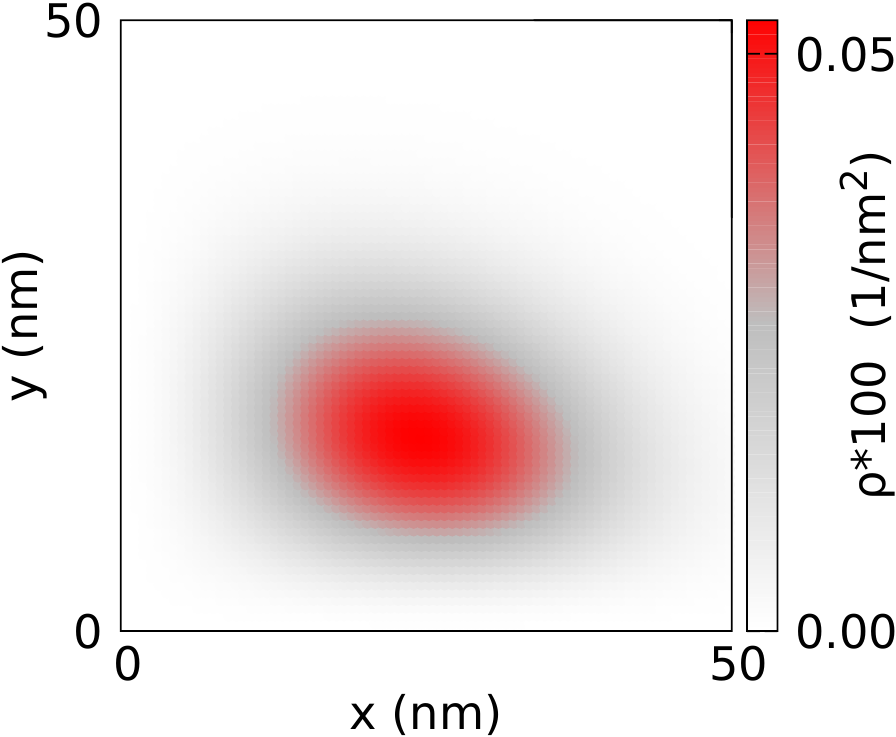} \put(-40,40) {$\nu=6$}&    \includegraphics[width=0.3\columnwidth]{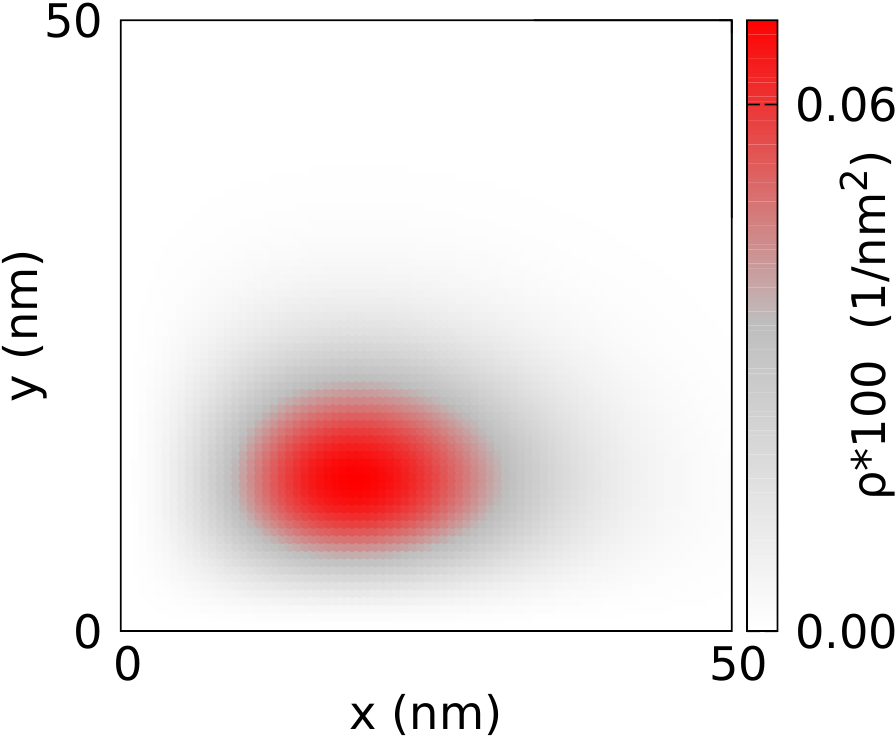}\put(-40,40) {$\nu=6$}\\
                        \includegraphics[width=0.3\columnwidth]{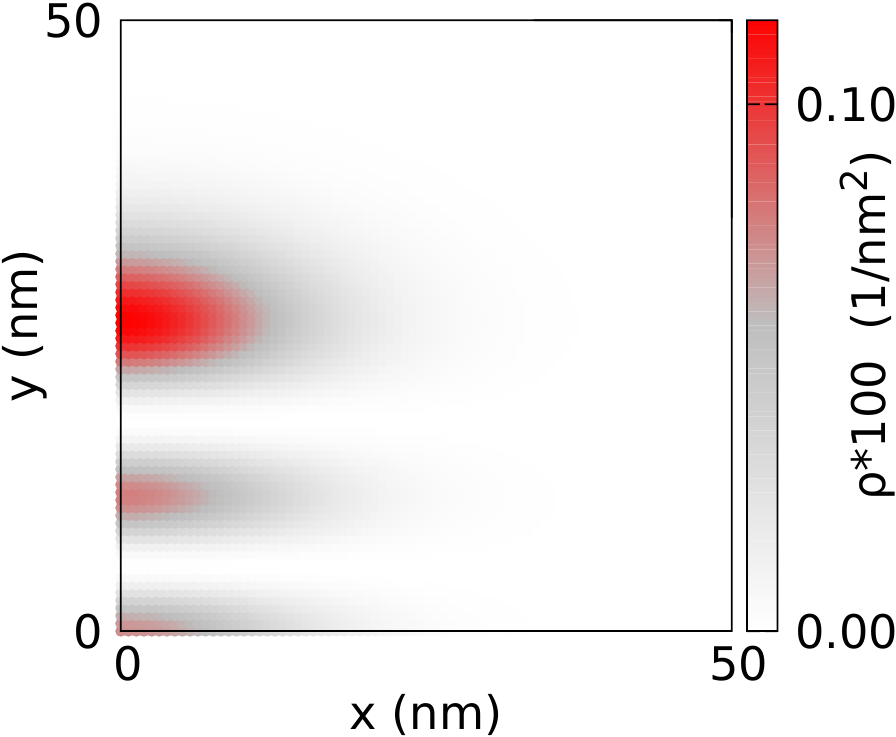}\put(-40,30) {$\nu=7$}&  \includegraphics[width=0.3\columnwidth]{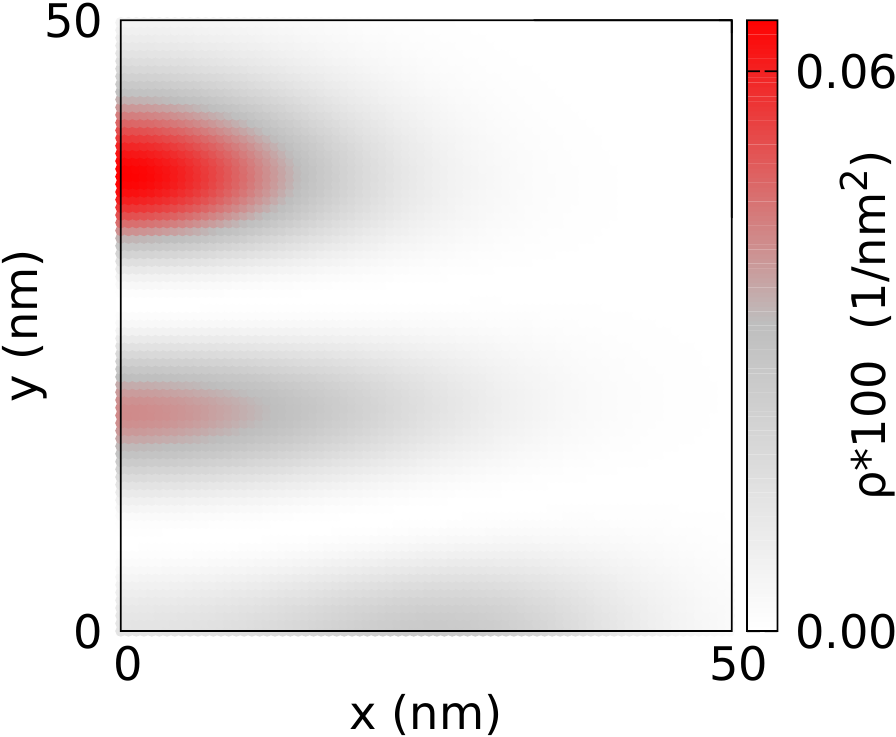} \put(-40,30) {$\nu=7$}&    \includegraphics[width=0.3\columnwidth]{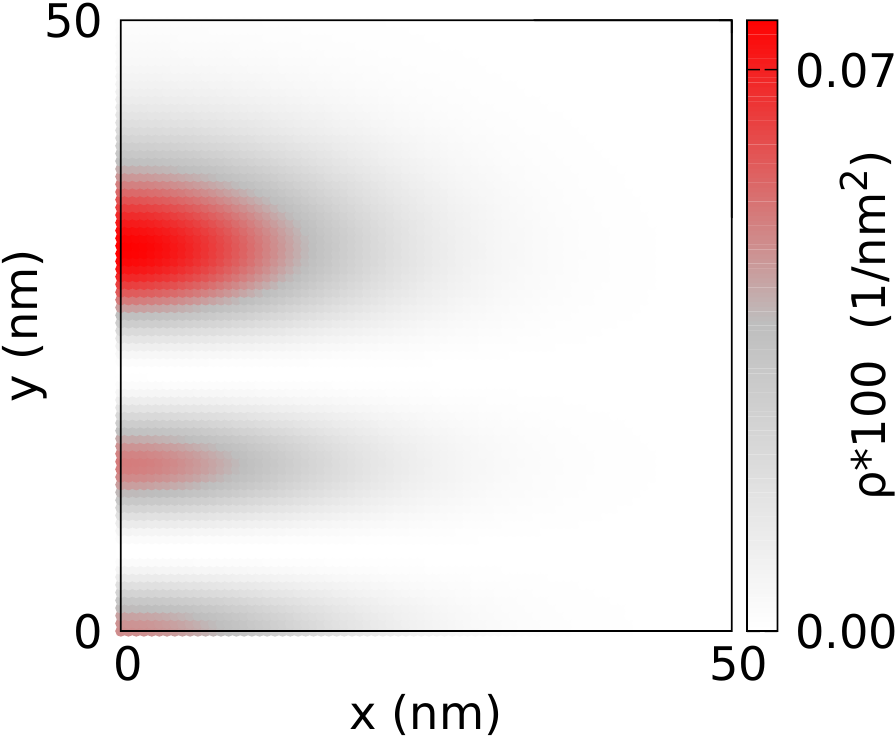}\put(-40,30) {$\nu=7$}\\
                                    \includegraphics[width=0.3\columnwidth]{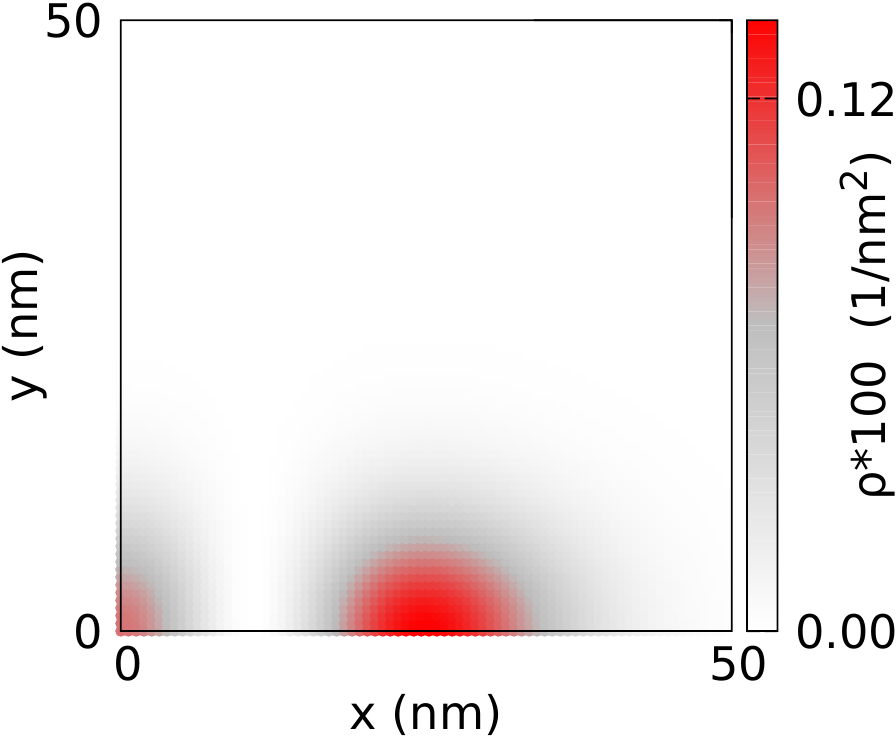}\put(-40,40) {$\nu=8$}&  \includegraphics[width=0.3\columnwidth]{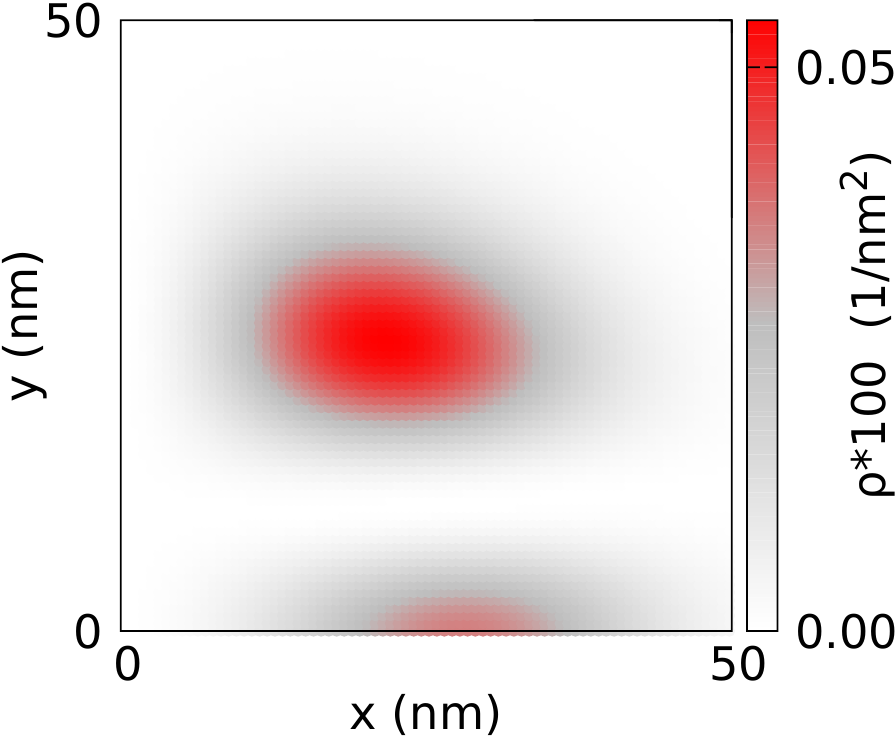}\put(-40,40) {$\nu=8$} &    \includegraphics[width=0.3\columnwidth]{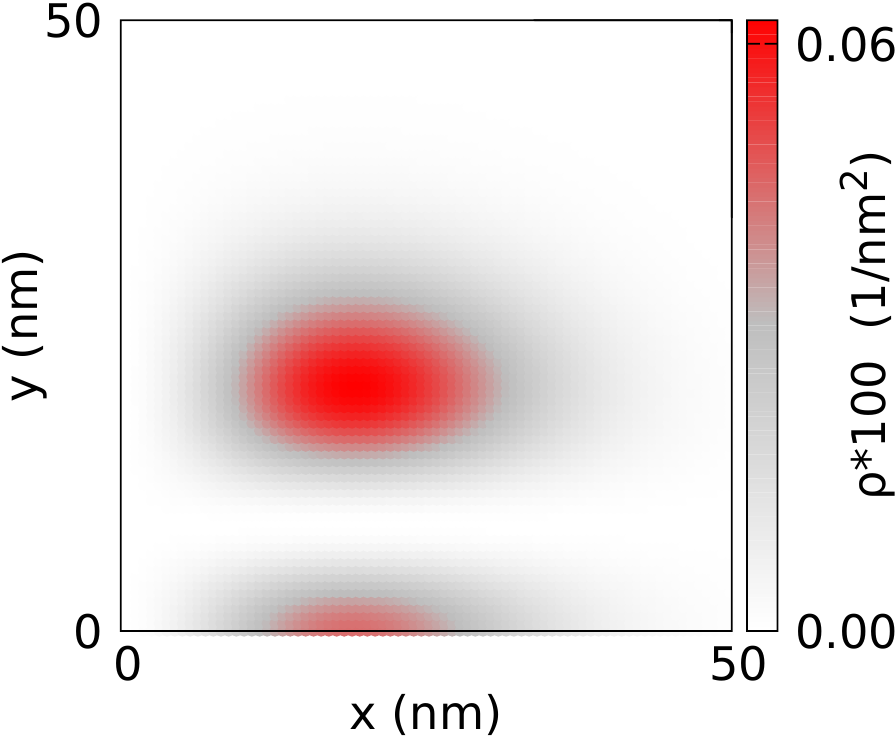}\put(-40,40) {$\nu=8$}\\
                                    \includegraphics[width=0.3\columnwidth]{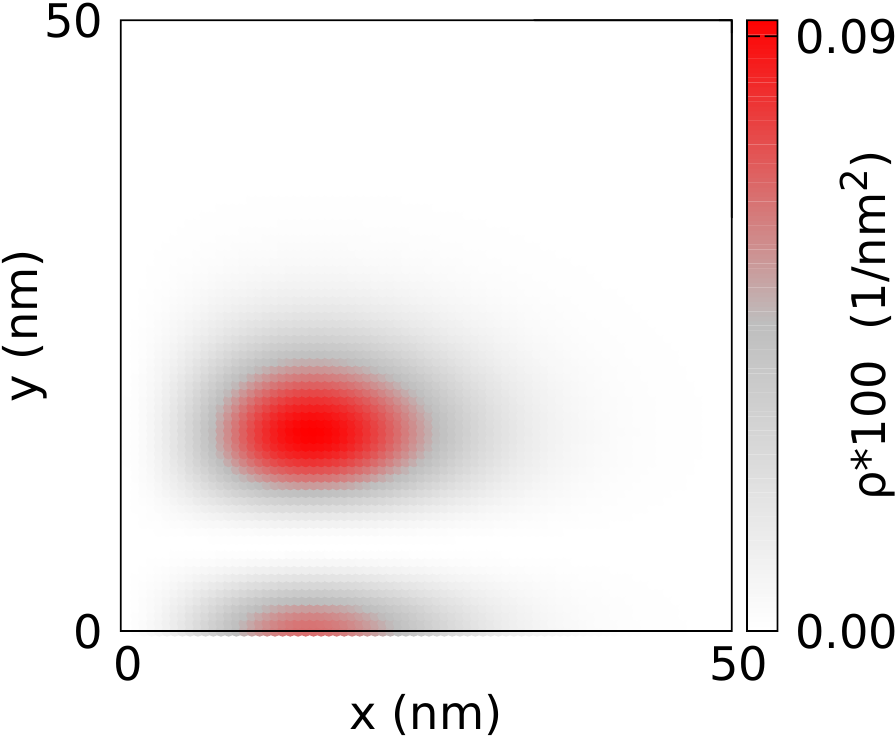}\put(-40,30) {$\nu=9$}&  \includegraphics[width=0.3\columnwidth]{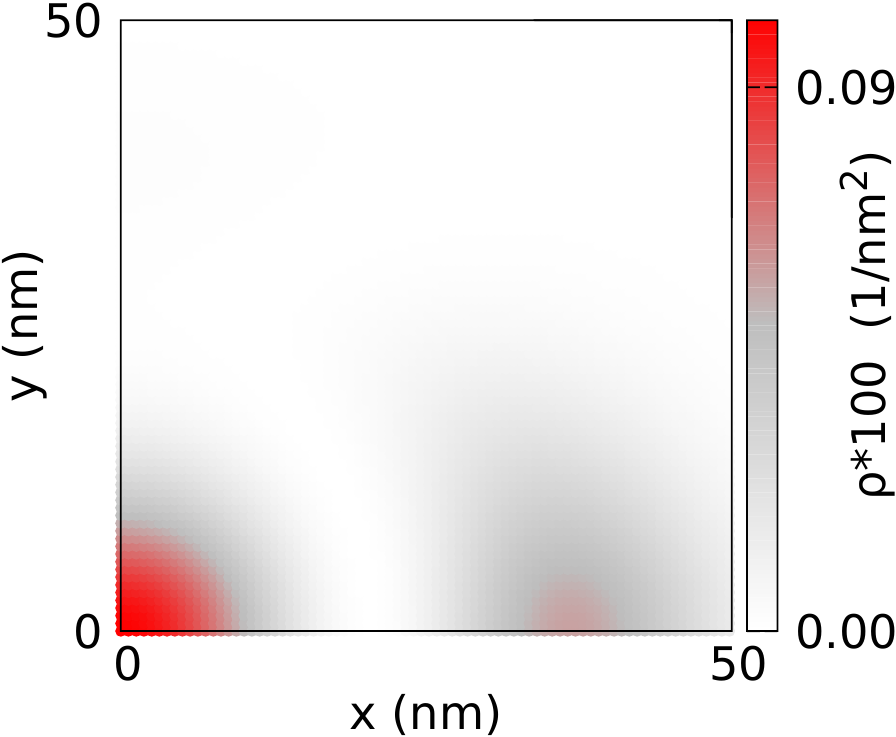}\put(-40,30) {$\nu=9$} &    \includegraphics[width=0.3\columnwidth]{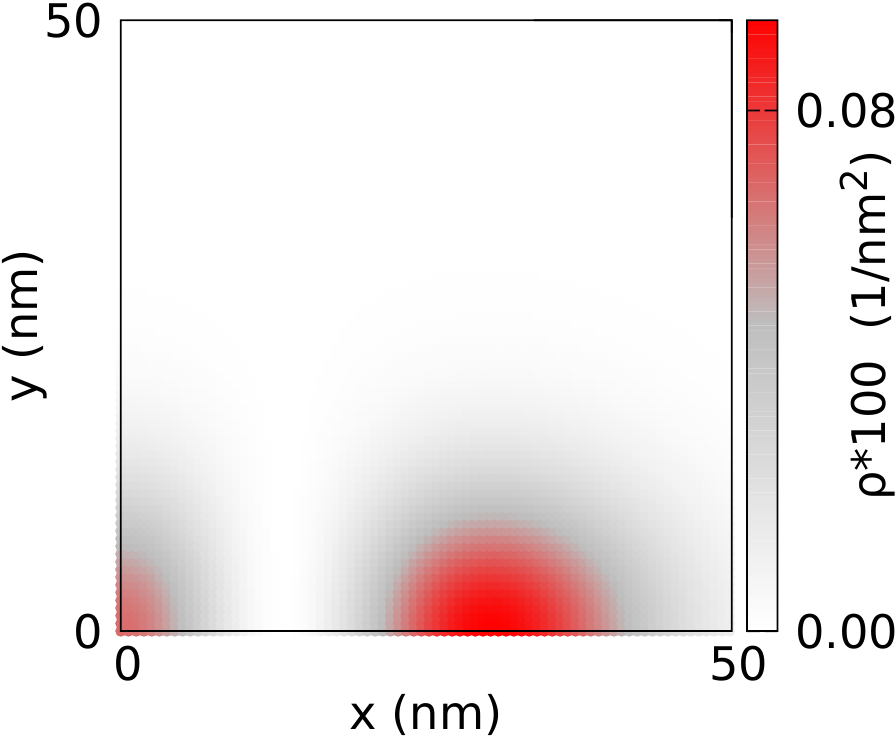}\put(-40,30) {$\nu=9$}    \end{tabular}
\caption{Probability density of spin-down single-electron states ordered from the ground-state ($\nu=1$ -- the first row) to 8-th excited spin-down energy level ($\nu=9$ -- the last row). The left column corresponds to the 
states of the bare circular confinement with potential $W$ (Hamiltonian $H_0$). The central column to the potential $W'$ -- with
an added central Gaussian with 8 meV, 35 meV (Hamiltonian $H_0'$). The right column correspond to states
obtained with potential $W''=0.455\times W$ (Hamiltonian $H_0''$). The parameters of the central and the right column
are optimized for diagonalization of the 4-electron system.} \label{wf1a}

 \end{figure}
Figure 2 presents the convergence of the CI method
the single-electron Hamiltonians $H_0$, $H_0'$ and $H_0''$ using
the bare QD potential $W$, the potential with the central repulsive Gaussian ($W'$)
and the reduced potential $W''$, respectively.
The low-energy single-electron wave functions are plotted in Fig. \ref{wf1a} for the three Hamiltonians
starting from the ground-state in the first row, and the subsequent excited states presented in the lower
panels of Fig. \ref{wf1a}. The seven lowest-energy states have the same character for all Hamiltonians
only with states for $H_0'$ (central column of plots) and $H_0''$ (right column) covering
a larger area than the ones for $H_0$, which turns out to speed-up the convergence of the CI method
for the system of size increased by the strong electron-electron interaction.

\subsection*{Choice of the computational box}

\begin{figure}
  \includegraphics[width=0.5\columnwidth]{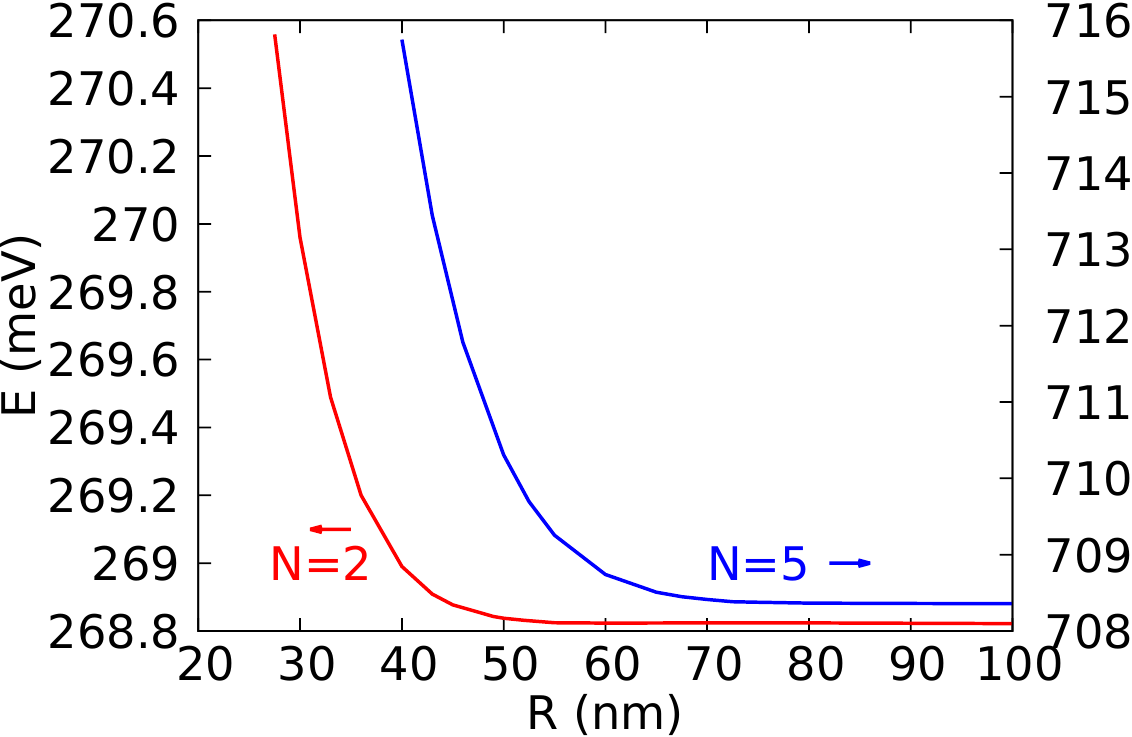} 
\caption{The ground-state energy for a circular quantum dot holding two (red line, left energy axis)
and five electrons (blue line, right energy axis) as a function of the radius of the computational box for $60$ single-electron
eigenstates of hamiltonian $H_0'$ used in construction of the basis for configuration interaction.}  \label{size}
\end{figure}

The CI method uses the single-electron wave functions that are obtained with the finite difference
approach in a finite computational box with the boundary conditions of vanishing wave function at the edges
of the box. The edges of the box form effectively an infinite quantum well that has to be chosen large enough to contain the few-electron system without perturbation to the low-energy states.
The influence of the size of the box on the results is given in Fig. \ref{size} for the circular quantum dot.
For the circular quantum dot we use the square grid of points of the spacing of $\Delta x=R/nx$
where $R$ is the radius of the computational box and we take $n_x=111$. 
The red (black) line in Fig. \ref{size} shows the ground-state energy for $N=2$ ($N=5$).
The growth of the energy for small $R$ is due to the finite-size effect with the quantum well ground state
changing as $1/R^2$. For calculations for $N=2$ the radius of $R=50$ nm is large enough
while for $N=5$ is has to be taken as large as $R=80$ nm.

\subsection*{Spectra without the spin Zeeman interaction}
\begin{figure*}
 \begin{tabular}{lll}
  \includegraphics[width=0.5\columnwidth]{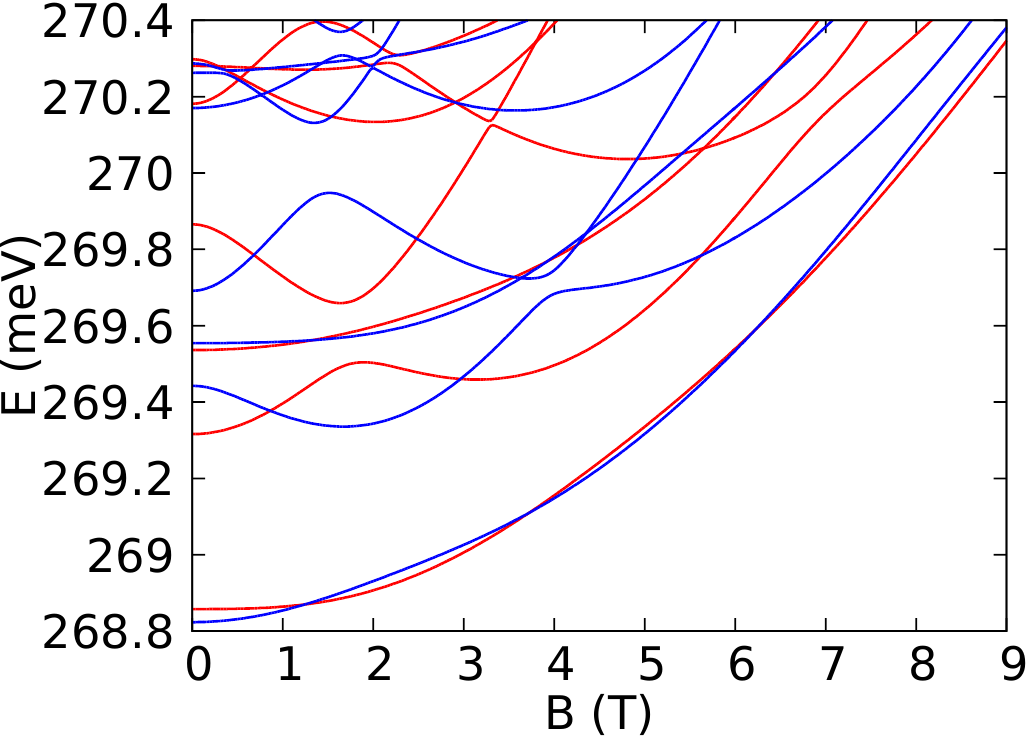} \put(-20,18){(a)} \put(-65,15){ N=2} 
  \includegraphics[width=0.5\columnwidth]{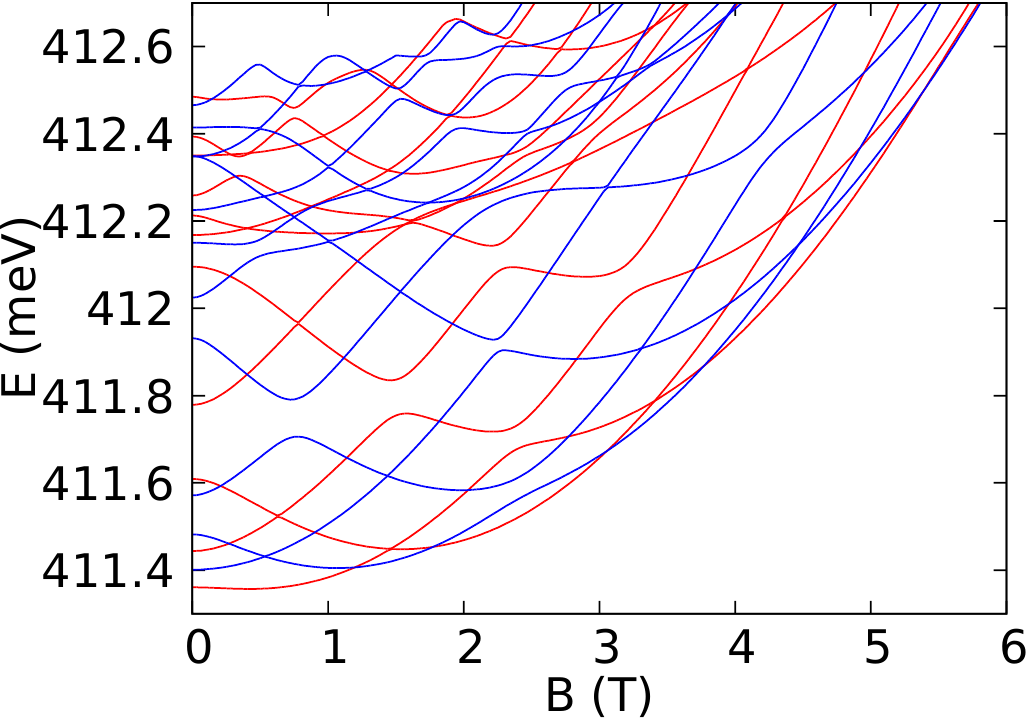} \put(-20,18){(b)} \put(-65,15){ N=3}
    \includegraphics[width=0.5\columnwidth]{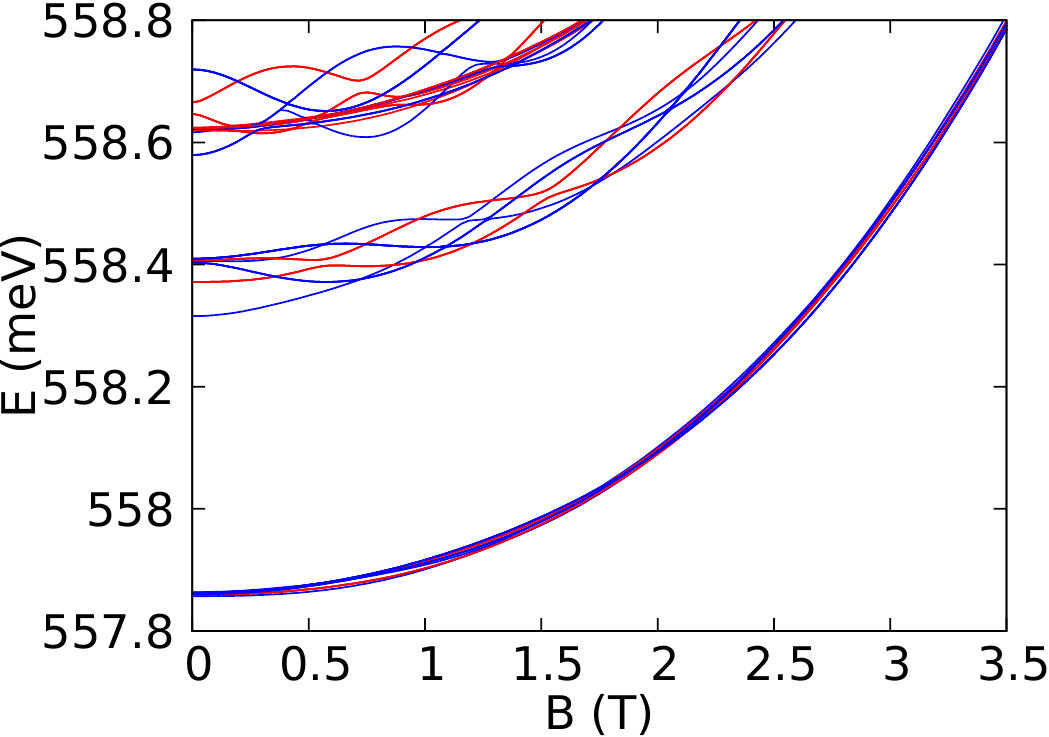} \put(-20,18){(c)} \put(-70,35) {N=4}
 \end{tabular}
\caption{Spectra of $N=2$, 3, and 4  electron systems in circular potential [as in Fig. 3] but without
the spin Zeeman interaction. The blue (red) lines show the energy levels of even (odd) parity.}  \label{noz}
\end{figure*}
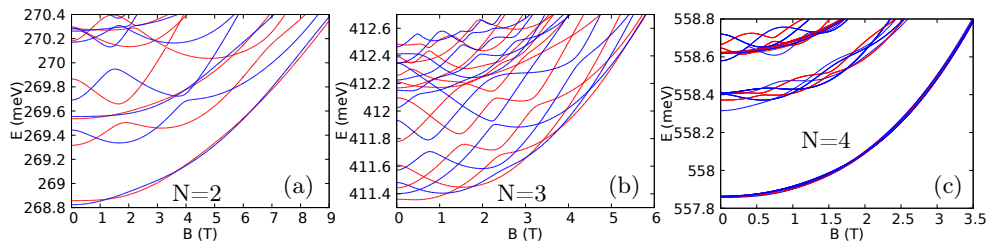

A striking difference in the energy spectra states with or without the Wigner molecule
in the ground state 
is revealed for the circular potential once the spin Zeeman interaction is removed.
This is illustrated in Fig. \ref{noz} which reproduces the data from Fig. \ref{scirc} for $g=0$.
The systems with Wigner form of the charge density,
i.e. the single-electron islands in the charge density),
 $N=2$ (Fig. \ref{noz}(a)) and $N=4$ (Fig. \ref{noz}(c)) contain a nearly degenerate ground state with energy levels
of different parity that interlace in increasing $B$. The separation of the electron charges
in the separate, weakly coupled, maxima produces a small energy difference due to the parity,
which is similar to the nearly degenerate ground state for identical, weakly coupled quantum
dots. The single electron islands are formed by a relatively strong electron-electron interaction 
and the weakness of the electron tunnelling between the islands can be deduced from the 
charge density plots of Fig. \ref{sd}(a,b) and Fig. \ref{sd}(e,f) for $N=2$ and $N=4$.
In both cases, the bunch of energy levels of the ground state is separated by a distinct energy gap 
from the excited states.

For $N=3$ [Fig. \ref{6}(b)] the parity of the ground state changes in growing $B$
as for even $N$ but the spacing between the even and odd parity energy levels near the ground state is much larger than for even $N$ and no energy gap  is observed between the nearly degenerate ground state and the rest of the spectrum.

\newpage{\pagestyle{empty}\cleardoublepage}


\begin{thebibliography}{100}
\bibitem{dostajemy} For the circular potential of Fig. 1(a) the optimal 
$(V_0,d)$ parameters for $N=3$, 4 and 5 electrons are equal 
to (4 meV, 35 nm), (8 meV, 35 nm) and (10 meV, 40 nm), respectively.
\end{thebibliography}
\end{document}